\documentclass[10pt, 
superscriptaddress,
frontmatterverbose, 
nofootinbib,
 amsmath,amssymb,
 aps,
prb,twocolumn
]{revtex4-2}

\usepackage{graphicx}% Include figure files
\graphicspath{ {images/}}
\usepackage{dcolumn}% Align table columns on decimal point
\usepackage{bm}% bold math
\usepackage[	
citecolor=blue,
colorlinks,
linkcolor=blue,
urlcolor=blue,
]{hyperref}
\usepackage{braket}
\usepackage{multirow}
\usepackage{appendix}
\usepackage{wasysym}
\usepackage{slashed}
\usepackage{xcolor}
\usepackage{subfigure}
\usepackage{soul}
\usepackage{dutchcal}
\usepackage{calligra}
\usepackage{cancel}

\newcommand{\mcl}[1]{\mathcal{#1}}
\newcommand{\mbf}[1]{\mathbf{#1}}
\newcommand{\mbb}[1]{\mathbb{#1}}

\begin{document}

\title{Superconductivity in Spin-Orbit coupled \texorpdfstring{ $SU(8)$}{}  Dirac Fermions on Honeycomb lattice}
 \author{Ankush Chaubey}
 \email{ankush.chaubey@icts.res.in}
 \affiliation{International Centre for Theoretical Sciences, Tata Institute of Fundamental Research, Bengaluru 560089, India.}
 \author{Basudeb Mondal}
 \email{basudeb.phys@gmail.com}
 \affiliation{Department of Physics, Hong Kong University of Science and Technology, Clear Water Bay, Hong Kong, China.}
 \author{Vijay B. Shenoy}
 \email{shenoy@iisc.ac.in}
 \affiliation{Centre for Condensed Matter Theory, Department of Physics, Indian Institute of Science, Bengaluru 560012, India.}
 \author{Subhro Bhattacharjee}
 \email{subhro@icts.res.in}
 \affiliation{International Centre for Theoretical Sciences, Tata Institute of Fundamental Research, Bengaluru 560089, India.}
\date{\today}

\begin{abstract}
We study superconducting (SC) phases that are naturally proximate to a spin-orbit coupled $SU(8)$ Dirac semi-metal on a honeycomb lattice. This system, which offers enhanced low-energy symmetries, presents an interesting platform for realising unconventional superconductivity in $j=3/2$ electrons. In particular, we find 72 superconducting charge-$2e$ fermion bilinears which, under classification of microscopic symmetries, lead to 12 different SCs -- {\it four} singlets, {\it two} doublets, and {\it six} triplets -- 7 of them are gapped and 5 are symmetry-protected nodal SCs. The strong spin-orbit coupling leads to locking of the spin of the Cooper pairs with real-space direction -- as is evident from the structure of the Cooper pair wave-functions -- leading to unusual non-unitary superconductors (even singlets), and with finite momentum pairing (for the triplets). This results, in many cases, in the magnitude of multiple pairing gaps being intricately dependent on the direction of the SC order-parameter. The present classification of SCs along with normal phases (\href{https://journals.aps.org/prb/abstract/10.1103/PhysRevB.108.245106}{Phys. Rev. B 108, 245106 (2023)}) provides the complete list of naturally occurring phases in the vicinity of such a SU(8) Dirac semi-metal. This study allows for understanding the global phase diagram of such systems, stimulating further experimental work on candidate materials such as metallic halides (MX$_3$ with M=Zr, Hf, and X=Cl, Br). Further, it provides the starting point for the exploration of unconventional phase transitions in such systems.  
\end{abstract}
%%%%%%%%%%%%%%%%%%%%

\maketitle
%%%%%%%%%%%%%%%%%%%%

\section{Introduction}

Superconductivity in Dirac systems (DSC)~\cite{boyack2021quantum,PhysRevLett.100.246808,PhysRevB.71.184509,PhysRevLett.86.4382,PhysRevLett.102.109701,PhysRevLett.97.146401,PhysRevB.82.035429} is different from their conventional weak-coupling BCS counterparts (those arising from the instability of a Fermi surface~\cite {tinkham2004introduction}). Starting with the fact that the superconductivity in Dirac semi-metals occur at finite interaction strength~\cite{wilson2024discretemodelgellmannmatrices, PhysRevLett.100.246808,PhysRevB.71.184509,PhysRevLett.86.4382,PhysRevLett.102.109701,PhysRevLett.97.146401,PhysRevB.82.035429} as opposed to the conventional superconductors, the Dirac nature of the underlying fermions clearly distinguishes, among other things, the properties of superconducting junctions involving a DSC~\cite{PhysRevLett.97.067007,PhysRevB.76.184514,PhysRevLett.97.217001,huang2019proximity}, from their BCS counterparts. Further, superconductors realised on the surface of the DSMs of topological insulators harbour a plethora of unconventional properties~\cite{PhysRevB.81.184502,huang2019proximity,PhysRevLett.115.187001,zhao2015emergent}. The nature of the superconductors realised in such Dirac systems is intricately related to the underlying symmetries~\cite{Ryu_graphene_masses}.  Therefore, the possibility of realising unconventional superconductivity in Dirac systems due to non-trivial implementation of microscopic symmetries arising from strong spin-orbit coupling offers an interesting direction of exploration apropos candidate materials~\cite{huang2019proximity,PhysRevLett.115.187001,zhao2015emergent}.

The study of these materials, particularly the metallic halides (MX$_3$ with M=Zr, Hf and X=Cl, Br) indicates~\cite{Masaki_su4, basusu8} that strong spin-orbit coupling (SOC) can lead to a  SU(8) Dirac semi-metal (DSM) in these layered honeycomb lattice materials with active $j=3/2$ electrons at $1/4$th filling. The low-energy description of such a DSM is captured by the  free Dirac Hamiltonian~\cite{basusu8}
\begin{align}
    H_D=-iv_F\int d^2{\bf x}~\boldsymbol{\chi}^\dagger({\bf x})\left[\bar{\boldsymbol{\alpha}}\cdot\boldsymbol{\partial}\right]\boldsymbol{\chi}({\bf x})
    \label{eq_dirac_intro}
\end{align}
where $v_F$ is the Fermi-velocity, ${\boldsymbol{\chi}}({\bf x})$ is a 16-component Dirac spinor and $\bar{\boldsymbol{\alpha}}=(\bar{\alpha}_1,\bar{\alpha}_2)$ are two $(16\times 16)$ anti-commuting Dirac matrices.

The above effective low-energy (IR) free theory has an enlarged global SU(8) symmetry, and the microscopic (UV) symmetries, including time reversal (TR), are embedded non-trivially on the low-energy fields due to the underlying SOC. A related outcome of the strong SOC is the position of the Dirac points in the Brillouin zone (BZ) -- unlike monolayer graphene, where there are two in-equivalent Dirac cones at the BZ corners (the so-called ${\bf K}'$ and {\bf K} points)~\cite{PhysRev.71.622} -- here, there are four Dirac points: one at the BZ centre ($\Gamma$) and three at the middle of BZ boundary ($M$) (see Fig. \ref{fig_BZ}(b)). Under UV symmetries, these four Dirac valleys break up as a singlet $(\Gamma)$ and a triplet $(M_1, M_2, M_3)$ along with locking of the real and spin spaces. This results in a plethora of phases proximate to the DSM obtained by breaking the SU(8) and/or TR symmetry by partially/fully gaping out the Dirac fermions due to condensation of fermion bilinears of the form $\langle\boldsymbol{\chi}^\dagger(\cdots)\boldsymbol{\chi}\rangle\neq 0$ (where $(\cdots)$ is a $16\times 16$ matrix that anti-commute with Dirac matrices $\bar\alpha_1$ and $\bar\alpha_2$ in Eq. \ref{eq_dirac_intro}). In particular, Ref. \cite{basusu8} analyzed 64 such bilinears comprised of the SU(8) singlet and the adjoint multiplet to obtain 24 distinct phases describing various charge/spin density waves as well as symmetry protected topological phases-- all of which conserve the global U(1) symmetry
\begin{align}
    \chi\longmapsto e^{i\theta}\chi,
    \label{eq_u1charge}
\end{align}
related to the electron charge conservation. {Due to the distinction in the position (and number) of the Dirac valleys, the larger IR symmetry and the implementation of the UV symmetries, the resultant phases are quite different compared to monolayer graphene~\cite{Ryu_graphene_masses}.}

In this paper, we {explore the consequences of the above symmetry implementation in the SU(8) DSM to understand the nature of }various superconducting instabilities that are naturally proximate to the SU(8) DSM and can be obtained at finite interaction strengths~\cite{PhysRevLett.100.246808,PhysRevB.71.184509,PhysRevLett.86.4382,PhysRevLett.102.109701,PhysRevLett.97.146401,PhysRevB.82.035429}. Such superconductors (SCs) spontaneously break the above $U(1)$ symmetry via condensation of charge $2e$ Cooper pairs, {\it i.e.},  $\langle\boldsymbol{\chi}^\dagger(\cdots)\boldsymbol{\chi}^\dagger\rangle\neq 0$, where $(\cdots)$ denote an appropriate matrix that decides the pairing symmetry. On generic grounds~\cite{kim2018beyond}, there are 16 different pairing channels for the $j=3/2$ electrons with total angular momentum of each Cooper pair ranging from $J_T=0$ (singlet) to $J_T=3$ (septet) with the even (odd) $J_T$ corresponding to spatial parity even (odd) pairing (see Eq. \ref{eq_SU_2decomp}). We show that the presence of many bands~\cite{anderson1984structure,ramires2022nonunitary,RevModPhys.81.1551,RevModPhys.74.235,PhysRevB.94.174513,Ueda_superconductivity,volovik1985superconducting,yarzhemsky1992space} (resulting in different flavours of Dirac fermions) and spin-orbitally locked higher ($j=3/2$) representation opens up the possibility to realise a new set of unconventional SCs via specific symmetry allowed combination of the larger number of available pairing channels for $j=3/2$ electrons~\cite{kim2018beyond} and characterize them systematically using space-group representations~\cite{yarzhemsky1992space,volovik1985superconducting,Ueda_superconductivity,PhysRevB.32.2935,PhysRevB.39.4145,Ryu_graphene_masses}. The present work, along with the study of the normal phases discussed in Ref. \cite{basusu8}, provides a comprehensive catalog of phases proximate to SOC induced SU(8) DSM.

Due to the intertwining of the spin and real spaces by SOC, the irreducible representations of the superconducting order parameters have a mixed structure that generically results in the direction dependence of the angular momentum of the Cooper pairs. This locking of angular momentum with direction naturally gives rise to non-unitary SCs~\cite{ramires2022nonunitary} with possible finite momentum pairing in the absence of magnetic order or external magnetic field. Indeed, the non-trivial interplay of the higher-dimensional spin representation and the superconducting pairing symmetry is central to the rich variety of SCs that can be realised in this system. 

Our classification shows that there are 72 superconducting fermion bilinears that form real and imaginary components of 36 pairing amplitudes. This gives rise to 12 spin-orbit coupled SCs that are distinct from each other under microscopic symmetries -- {\it four} singlets, {\it two} doublets and {\it six} triplets. Most of the SCs -- including several singlet ones--  are non-unitary and hence show a multi-gap structure. Five of these are nodal SCs (the smallest gap is zero), which are protected by some subgroup of the $SU(8)$. For the other seven, we generically obtain gapped superconductivity except in some cases (such as the odd parity doublets, $\mcl{E}_u$, in Sec. \ref{sec_gapped_doublets} and $\mcl{T}_g$ and $\mcl{T}_{u}$ triplets in Sec. \ref{sec_tripletsc}), where the smaller of the two gaps collapses on sub-space of the superconducting order-parameter manifold, giving rise to nodal SCs. The SOC-mediated interlocking of the spin and real spaces is manifest in the structure of the Cooper-pair wave functions of the respective SCs, which also results in multi-gap structures and possible Leggett modes~\cite{leggett1966number} -- even for the singlets due to the higher $j=3/2$ representation and SOC. Another reflection of the spin-orbital interlocking is seen in the variation of the multiple gaps as a function of the SC order-parameter space, including the collapse of the gap at isolated points or on extended order-parameter subspaces (only for triplets). Further, while the singlet and the doublet SCs result from zero momentum pairing, this is not the case for the triplets, which then give rise to pair density-wave (PDW) SCs.

The remainder of the paper is organised as follows. We start with an overview of the different superconducting phases obtained in this work in Section \ref{sec_overview}. Sec.~\ref{sec_su(8)_dirac_fermions} summarises the SU(8) Dirac theory of SOC fermions, which is then used in sec.~\ref{sec_nambubasis} to cast the theory in the Nambu formulation to aid the classification of superconducting masses. Sec.~\ref{sec_classification} discusses the classification of various masses under microscopic symmetries, which results in 12 SC phases. The four singlet phases are discussed in detail in Sec.~\ref{sec_singletsc}, the two doublet phases in Sec.~\ref{sec_doubletsc}, and the six triples are detailed in Sec.~\ref{sec_tripletsc}. The paper is concluded in Sec.~\ref{sec_summary} with a discussion and perspective on the results. Appendices \ref{appen_global}-\ref{sec_su2global}, in addition to containing useful technical details, also discusses a Majorana representation (Appendix \ref{appen_majorana}) and the study of a honeycomb system with SOC $j=1/2$ states (Appendix~\ref{sec_su2global}) to be contrasted with richer physics obtained in the $j=3/2$ system discussed in the main text. 
%%%%%%%%%%%%%%%%%%%%%%%%%%%%%%%%%%%

\section{Overview of results}
\label{sec_overview}

The $SU(8)$ DSM described by Eq. \ref{free dirac theory} (or Eq. \ref{eq_dirac_intro}) is obtained in indirect hopping model of $j=3/2$ electrons on honeycomb lattice at $1/4$th filling due to the two valleys leading to the 16-component Dirac fermions, $\boldsymbol{\chi}$ (Eq. \ref{Eq_Dirac_spinor_Xi}),  in local basis or equivalently four 4-component Dirac fermions, $\chi_g$ (Eq. \ref{eq_globalspinor}), in global basis, as shown in Fig. \ref{fig_BZ}. To analyze all the phases -- both normal and superconducting -- on equal footing, it is useful to describe the theory in terms of the Majorana representation whence the free Dirac theory has a manifest $SO(16)$ symmetry (Appendix \ref{appen_majorana}), such that there are 136 fermion bilinears corresponding to masses for the fermions. These correspond to one SO(16) singlet and 135 symmetric rank-2 tensors. Out of the 136 bilinears, 64 correspond to normal (non-superconducting)  masses that correspond to 24 phases studied recently in Ref. \cite{basusu8}. These break various microscopic symmetries and correspond to different symmetry broken as well as SPT phases. While the remaining 72 bilinears, the topic of the present work, correspond to various superconducting phases proximate to the SU(8) DSM.

\begin{figure}%[h!]
\includegraphics[scale=.32]{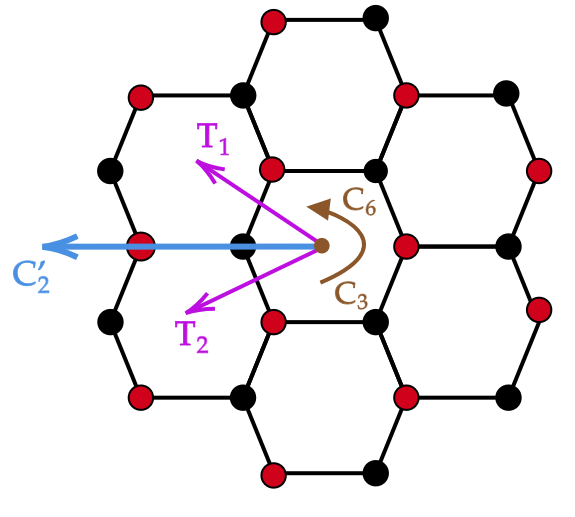}
\caption{Lattice transformation: Here, the honeycomb lattice lies in a plane perpendicular to [111] planes in Cartesian coordinates (also discussed in Ref.~\cite{basusu8, Nanda_2020}). Transformation ${\bf C_2^{\prime}}$ is the rotation by $\pi$ amount about the axis shown in cyan. $\bf T_1, T_2$ are translations by honeycomb lattice vectors. ${\bf C_6 (C_3)}$ corresponds to rotation by $2\pi/6(2\pi/3)$ about the center of Honeycomb. $\sigma_h$ is the reflection about the Honeycomb plane so that the combined transformation ${\bf S_6 = C_6\sigma_h}$ is the symmetry of the lattice.}
\label{fig_transformation}
\end{figure}

These 72 bilinears correspond to the real and imaginary components of 36 superconducting pairing amplitudes, which, under microscopic symmetries, divide up to give 12 distinct SCs comprising up of four singlets ($\mcl{A})$, four doublets ($\mcl{E}$), and eight triplets ($\mcl{T}$), and cataloged in Tables \ref{tab_singlets}, \ref{tab_doublets} and \ref{tab_triplets}. The details of the symmetry transformations for the superconducting bilinears are given in Appendix \ref{app:C}. The further subdivision of the different Irreps is given by 
\begin{itemize}
    \item 4-singlets : $\mcl{A}_{1g}^{I},\mcl{A}_{1g}^{II}, {\color{red}\mcl{A}_{1u}}, {\color{red}\mcl{A}_{2u}}$
    \item 2-doublets : $(3)\mcl{E}_{u}, {\color{red}\mcl{E}_{g}}$
    \item 6-triplets : $\mcl{T}_{1g}^{I},\mcl{T}_{1g}^{II}, {\color{red}\mcl{T}_{1g}^{III}}, (2)\mcl{T}_{1u}, {\color{red}\mcl{T}_{2g}}, (2) \mcl{T}_{2u}$.
\end{itemize} 
where, similar to Ref. \cite{basusu8}, we use the subscripts $1(2)$ and $g(u)$ to denote even (odd) under $\pi$-rotation (${\bf C}_2'$) and inversion, {\bf I}, respectively (see Fig. \ref{fig_transformation}). Note that the number within $()$ indicates the multiplicity of the Irreps. Particularly for doublet and triplet SCs, some of the Irreps with higher multiplicities (namely, $\mcl{E}_{u}$, $\mcl{T}_{1u}$ and $\mcl{T}_{2u}$) do not give rise to distinct SCs. Out of the 12 different SCs listed above, 7(denoted in black colour) are gapped and 5 (in red) are gapless (nodal). Further, while the singlets are generically time-reversal (TR) invariant, the doublets and the triplets generically break it spontaneously. The above list of SCs realised in the $j=3/2$ system, as described below, is much richer than their $j=1/2$ counterparts -- briefly summarised in Appendix \ref{sec_su2global}. Indeed, the $j=1/2$ system realises only five superconducting phases that have analogues in the $j=3/2$ system.

The real and imaginary components of the $j=3/2$ superconducting bilinears have a generic form 
\begin{align}
    \langle\chi^\dagger (\Sigma_\beta\tau_\gamma\sigma_\delta)(\chi^{\dagger})^T\rangle +{\rm h.c.}, ~-i\langle\chi^\dagger(\Sigma_\beta\tau_\gamma\sigma_\delta)(\chi^{\dagger})^T\rangle+{\rm h.c.}
\end{align}
(as discussed in Eq. \ref{Mass_in_dirac_spinor}) where $\Sigma_\beta$ are $4\times4$ matrices that act in the SU(4) flavour space, while, as mentioned before, $\tau_\gamma$ and $\sigma_\delta$ are Pauli matrices that act in the Dirac valley and band spaces, respectively. 

\begin{table}
\begin{tabular}{|c|c|c|c|c|}\hline
 & Broken  & & \\ 
Irrep & Microscopic  & $\Delta\cdot\Delta^\dagger$ & Brief Description\\ 
 & Symmetries  &  & \\ \hline\hline
     &  &  & Gapped. $j=3/2$ singlet.\\ 
  $\mathcal{A}^{I}_{1g}$ & None & U & On-site $s$-wave pairing.\\ 
  & & & Possible inversion-odd Leggett\\
  & & & mode. (Sec. \ref{sec_a1g_singlets})\\\hline
  $\mathcal{A}^{II}_{1g}$   & None &  & Double gap. $j=3/2$ singlet.\\ 
    & & N. U. & Extended $s$-wave (NNN pairing).\\ 
    & &  &  (Sec. \ref{sec_a1g_singlets})\\ \hline
  $\mathcal{A}_{1u}$   & None &  & Both gapped and gapless (nodal)\\
  && N. U. & branches of excitations with\\
  &&& nodes at $\Gamma$-point protected by\\
  &&&  $SO(4)$ symmetry. NNN pairing. \\
  &&& Cooper pair wave function\\
  &&& symmetric (anti-symmetric) in\\
  &&&  space (spin). (Sec. \ref{sec_gapless_singlets})\\ \hline
  $\mathcal{A}_{2u}$   & None & & Similar to $\mcl{A}_{1u}$ but distinct\\ 
  &&N. U. & Irrep. (Sec. \ref{sec_gapless_singlets})\\\hline
\end{tabular}
\caption{Summary of the singlet SCs discussed in Sec. \ref{sec_singletsc}. NNN refers to next nearest neighbor. $g (u)$ denotes the Irrep is even (odd) under parity. U (N.U.) stands for unitary (non-unitary) pairing.}
\label{tab_singlets}
\end{table}

Out of the four singlet SCs listed in Table \ref{tab_singlets}, two belong to $\mathcal{A}_{1g}$, and the other two respectively to $\mcl{A}_{1u}$ and $\mcl{A}_{2u}$, comprising four different SCs. Notably, the first two even-parity pairings lead to distinct gapped SCs, even though they correspond to the same Irrep, $\mcl{A}_{1g}$, and both are $j=3/2$ singlets. They respectively correspond to on-site $s$-wave and next nearest neighbour (NNN) $s$-wave SCs, with the former being unitary and the latter being a rather rare example of a non-unitary singlet SC with a double gap excitation spectrum (Fig. \ref{symmetric gapped singlet spectrum}). The distinction arises due to the SU(4) flavour space structure. Interestingly, the unitary $\mcl{A}_{1g}$ SC can be viewed as coexisting two $s$-wave condensates-- one made up of pairing between $J_z=\pm 3/2$ orbitals and the other the same between the $J_z=\pm 1/2$ orbitals (Eq. \ref{eq_aigunitarysinglet}) with a relative (inversion) symmetry enforced $\pi$-phase difference between the two condensates. This leads to a possibility of a gapped Leggett mode~\cite{leggett1966number} that is odd under inversion. The two odd parity SCs, on the other hand, are both non-unitary with the smaller of the two gaps being zero, leading to nodal Bogoliubov excitations (Fig. \ref{Fig_Gapless_singlet}) that are protected by an $SO(4)$ subgroup. The mean field lattice Hamiltonians for the unitary (non-unitary) SCs have on-site (next nearest neighbour) pairings. Finally, even though none of the singlets break any microscopic symmetries, a linear superposition of any two of them generically breaks one or more symmetries.

\begin{table}
\begin{tabular}{|c|c|c|c|c|}\hline
 & Broken  & \\ 
Irrep& Microscopic  & Brief Description\\ 
 & Symmetries  & \\ \hline\hline
 && Staggered on-site pairing (fig. \\
      & &\ref{Fig_lattice_model_E_u}) for $\mathcal{E}_u^I$ and NNN pairing \\
&&      for $\mathcal{E}_u^{II}$ and $\mathcal{E}_u^{III}$. Double gap. \\
$\mathcal{E}_u^I$&& $\mathcal{E}^{I}_{u}$ reduces to a single gap in the   \\
  $\mathcal{E}^{II}_{u}$   & $({\bf C_2', C_3, S_6},\sigma_d)$ & TRI sub-manifold. For all  \\ 
  $\mathcal{E}^{III}_{u}$&& doublets on isolated points   \\ 
  && in the TRB manifold,    \\ 
  && the lower of the two gaps \\
  && become zero (Eq. \ref{eq_isonode}) \\ 
  && giving rise to a nodal SC.  \\ 
  && Order parameter manifold\\ 
  && is $(S^1\times S^2)/Z_2$ which  \\
  &&  reduces to $(S^1\times S^1)/Z_2$ \\
  &&  in the TRI sub-manifold.\\
  && (Sec. \ref{sec_gapped_doublets})\\ \hline
  && Anisotropic NNN pairing.\\
  && (Fig. \ref{Fig_lattice_model_E_u_12}). Three gaped branches \\
  && and one gapless branch of\\
  && excitation, each $4$-fold\\
  $\mathcal{E}_{g}$   & $({\bf C_2', C_3, S_6},,\sigma_d)$ & degenerate (Fig. \ref{fig_gapless_doublet_spectrum}(a)).\\ 
  &&The nodes are at $\Gamma$-point\\ 
  && and their degeneracy \\
  && changes (Fig. \ref{fig_gapless_doublet_spectrum}(b)). \\
  && (Sec. \ref{sec_doubleteg}).\\ \hline
\end{tabular}
\caption{Summary of the doublet SCs discussed in Sec. \ref{sec_doubletsc}. TRI (TRB) stands for TR invariance (TR breaking). Note that doublets are generically non-unitary and break TR.}
\label{tab_doublets}
\end{table}

There are four doublets corresponding to two distinct non-unitary SCs summarised in Table \ref{tab_doublets}-- all with spin anti-symmetric pairing. The three odd parity doublet SCs (denoted as $\mcl{E}^I_u, \mcl{E}^{II}_u$ and $\mcl{E}^{III}_u$ in Table \ref{tab_doublets}) obtained from a combination of $J_T=2 ~(m_T=0,\pm 2)$ angular momentum states,  are adiabatically connected without breaking any microscopic symmetries (appendix \ref{sec_E_u_doublet_II_III}). Together, they give rise to a single non-unitary SC with two pairing gaps. On the TRI sub-manifold, the two gaps may coincide for some combination of pairings. At other combinations of pairing, generally breaking TR, the lower of the two gaps may vanish at isolated points on the order-parameter manifold giving rise to nodal SCs (Eq. \ref{eq_isonode}) making them rather unconventional where the direction of the superconducting order parameter can be changed in the order parameter manifold to obtain a gapped or a nodal SC. The truly gapless (nodal) SC is obtained from the pairing of the ($J_T=0$) spin singlet; on the other hand, in the parity even $\mcl{E}_g$ doublet, where the superconductivity originates from anisotropic NNN pairing as shown in Fig. \ref{Fig_lattice_model_E_u_12}. The excitation spectrum in this case (Fig. \ref{fig_gapless_doublet_spectrum}(a)) generically contains three four-fold degenerate gapped branches and one gapless nodal branch. The origin of the nodal branch can be traced to the Dirac node at the centre of the BZ in the global basis (Appendix  \ref{sec_symmetry_analysis_gapless_modes}) that is protected by a SO(4) subgroup. The order parameters for all the doublets generically span a $(S^1\times S^2)/Z_2$ manifold, which reduces to $(S^1\times S^1)/Z_2$ within the TRI sub-manifold. Lattice symmetries do allow (see for example Eqs. \ref{eq_E_g_doublet_secondary_OP} and \ref{eq_anisotropysecondaryeg}) energetic distinction between the TRB and TRI sub-manifolds leading to the possibility of obtaining unconventional vortices in the TRI manifold since $\pi_1\left(\frac{S^1\times S^1}{Z_2}\right)=Z\times Z$~\cite{PhysRevLett.87.080401,Ueda_2014,subrotopaper}. 

\begin{table}
\begin{tabular}{|c|c|c|c|}\hline
 & Broken  & \\ 
Irrep& Microscopic  & Brief Description\\ 
 & Symmetries  &   \\ \hline\hline

 && Two gaps which become equal\\
 && in the TRI sub-manifold. However,\\
   $\mathcal{T}^{I}_{1g}$   & $(\bf T_1, T_2, C_2',$ & on sections of TR breaking   \\  
   & ${\bf  C_3 ,S_6},\boldsymbol{\sigma}_d)$& sub-manifold, the smallest gap    \\ 
   && goes to zero leading to a nodal SC.\\
   && Onsite finite momentum pairing\\
   && with {\it stripy} pattern (fig. \ref{fig_Triplet_lattice_model_T_1G_onsite}).\\
   && (Sec. \ref{sec_t_1g}).\\ \hline
   && Very similar to the triplet above\\
   && but arises from flavour symmetric\\
   $\mathcal{T}^{II}_{1g}$&$(\bf T_1, T_2, C_2',$&pairing and is a distinct SC. \\
      & ${\bf  C_3 ,S_6},\sigma_d)$  & The {\it stripy} pairing is NNN (Fig. \ref{fig_Triplet_lattice_model_T_1G_2_onsite}).\\
   & & (Sec. \ref{sec_t_1g}).\\ \hline
    &   & Nodal SC with NNN finite \\ 
   & & momentum pairing (Fig. \ref{fig_Triplet_lattice_model_T_1g}(b)).   \\ 
  $\mathcal{T}^{III}_{1g}$ &$(\bf T_1, T_2, C_2',$&   Dirac nodes at the $\Gamma$-point.   \\
   &${\bf  C_3 ,S_6})$& The degeneracy of the nodes change \\
   && with the direction of the \\ 
   && order parameter (Fig. \ref{triplet parameter space}).\\ 
   && (Sec. \ref{sec_t_1g})\\\hline
   && Multiply gapped (Fig. \ref{fig_Triplet_lattice_model_T_2g}(a)) along\\
&& with gapless nodes at M-points\\
   $\mathcal{T}_{2g}$   & $(\bf T_1, T_2, C_2',$& (Eq. \ref{Eq_triplet_in_global_basis}) leading to a PDW SC obt-\\ 
   &${\bf  C_3 ,S_6},\sigma_d)$& ained from NNN pairing (Fig. \ref{fig_Triplet_lattice_model_T_2g}(b)).\\
   && (Sec. \ref{sec_t2g}).\\ \hline
   && Multiple gaps with the lowest gap\\
   && reduce to zero on a  \\
   $\mathcal{T}^{I}_{2u}$ & $(\bf T_1, T_2, C_2',$ & sub-manifold of the order-parameter\\
   &&space giving rise to a nodal SC.\\
   && {The zero modes correspond to linear} \\
   && {combination of Dirac modes at }\\
   $\mathcal{T}^{II}_{2u}$ & ${\bf  C_3 ,S_6},\sigma_d)$ &   different valleys. The two triplets can \\
   &&  be adiabatically connected without  \\
   && breaking any further microscopic \\
   && symmetries and are not distinct SCs. \\
   && NNN finite momentum pairing \\
   && (Fig. \ref{fig_Triplet_lattice_model}). (Sec. \ref{sec_t2u}).\\\hline
   $\mathcal{T}^{I}_{1u}$   & $(\bf T_1, T_2, C_2',$ & Spectrum is similar to the above.\\ 
$\mathcal{T}^{II}_{1u}$ & ${\bf  C_3 ,S_6},\sigma_d)$& Two triplets represent the same SC. \\
&& NNN finite momentum pairing\\
&& (Fig. \ref{fig_Triplet_lattice_model}).\\
&& (Sec.~\ref{sec_t1u}).\\
\hline
\end{tabular}
\caption{Summary of the Triplet SCs. Like the doublets, all the triplets SCs are generically non-unitary and break TR. Finally, generically the triplets have a finite momentum pairing resulting in breaking of translation symmetries in the resultant pair density wave (PDW) SCs.}
\label{tab_triplets}
\end{table}

Finally, there are eight triplets giving rise to six distinct SCs summarised in Table \ref{tab_triplets}. These SCs generically correspond to finite momentum pairing and hence break translation symmetry in addition to other point group symmetries, as well as TR. Hence, they generically correspond to pair-density-wave (PDW) SCs~\cite{PhysRevLett.88.117001,PhysRevLett.99.127003,agterberg2020physics} with the order-parameter manifold being $(S^1\times CP^2)/Z_2$~\cite{massey_1973,kuiper_1974} which reduces to $(S^1\times S^2)/Z_2$ in the TR invariant sub-space. The associated Cooper-pair wave functions are formed by a mixture of $J_T=1, 2,$ and $3$ sectors that are intricately interlocked with the real space directions. Although the three $\mcl{T}_{1g}$ triplets (denoted by $\mathcal{T}_{1g}^{I}, \mathcal{T}_{1g}^{II}$ and $\mathcal{T}_{1g}^{III}$ in Table \ref{tab_triplets}) belong to the same Irrep, they have somewhat different properties. Further, the TR invariant sub-manifolds of these three SCs cannot be adiabatically connected without breaking further lattice symmetries and/or TR. Finally, while the two of them are gapped, the third one is a nodal SC. Among the other three, the $\mathcal{T}_{2g}$ triplet is a nodal SC, while the other two ($\mathcal{T}_{1u}$ and $\mathcal{T}_{2u}$) are generically gapped, except at sub-parts of the order-parameter manifold when the smallest of the multiple gaps closes. 

%%%%%%%%%%%%%%%%%%%

\section{SU(8) Dirac fermions on honeycomb lattice.}\label{sec_su(8)_dirac_fermions}

\begin{figure*}
    \centering
    \includegraphics[scale=.3]{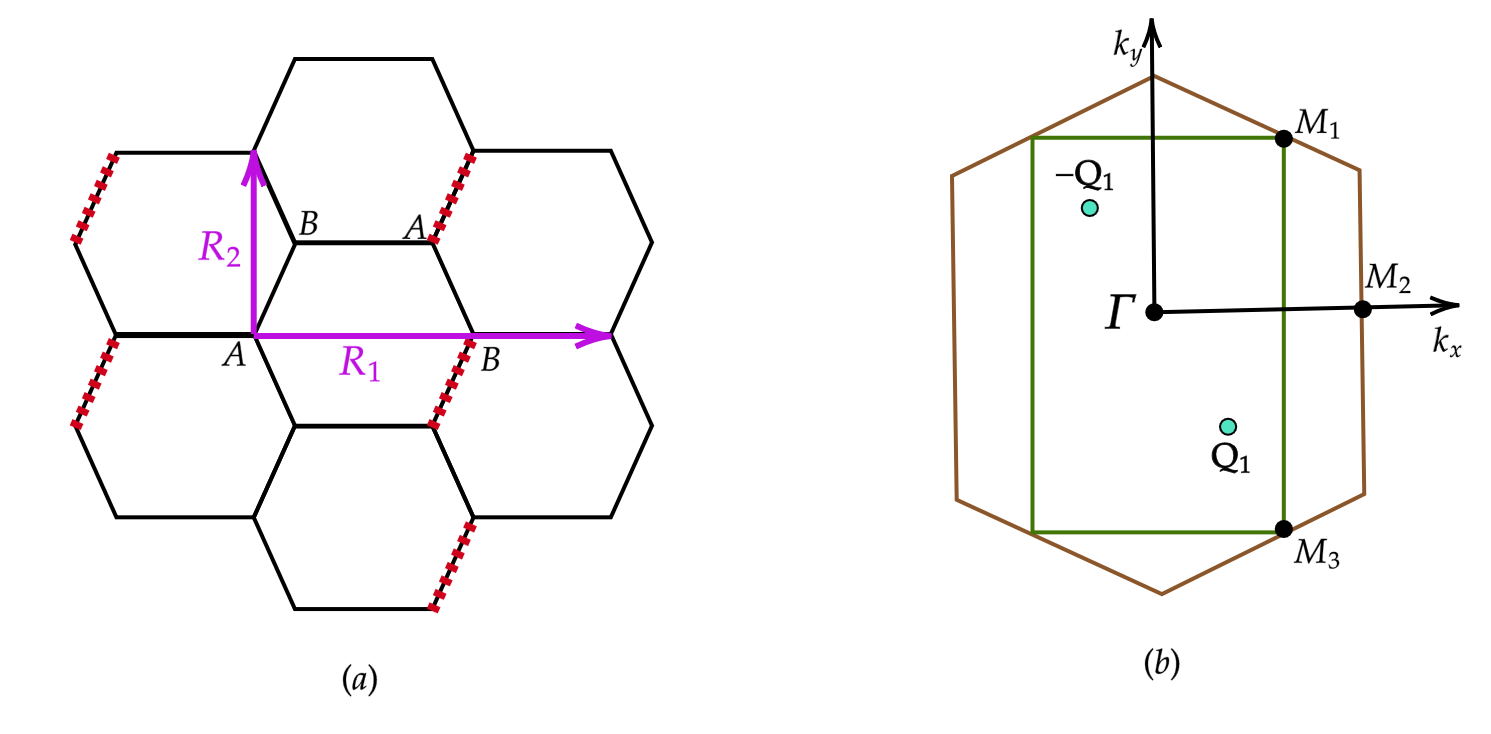}
    \caption{(a) Honeycomb lattice with a particular realization $\eta_{\bf rr^{\prime}}=+(-1)$ on black(red) bonds that implements $\pi$-flux through each honeycomb plaquette. There are two choice of unit cell --(1) the {\it global basis} that uses the primitive 2-site unit cell of the honeycomb lattice consisting of (A, B) (Eq. \ref{eq_masaki_ham}), or (2) the {\it local basis} that has a 4-site unit cell (Eq. \ref{eq_phi_definition} with lattice vectors $(\bf R_1, R_2)$). (b) The primitive honeycomb BZ (Brown)  with Dirac cones at $\Gamma, M_1, M_2, M_3$ points. For the local basis with a 4-site unit cell as shown in (a), we have a rectangular BZ (green) with Dirac cones as $\pm {\bf Q}_1=\pm[\pi/6, -\pi/2\sqrt{3}]$ points~\cite{basusu8}.  }
    \label{fig_BZ}
\end{figure*} 

We start by summarizing the essential microscopic details from Ref.~\cite{basusu8} (using the same notations) leading to the low-energy Dirac theory of Eq.~\ref{eq_dirac_intro},  which is realised in $j=3/2$ SOC coupled honeycomb systems with $d^1$ electronic configuration in the nearest neighbor in-direct hopping model given by~\cite{basusu8,Masaki_su4,PhysRevResearch.5.043219}
	\begin{eqnarray}\label{eq_masaki_ham}
		H = -\frac{t}{\sqrt{3}}\sum_{\braket{\mbf{rr'}}} \Psi^\dagger(\mbf{r}) \mcl{U}^{global}_{\mbf{rr'}}\Psi(\mbf{r'}) + {\rm h.c.},
	\end{eqnarray}
where   
\begin{align}
    \Psi(\mbf{r}) = [\Psi_{1/2},\Psi_{-1/2},\Psi_{3/2},\Psi_{-3/2}]^T,
    \label{eq_j3/2orb}
\end{align} is the four-component electron annihilation operator corresponding to the $j=3/2$ orbitals at the lattice site at $\mbf{r}$ and $\mcl{U}^{global}_{\mbf{rr'}}$ are $4\times 4$ hopping matrices. Various lattice symmetries are indicated in Fig. \ref{fig_transformation}. These, along with the time reversal (TR), $\mathbb{T}$ (with $\mathbb{T}^2 = -1$), generate the microscopic (UV) symmetry group of the system.

The directed product of the hopping matrices on the honeycomb plaquette, {\it i.e.},  $\prod_{\braket{\mathbf{rr^{\prime}}\in \hexagon}} \mcl{U}^{global}_{\mathbf{rr^{\prime}}} = -\Sigma_0$ (with $\Sigma_0$ being the $4\times 4$ identity matrix). This implies that the electrons have a SU(4) symmetry, which can be made manifest via a site-dependent unitary transformation on the atomic orbitals via 
\begin{eqnarray}\label{eq_phi_definition}
		\phi'(\mbf{r}) = g(\mbf{r})^\dagger \Psi(\mbf{r}),
	\end{eqnarray}
where $g^{\dagger}({\bf r})~\mcl{U}^{global}_{\mathbf{rr^{\prime}}}~ g({\bf r^{\prime}})= \eta_{\bf r r^{\prime}}\Sigma_0$, with $g(\mbf{r})$ being a $4\times 4$ unitary matrix at the lattice site $\mbf{r}$, $\phi'(\mbf{r})$ being fermion orbitals (flavours) in the {\it local basis}, and $\eta_{\bf rr'}=\pm 1$ being an Ising gauge field. In this local basis, the problem reduces to four decoupled flavours (from the underlying SU(4) symmetry) of fermions hopping on the honeycomb lattice in $\pi$-flux, whence, choosing a suitable magnetic unit-cell ({Fig.~\ref{fig_BZ}(a)}), the band Hamiltonian is diagonalized to obtain two $4$-fold (flavour) degenerate Dirac nodes/valleys at momenta $\pm {\bf Q}_1$, as shown in Fig. \ref{fig_BZ}(b). Expanding about the two valleys ($\pm$), we get the continuum Dirac Hamiltonian given by 
{
 \begin{eqnarray}\label{free dirac theory}
    &H_{D} &=  -i v_F \sum_{f=1}^4\int d^2\mbf{x}\,\,
    \chi_{f}^{\dagger}\left( \mbf{x}\right) \left[ \alpha_1\partial_1+\alpha_2\partial_2 \right]\chi_{f}\left( \mbf{x}\right),
\end{eqnarray}
where there are four flavours (for $f=1,2,3,4$) of  4-component Dirac spinor, $\chi_{f}$, {\it i.e.},
 \begin{eqnarray}
    &&\chi_f= \left( \chi_{f1+}, \chi_{f2+}, \chi_{f1-}, \chi_{f2-} \right)^T,
    \label{Eq_Dirac_spinor_Xi_f} 
\end{eqnarray}
with the subscript $1(2)$ denoting the two bands that touch at each of the two Dirac valleys ($\pm$), in the local basis, and 
\begin{eqnarray}
    \alpha_1=\tau_3\sigma_1,\qquad \alpha_2=\tau_0\sigma_2, 
    \label{eq_alphamat}
\end{eqnarray}
are the two anti-commuting Dirac matrices with $\tau_i,\sigma_i$ (for $i=0,1,2,3$) being $2\times 2$ Identity and the Pauli matrices that act on the valley and band spaces, respectively.

This can be written in a more compact notation given in Eq. \ref{eq_dirac_intro} with
\begin{align}
    \bar{\alpha}_i=\Sigma_0\alpha_i~~~~~(i=1,2),
\end{align}
and
\begin{equation}\label{Eq_Dirac_spinor_Xi}
     \boldsymbol{\chi}=(\chi_1,\chi_2,\chi_3,\chi_4)^T,
 \end{equation}
 being the 16-component spinor obtained by stacking the different flavours with $\Sigma_0$ acting in the flavour space.

}

The flavour SU(4) symmetry is manifest in the local basis and is generated by the fifteen $\Sigma_i$ matrices (see Ref.~\cite{basusu8} for their explicit form). In addition, the above free Dirac theory has an emergent SU(2) symmetry generated by
\begin{eqnarray}
    \{\tau_3\sigma_0/2,\tau_1\sigma_2/2, \tau_2\sigma_2/2\} \equiv \{\zeta_1,\zeta_2,\zeta_3\}/2, 
    \label{eq_chiralsu2}
\end{eqnarray}
such that it (Eq. \ref{eq_dirac_intro}) is invariant under the covering group of $SU(4)\otimes SU(2)$, {\it i.e.}, $SU(8)$ generated by~\cite{basusu8}
\begin{eqnarray}
    \mcl{P}_a = \Sigma_i\zeta_j.
\end{eqnarray}
It is useful to note that as discussed in Ref. \cite{basusu8}, the above low-energy theory can also be derived in the {\it global basis} of $\Psi$ (Eq. \ref{eq_masaki_ham}), using the 2-point honeycomb unit cell (Fig. \ref{fig_BZ}). In this basis, the SU(4) symmetry is not manifest, and we get four bands (each two-fold Kramers degenerate) and have 4 Dirac points at 1/4th filling. The location of the four Dirac points in the global basis is also shown in Fig. \ref{fig_BZ}, with one being located at the primitive BZ centre ($\Gamma$ point) while the other three being located at the three inequivalent mid-points of the BZ boundary-- the so-called $M$ points.  The details are briefly discussed in Appendix \ref{appen_global} for completeness, while further details are given in Ref. \cite{basusu8}. Throughout the paper, we shall mostly work in the local basis unless explicitly stated. However, occasionally we shall refer to the global basis when it proves more insightful.

\section{Superconductivity in Dirac fermions}
\label{sec_nambubasis}

The above free Dirac theory needs to be supplemented with short-range four-fermion interactions amongst the electrons of the generic form 
\begin{equation}\label{eq_four_fermion_interation}
H_{int}=\int d^2\mathbf{x} d^2\mathbf{x'}~ V_{ijkl}~ \chi^{\dagger}_i(\mathbf{x}) \chi^{\dagger}_j(\mathbf{x'}) \chi_k(\mathbf{x}) \chi_l(\mathbf{x'}).
\end{equation}
The Dirac semi-metal is perturbatively stable to such short-range four-fermion interaction~\cite{PhysRevLett.100.246808,PhysRevB.71.184509,PhysRevLett.86.4382,PhysRevLett.102.109701,PhysRevLett.97.146401,PhysRevB.82.035429}, but strong interaction generically leads to condensation of fermion bilinears, {\it i.e.}, 
\begin{equation}\label{condensed_bilinear}
\langle \chi_{f\tau\sigma}^{\dagger}\chi_{f'\tau'\sigma'}\rangle\neq 0 , \,\,\,\text{or}\,\,\,\ \langle\chi_{f\tau\sigma}\chi_{f'\tau'\sigma'}\rangle\neq 0.
 \end{equation} 
While condensation of both these sets of bilinears spontaneously break the symmetries of the free theory as well as possibly other microscopic symmetries with concomitant gaping out of all/fraction of the Dirac fermions, a major classifying difference between them is -- the first set is electronic charge neutral and the second set carry charge $2e$.

The mean field Hamiltonian for both the normal and the superconductors is obtained by decoupling the four-fermion interactions in the different particle-hole and particle-particle channels, respectively, to obtain
 \begin{align}\label{eq_sc_mean_field}
H_{MF}=&H_D \,\,+\,\,\tilde \Delta\int d^2\textbf{x}\,[\boldsymbol{\chi}^{\dagger}\,\,\tilde m\,\,\boldsymbol{\chi}]\,\, \nonumber \\
&+\,\, \Delta \int d^2\mathbf{x}\,\big[{\boldsymbol{\chi}^\dagger}\,\,m\,\,(\boldsymbol{\chi}^{\dagger})^T\big]  +h.c., 
\end{align}
with $H_D$ is given by Eq. \ref{eq_dirac_intro} or Eq. \ref{free dirac theory} and
\begin{eqnarray}
 \tilde \Delta = \langle \boldsymbol{\chi}^{\dagger} \tilde m \boldsymbol{\chi}\rangle \,\, , \,\,
\Delta = \braket{\boldsymbol{\chi}^T~m^\dagger~\boldsymbol{\chi}}, 
\end{eqnarray}
where $\tilde m$ is a 16-dimensional Hermitian matrix and $m$ is a 16-dimensional anti-symmetric matrix (see below). Further details for diagonalising Eq. \ref{eq_sc_mean_field} are given in Appendix \ref{appen_bdg}.
%%%%%%%%%%%%%%%%%%

\subsection{Hamiltonian in Nambu basis}
To study these orders, it is useful to use the Nambu spinor representation~\cite{Ryu_graphene_masses}. We define the 32-component Nambu spinor as
\begin{equation}\label{eq:Nambu spinor in real space}
\chi_N(\mathbf{x})=[\chi(\mathbf{x}),{\chi^{\dagger}}^T(\mathbf{x})],
\end{equation}
or equivalently in momentum space
\begin{equation}\label{eq:Nambu spinor in momentum space}
\chi_N(\mathbf{q})=[\chi(\mathbf{q}),{\chi^{\dagger}}^T(-\mathbf{q})].
\end{equation}
Note that this is slightly different from the conventional Nambu spinor (Eq. \ref{eq_nambuspinor}), where the TR partners are stacked~\cite{PhysRevLett.97.067007,PhysRevLett.97.217001}, as discussed in Appendix \ref{appen_bdg}. However, we find the representation in Eq. \ref{eq:Nambu spinor in real space} more convenient to implement the symmetries in the present case. The transformation between the two bases is straightforward as discussed in Appendix \ref{appen_bdg}.
The free Dirac Hamiltonian in Eq.~\ref{eq_dirac_intro} can then be written, in momentum space, as
\begin{equation}\label{eq:Hamiltonian in nambu basis}
H_{D}=\frac{v_F}{2}\int\text{d}^2\textbf{q}\,\,\,\,\chi^{\dagger}_N(\textbf{q})\Big(\tilde{\alpha}_1\,q_x+\tilde{\alpha}_2\,q_y\Big)\chi_N(\textbf{q}),
\end{equation}
where, for the rest of the calculations, we set $v_F=1$ and 
\begin{eqnarray}\label{Eq_alpha_tilda_matrices}
\tilde{\alpha}_1
=\text{$\mu$}_0\bar{\text{$\alpha$}}_1,~~~~~
\tilde{\alpha}_2
= \text{$\mu_3$}\bar{\text{$\alpha$}}_2.
\end{eqnarray}
Here, $\mu_0$ is $2\times 2$ the identity matrix, and $\mu_i$ for $i=1,2,3$ are the Pauli matrices that act on the Nambu sector. 

The mean-field Hamiltonian in Eq.~\ref{eq_sc_mean_field} in the Nambu basis has the following form,
\begin{eqnarray}
H_{MF} = H_{D} + \int d^2\mathbf{x}~\big[\chi_N^\dagger~M_{general}~\chi_N\big] .
\label{eq:general term in Nambu basis}
\end{eqnarray}

Here, $H_{D}$ is given by Eq.~\ref{eq:Hamiltonian in nambu basis}  and $M_{general}$ is $32$-dimensional Hermitian matrix which has the following generic form
\begin{equation}\label{eqn:15}
    M_{general} = \begin{bmatrix}
    M_{ph} & M_{pp}\\
    M_{pp}^{\dagger} & -M^{T}_{ph}
    \end{bmatrix},
\end{equation}
where $M_{ph} (= M_{ph}^\dagger)$ and  $M_{pp}(=-M_{pp}^T)$  are $16\times 16$ matrices respectively in the particle-hole (normal) and particle-particle (superconducting) sectors. Note that the anti-symmetry of $M_{pp}$ is due to fermionic statistics and ensuing Pauli exclusion principle~\cite{Ueda_superconductivity} such that the mass matrix obeys
 \begin{equation}\label{eq:constraint on Nambu hamiltonian}
     M_{general}=-(\mu_1~M_{general}~\mu_1)^T.
     \end{equation}
Thus, to obtain the different masses, we consider all possible independent matrices $M_{general}$ which anti-commute with the matrices $\tilde\alpha_1$ and $\tilde \alpha_2$, and follow the above constraint (Eq.~\ref{eq:constraint on Nambu hamiltonian}). To this end, we write
\begin{align}
    M_{general}=\mathcal{M}_{\alpha\beta\gamma\delta}X_{\alpha\beta\gamma\delta},
\end{align}
where $\mathcal{M}$ is the amplitude and 
\begin{equation}\label{form_of_mass_term}
    X_{\alpha \beta \gamma \delta} = \mu_{\alpha}\Sigma_{\beta} \tau_{\gamma}\sigma_{\delta},
\end{equation}
forms the set of linearly independent $32$-dimensional matrices formed from the matrices acting in the different spaces mentioned above. There are possible 1024 such $X_{{\alpha \beta \gamma \delta}}$ matrices, of which 256 anticommute with $\tilde{\alpha_1}$ and $\tilde{\alpha_2}$ (Eq.~\ref{Eq_alpha_tilda_matrices}), and only 136 of them are allowed by the antisymmetry constraint of  Eq.~\ref{eq:constraint on Nambu hamiltonian}. Out of these 136, 64 are normal masses of the form
\begin{align}
    \chi_N^\dagger (X_{0\beta\gamma\delta}) \chi_N, \,\,\text{and}\,\, \chi_N^\dagger (X_{3\beta\gamma\delta}) \chi_N,  
    \label{eq_normnambu}
\end{align}
which correspond to the SU(8) singlet and the adjoint multiplet, whose properties were analysed in Ref.~\cite{basusu8}. 

The rest of the 72 matrices have a generic form of 
\begin{align}
    \chi_N^\dagger (X_{1\beta\gamma\delta}) \chi_N, \,\,\text{and}\,\, \chi_N^\dagger (X_{2\beta\gamma\delta}) \chi_N,  
    \label{eq_scnambu}
\end{align}
and correspond to U(1) breaking superconducting masses. In the rest of the work, we study the properties of these superconducting masses. Upon expanding the two expressions above, we have, in terms of the 16-component Dirac spinor (Eq. \ref{Eq_Dirac_spinor_Xi}) 
\begin{eqnarray} \label{Mass_in_dirac_spinor}
\chi_N^\dagger (X_{1\beta\gamma\delta}) \chi_N =  \boldsymbol{\chi}^{\dagger}~\left(\Sigma_{\beta}\tau_{\gamma}\sigma_{\delta}\right)~ (\boldsymbol{\chi}^{\dagger})^T +h.c., \nonumber \\ 
\chi_N^\dagger (X_{2\beta\gamma\delta}) \chi_N  =  -i~\left(\boldsymbol{\chi}^{\dagger}~\left(\Sigma_{\beta}\tau_{\gamma}\sigma_{\delta}\right)~ (\boldsymbol{\chi}^{\dagger})^T\right) +h.c., \nonumber \\
\end{eqnarray}
where the constraint in Eq. \ref{eq:constraint on Nambu hamiltonian} now demands that
\begin{align}
    \left(\Sigma_\beta\tau_\gamma\sigma_\delta\right)^T=-\Sigma_\beta\tau_\gamma\sigma_\delta,
    \label{eq_diracanti}
\end{align}
{\it i.e.}, they are anti-symmetric. Eq. \ref{Mass_in_dirac_spinor} for the masses has a simple interpretation.  In terms of Dirac spinor $\boldsymbol{\chi}$, there are 36 independent bilinears of the form $\boldsymbol{\chi}^T\,\left(\Sigma_{\beta}\tau_{\gamma}\sigma_{\delta}\right)\,\boldsymbol{\chi}$,  which, when expressed in terms of the Nambu spinors, give the real and imaginary parts of the superconducting amplitude resulting in 72 superconducting masses with pairing amplitudes
\begin{align}
    \Delta_{\beta\gamma\delta} = \langle \boldsymbol{\chi}^T \left(\Sigma_{\beta}\tau_{\gamma}\sigma_{\delta} \right) \boldsymbol{\chi} \rangle.
    \label{eq_scbil}
\end{align}
Thus, in analogy with graphene~\cite{Ryu_graphene_masses,PhysRevB.108.L161108}, the 136 bilinears break up as Irreps of SU(8) as
\begin{align}
    1\oplus 63\oplus 36 \oplus 36^*
\end{align}
where the $1$ and $63$ are the normal orders corresponding to SU(8) singlet and adjoint multiplet, respectively, while the $36$ and $36^*$ correspond to real and imaginary components of the superconducting orders. 
The global symmetry of the free Dirac theory and the constraints on the mass matrix (Ref. \ref{eq:constraint on Nambu hamiltonian}) become manifest in the Majorana representation of the $\boldsymbol{\chi}$ fermions, as shown in Appendix \ref{appen_majorana}, which has a manifest $SO(16)$ symmetry. The 136 masses then are given by the $SO(16)$ singlet and the 135-dimensional symmetric rank-2 tensors (see Appendix \ref{appen_majorana})~\cite{PhysRevB.108.L161108}. Therefore, the Majorana representation naturally unifies the normal and the superconducting masses. 
%%%%%%%%%%%%%%%%%%%%%%% 

\subsection{Microscopic symmetries}

To classify the different SCs, it is useful to understand the transformations of the superconducting bilinears (Eq. \ref{eq_scbil}) under the microscopic symmetries-- lattice transformations and TR. This, of course, is related to the symmetry transformation of the Dirac fermions~\cite{basusu8}. Due to the underlying SOC, the microscopic transformation leads to mixing of the SU(4) flavours. Here we summarise these transformations for completeness and refer the reader to Ref. \cite{basusu8} for further details.

The symmetry group of lattice transformations, $\mathbb{S}$,  is generated by $\{{\bf T_1},{\bf T_2},{\bf C_3},{\bf C_6\sigma_h},{\bf C_2},{\bf \sigma_d},{\bf I}\}$ where $\{{\bf T_1},{\bf T_2}\}$ correspond to the lattice translations of honeycomb lattice, $\{{\bf C_2},{\bf  C_3},{\bf C_6}\}$ are two, three and six-fold lattice rotations and $\{\sigma_h,\sigma_d,{\bf I}\}$ are reflections and inversions as explained in Ref. \cite{basusu8} {(also shown in fig.~\ref{fig_transformation})}.  Under the action of $\mathbb{S}$, the Dirac spinors transform as
\begin{equation}\label{eq_transformation of chi under lattice}
\chi(\mathbf{x}) \xrightarrow{\mbb{S}} \Omega_{\mathbb{S}}~\chi (\mathbb S^{-1}\mbf{x}),
\end{equation}
where $\Omega_{\mathbb S}$ is a $16\times16$ dimensional matrix representation of $\mathbb S$ whose explicit forms are given in Ref.~\cite{basusu8}. This, for Nambu spinors, $\chi_N(\mbf{x})$, leads to
\begin{eqnarray}\label{eq:tranformation_of_Nambu_spinor}
&&\chi_N(\mbf{x}) \xrightarrow{\mbb{S}}  \left[~\Omega_{\mbb{S}}~\chi(\mbb{S}^{-1}\mathbf{x}),~\left(\chi^{\dagger}(\mbb{S}^{-1}\mathbf{x})\,\Omega_{\mbb{S}}^{\dagger}\right)^T~\right]^T \nonumber \\
&& \qquad =  \frac{1}{2}\Big[\mu_0\otimes(\Omega_{\mbb{S}}+\Omega_{\mbb{S}}^*)+\mu_3\otimes(\Omega_{\mbb{S}}-\Omega_{\mbb{S}}^*)\Big]\chi_N(\mbb{S}^{-1}\mathbf{x}). \nonumber\\
\end{eqnarray}

In addition, under the microscopic TR symmetry, the Dirac fermions transform such that TR$^2=-1$ with the detailed form of the transformation matrix also given in Ref. \cite{basusu8}. Finally, the microscopic charge conservation is described by Eq. \ref{eq_u1charge}.
%%%%%%%%%%%%%%%%%%%%%%%%%%

\section{Classification of Superconducting masses}\label{sec_classification}

Given the microscopic symmetries, we now turn to the classification of the superconducting masses. The resultant structure of the mass matrix is obtained by setting the normal masses to zero, whence Eq. \ref{eqn:15} reduces to 
\begin{eqnarray}\label{Eq_superconducting_mass_matrix}
  &  M^{Sc} =& \begin{bmatrix}
   0_{16\times16} && \Delta_{\beta\gamma\delta}(\Sigma_{\beta}\tau_{\gamma}\sigma_{\delta})\ \\
\Delta_{\beta\gamma\delta}^*(\Sigma_{\beta}\tau_{\gamma}\sigma_{\delta})   && 0_{16\times16} 
 \end{bmatrix} \nonumber \\ 
 && =
\sum_{i=1}^d\ \begin{bmatrix}
   0_{16\times16} &&\Delta_{i} \tilde{\text{M}}_{i} \\ \Delta_{i}^* \tilde{\text{M}}_{i}    && 0_{16\times16} 
 \end{bmatrix}, 
\end{eqnarray}
where $0_{16\times 16}$ is the $16\times 16$ zero matrix. The second expression concretely applies for an irreducible representation (Irrep) of dimension $d$, with pairing amplitude $\mathbf{\Delta}=(\Delta_1, \Delta_2, \cdots, \Delta_d)$, with $\Delta_i$ being complex numbers and $\tilde{M}_i$ denoting corresponding matrices which are given by particular combination of $\Sigma_\beta\tau_\gamma\sigma_\delta$ specific for the Irrep. For our case, $d\in\{1,2,3\}$, which corresponds to singlet, doublet and triplet representations, respectively.

At the outset, we would like to point out the broad classification of such masses under the microscopic TR symmetry. Under TR, 
\begin{equation}
    \tilde{M}_{i}\to \pm \tilde{M}_{i}\,\,\,\,\, \forall ~i.
\end{equation}

Now we can define 
\begin{align}
\text{T}^{+}_i = (\mu_1 + i \mu_2)\otimes \tilde{\text{M}}_{i}
~\text{and}~
\text{T}^{-}_i = (\mu_1 - i \mu_2)\otimes \tilde{\text{M}}_{i} 
\end{align}
such that, 
\begin{equation}
    M^{Sc}=\sum_{i=1}^{d} (\Delta_i \text{T}^{+}_i + \Delta_i^*\text{T}^{-}_i ) = \sum_{i=1}^{d} \vert\Delta_i\vert(\text{e}^{i\tilde{\delta}_i}\text{T}_{i}^{+} + \text{e}^{-i\tilde{\delta}_i}\text{T}_{i}^{-}).
\label{eq_trtrans}
\end{equation}
Under charge U(1) (Eq. \ref{eq_u1charge}),
\begin{align}
\chi_{i} \to \text{e}^{i\theta}\chi_i,~\text{T}_{i}^{+} \to \text{e}^{-i2\theta} \text{T}_{i}^{+},~
 \text{T}_{i}^{-} \to \text{e}^{i2\theta} \text{T}_{i}^{-}.
\end{align}

Thus, if all the phases $\tilde{\delta}_i$ (in Eq. \ref{eq_trtrans}) for all the components in a particular Irrep are the same, then a uniform U(1) rotation can get rid of it. This is always possible for singlets ($d=1$), but is generically not possible for doublets and triplets ($d=2,3$). Hence, while the former are TR symmetric, the latter generically break it. However, a linear combination of different singlets can break TR.

With this, we now turn to the complete classification of the SCs under the microscopic symmetries. To this end, we provide a two-step classification by first considering only the internal U(4) symmetry of the Dirac bilinears defined in Eq. \ref{eq_scbil} (the Nambu bilinears can be then obtained via Eq. \ref{eq_scnambu}) before taking on the full classification under the microscopic lattice symmetries. 
%%%%%%%%%%%%%%%%%%%%%%%%%%%%%%%%%%%

\subsection{Classification under U(4)}
\label{sec_u4classify}
As mentioned before, the Hamiltonian in Eq.~\ref{eq_dirac_intro} has U(4) invariance, under which the Dirac bilinears in Eq.~\ref{eq_scbil} that transform as 
\begin{eqnarray}
    \chi^T~\Sigma_{\beta} \tau_{\gamma} \sigma_{\delta}~ \chi \xrightarrow{U(4)} \chi^T~ \left(\Omega_U^{(f)}\right)^T\Sigma_{\beta}\Omega_U^{(f)} \otimes \tau_{\gamma} \sigma_{\delta}~ \chi. \nonumber\\
    \label{eq_u4trans}
\end{eqnarray}
Here, $\Omega^{(f)}_U$ is a $4\times4 $ unitary matrix in flavour space as is evident from the superscript $f$. {The Dirac bilinear in Eq.~\ref{eq_u4trans} transforms as a $4\otimes4$ representation of U(4) which is reduced to}  
\begin{equation}\label{eq:41}
4\otimes4=6\oplus10, 
\end{equation} 
where $6$ and 10 correspond to 6-dimensional anti-symmetric and 10-dimensional symmetric Irreps of U(4). The above U(4) transformations do not mix the two classes -- the symmetric and anti-symmetric flavour matrices, $\Sigma_\beta$. Hence, we can classify the above superconducting masses into two groups
\begin{eqnarray}
    &\Sigma_\beta\in\Sigma_{Asy}~~~~&\text{\textit{i.e., }} \Sigma_\beta^T=-\Sigma_\beta, \nonumber\\
    &\Sigma_\beta\in\Sigma_{sy}~~~~&\text{\textit{i.e., }} \Sigma_\beta^T=\Sigma_\beta.
    \label{eq_flasyasy}
\end{eqnarray}
Eq. \ref{eq_diracanti} now demands that the first class of flavour matrices in Eq. \ref{eq_flasyasy} can only occur in combination with $\tau_\gamma\sigma_\delta$ such that this {combination of valley-band space matrices} is symmetric (in addition to the constraints discussed above) and there is precisely one such combination, given by 
\begin{align}
    \tau_1\sigma_0.
    \label{eq_symvb}
\end{align}
Similarly, for the second class of matrices in Eq.~\ref{eq_flasyasy}, $\tau_\gamma\sigma_\delta$ needs to be anti-symmetric, and there are three such  combinations, given by 
\begin{align}
    \tau_3\sigma_2, \tau_2\sigma_0, i\tau_0\sigma_2.
    \label{eq_asymvb}
\end{align}

These are summarised in Table \ref{valley_subband_decomp}.  Putting this back in the general form of the masses for the Nambu spinor consistent with Eq. \ref{eq:constraint on Nambu hamiltonian}, the 72 superconducting bilinears (Eq. \ref{eq_scnambu}) can then be classified into four groups as shown in Table \ref{Classification under U(4) transformation 2}.\\
\begin{table}%[h!]
\begin{tabular}{|c|c|c|c|}\hline
Irreps &Set of masses & Flavour Sector&  Dimension\\\hline
1 &\Big($\mu_1\Sigma_i\tau_1\sigma_0$ , $\mu_2\Sigma^i\tau_1\sigma_0$\Big)&$\Sigma_i\in \Sigma_{Asy}$ & $ (6,6)$ \\\hline
2 & \Big($\mu_1\Sigma_i\tau_0\sigma_2$ , $\mu_2\Sigma^i\tau_0\sigma_2$\Big) &$\Sigma_i\in \Sigma_{Sy}$ & (10, 10)\\\hline
3 & \Big($\mu_1\Sigma_i\tau_2\sigma_0$ , $\mu_2\Sigma^i\tau_2\sigma_0$ \Big)& $\Sigma_i\in \Sigma_{Sy}$ & (10, 10)\\\hline
4 & \Big($\mu_1\Sigma_i\tau_3\sigma_2$, $\mu_2\Sigma^i\tau_3\sigma_2$ \Big) & $\Sigma_i\in \Sigma_{Sy}$ & (10, 10)\\\hline
\end{tabular}
\caption{Classification under U(4) transformation}\label{Classification under U(4) transformation 2}
\end{table} 
\subsection{Classification under lattice symmetries}\label{Classification under lattice symmetries}

 The lattice symmetries, however, mix the flavour and the chiral spaces due to the SOC as realised in Ref. \cite{basusu8}, such that the structure of the Irreps and hence the resultant symmetry breaking becomes more involved. In particular, under lattice symmetry $\mbb{S}$, the  Dirac  bilinears transform as 
\begin{align}
    & \chi^T~\Sigma_{\beta} \tau_{\gamma} \sigma_{\delta}~ \chi \nonumber \\  
    & \quad \xrightarrow{\mbb{S}} \chi^T \left[\left(\Omega_\mbb{S}^{(f)}\right)^T\Sigma_{\beta}\Omega_\mbb{S}^{(f)}\right] \otimes \left[\left(\Omega_\mbb{S}^{(c)}\right)^T\tau_{\gamma} \sigma_{\delta} \Omega_\mbb{S}^{(c)}\right]~ \chi,
\label{eq_transformation_under_lattice}
\end{align}
where the superscripts $f$ and $c$ represent respectively the flavour and the valley-subband (chiral) spaces on which $\Omega_{\mbb{S}}^{(f)} $ and $\Omega_{\mbb{S}}^{(c)} $ act such that the transformation of the Dirac spinor under symmetry $\mbb{S}$ is given by
\begin{align}
\boldsymbol{\chi} \to \Omega^{(f)}_{\mbb{S}} \otimes \Omega^{(c)}_{\mbb{S}} \, {\boldsymbol{\chi}}.    
\end{align}
This direct product structure of the transformation matrices arises naturally from working in the local basis and is explicitly demonstrated in Ref.~\cite{basusu8}.

The four matrices (Eqs. \ref{eq_symvb} and \ref{eq_asymvb}) in the valley-band space break up into $1\oplus3$, as discussed above, to give
\begin{align}
    4 = \mcl{A}_{1g}  \oplus \mcl{T}_{2g}.
    \label{eq_valley_subband_decomp}
\end{align}
Here, we use the same notation for the Irreps as introduced in Ref. \cite{basusu8} and summarised in Appendix \ref{app:C}. The singlet is the symmetric matrix, and the triplet is composed of three anti-symmetric matrices and is summarised in Table \ref{valley_subband_decomp} for easy reference. 

\begin{table}%[h!]
    \begin{tabular}{|c|c|c|} \hline
    \text{Valley-Subband Sector} & \text{Irrep} & \text{Transposition}  \\ \hline\hline
$\{\tau_1 \sigma_0\}$ & $\mcl{A}_{1g}$ & \text{Symmetric}\\ \hline
\{$\tau_3\sigma_2,\,\,\,  \tau_2\sigma_0,\,\,\, i\,\tau_0\sigma_2$\} & $\mcl{T}_{2g}$ & \text{Anti-symmetric}\\ \hline
    \end{tabular}
    \caption{Decomposition of chiral (valley and sub-band) sector under lattice symmetries.}
    \label{valley_subband_decomp}
\end{table}

For the 16 flavour matrices (Eq. \ref{eq:41}), the six antisymmetric ones (Eq. \ref{eq_flasyasy}), under lattice symmetries, break up into $1\oplus2\oplus3$ with
\begin{align}
6=\mathcal{A}_{1g}\oplus\mathcal{E}_u\oplus\mathcal{T}_{1g},
\label{eq_asylat}
\end{align}
as summarized in the first three rows of Table \ref{Flavour_Decomp}, while the 10 symmetric ones (Eq. \ref{eq_flasyasy}) break up as $1\oplus3\oplus3\oplus3$, {\it i.e.},
\begin{align}
10=\mathcal{A}_{2g}\oplus\mathcal{T}_{2g}\oplus\mathcal{T}_{1u}\oplus\mathcal{T}_{2u}
\label{eq_sylat}
\end{align}
as denoted by the lower four rows of Table \ref{Flavour_Decomp}.

The full classification of the 36 superconducting bilinears for the Dirac fermions, $\boldsymbol{\chi}^T\Sigma_\beta \tau_\gamma \sigma_\delta\boldsymbol{\chi}$,  can now be obtained by combining the representations of the flavour and the chiral sectors, similar to Ref. \cite{basusu8}, but noting, as above, that only combinations of symmetric (anti-symmetric) flavour matrices are allowed with anti-symmetric (symmetric) chiral matrices.

\begin{table}%[h!]
\begin{tabular}{|c|c|c|} \hline
    \text{Flavour Sector} & \text{Irrep} & \text{Transpose}  \\ \hline\hline
\{$\Sigma_{13}$\}     &  $\mcl{A}_{1g}$ & \text{AntiSym} \\ 
$\{\Sigma_{24}, \Sigma_{25}\}$ & $\mcl{E}_u$  & \text{AntiSym} \\
$\{\Sigma_1, \Sigma_3, i\,\Sigma_{45}\}$ & $\mcl{T}_{1g}$  & \text{AntiSym}\\ \hline
$\{\Sigma_2\}$ & $\mcl{A}_{2g}$  & \text{Sym}\\
$\{\Sigma_{32}, i\,\Sigma_0, \Sigma_{12}\}$ & $\mcl{T}_{2g}$ & \text{Sym}\\
$\{\Sigma_{15}$,\, $i(\frac{\sqrt{3}}{2}\Sigma_5-\frac{1}{2}\Sigma_4),\, \frac{\sqrt{3}}{2}\Sigma_{34} - \frac{1}{2}\Sigma_{35}$ \} &  $\mcl{T}_{1u}$ & \text{Sym} \\
\{$\Sigma_{14},\, i(\frac{\sqrt{3}}{2}\Sigma_4+\frac{1}{2}\Sigma_5), \,-\frac{\sqrt{3}}{2}\Sigma_{35} - \frac{1}{2}\Sigma_{34}$ \}  &  $\mcl{T}_{2u}$ & \text{Sym}\\ \hline
\end{tabular}
\caption{Decomposition of flavour space matrices under lattice symmetries. AntiSym(Sym) specifies if matrix is anti-symmetric(symmetric) under transposition (see Eq. \ref{eq_flasyasy} and Table \ref{Classification under U(4) transformation 2}).} \label{Flavour_Decomp}
\end{table}

\textit{Flavour anti-symmetric masses :}
There are six anti-symmetric $\Sigma_\beta$ matrices (Table~\ref{Flavour_Decomp} and Eq. \ref{eq_asylat}) which can only combine with the symmetric singlet of the chiral space to give 
\begin{eqnarray}\label{eq_decomposition_in_antisym_1}
&&\mcl{A}_{1g} \otimes \mcl{A}_{1g} = \mcl{A}_{1g}, \\
\label{eq_decomposition_in_antisym_2}
&&\mcl{E}_{u} \otimes \mcl{A}_{1g} = \mcl{E}_{u}, \\
\label{eq_decomposition_in_antisym_3}
&&\mcl{T}_{1g} \otimes \mcl{A}_{1g} = \mcl{T}_{1g}. 
\end{eqnarray}

\textit{Flavour symmetric  masses :}
Similarly, the 10 flavour symmetric matrices can only combine with the anti-symmetric chiral triplet to give a total of 30 flavour symmetric direct product matrices that are divided into three singlets, three doublets, and seven triplets, as shown below 
\begin{eqnarray}\label{eq_decomposition_in_sym_1}
    &&\mcl{A}_{2g} \otimes \mcl{T}_{2g} = \mcl{T}_{1g}, \\ \label{eq_decomposition_in_sym_2}
    &&\mcl{T}_{2g} \otimes \mcl{T}_{2g} = \mcl{T}_{1g} \oplus \mcl{T}_{2g} \oplus \mcl{E}_{g} \oplus \mcl{A}_{1g}, \\ \label{eq_decomposition_in_sym_3}
    &&\mcl{T}_{1u} \otimes \mcl{T}_{2g} = \mcl{T}_{1u} \oplus \mcl{T}_{2u} \oplus \mcl{E}_{u} \oplus \mcl{A}_{2u}, \\ 
    \label{eq_decomposition_in_sym_4}
    &&\mcl{T}_{2u} \otimes \mcl{T}_{2g} = \mcl{T}_{1u} \oplus \mcl{T}_{2u} \oplus \mcl{E}_{u} \oplus \mcl{A}_{1u}. 
\end{eqnarray}

Finally, turning to the Nambu bilinears (Eq. \ref{eq_scnambu}), it is clear that we have 72 superconducting masses that are formed out of the real and imaginary components of the above 36 bilinears as denoted in the Table. \ref{Classification under U(4) transformation 2}. 
This leads to the classification of the superconducting masses under microscopic symmetries and leads to 
\begin{itemize}
    \item 4-singlets : $(2)\mcl{A}_{1g}, \mcl{A}_{1u}, \mcl{A}_{2u}$
    \item 4-doublets : $(3)\mcl{E}_{u}, \mcl{E}_{g}$
    \item 8-triplets : $(3)\mcl{T}_{1g}, (2)\mcl{T}_{1u}, \mcl{T}_{2g}, (2) \mcl{T}_{2u}$.
\end{itemize} 
each of which can be TR even or odd, corresponding to the real and imaginary components of the pairing amplitude. 

As remarked above, the two components of the singlet pairing can be combined along with charge $U(1)$ to give rise to four distinct TR even SCs. Further, while the $\mcl{A}_{1u}$ and the $\mcl{A}_{2u}$ singlets go to their negative under some point group transformation, this can be rectified via the U(1) phase and hence they do not break any microscopic symmetries. However, a linear combination of two singlets may break lattice and/or TR symmetries. The doublets and the triplets, on the other hand, generically break TR symmetry. While there are 12 of them, we find that some of the doublets and triplets can be adiabatically deformed into each other without breaking any further symmetries, thus leading to only 8 distinct SCs, as we shall discuss below. 
The above superconducting masses give rise to both gapped and gapless SCs, which we now discuss in detail. To this end, we divide the discussion into the singlet, doublet and triplet Irreps in the next three sections. 

It is useful to note the classification of SCs for a two-orbital per site, {\it i.e.}, an effective $j=1/2$ system, instead of four (Eq. \ref{eq_j3/2orb}), on the honeycomb lattice with $\pi$-flux at $1/4$th filling. The resultant low-energy theory corresponds to a different representation of the microscopic symmetries, now acting on the $j=1/2$ electrons. This is discussed in Appendix \ref{sec_su2global} and leads to {\it five} SCs -- two singlets, one doublet and two triplets. As is evident from the discussion presented in the section, the $j=3/2$ system discussed in the main text realizes a myriad of new phases that are not realized by the $j=1/2$ system. 
%%%%%%%%%%%%

\section{Singlet superconductors}
\label{sec_singletsc}

As discussed below Eq. \ref{Eq_superconducting_mass_matrix}, for singlets we can always perform a U(1) transformation on the masses to make them TR symmetric, resulting in  4 different TR symmetric singlet SCs-- two $\mcl{A}_{1g}$, and one each of $\mcl{A}_{1u}$ and $\mcl{A}_{2u}$.
%%%%%%%%

\subsection{The two gapped \texorpdfstring{$\mathcal{A}_{1g}$}{} singlets}
\label{sec_a1g_singlets}
The two $\mcl{A}_{1g}$ singlets (we denote them as $\mcl{A}_{1g}^{I}$ and $\mcl{A}_{1g}^{II}$) arise respectively from Eqs.~\ref{eq_decomposition_in_antisym_1} and \ref{eq_decomposition_in_sym_2}. These two singlets correspond to two different superconducting phases as they cannot be deformed into each other without closing the spectral gap for the Bogoliubov quasi-particles (see below) or breaking further symmetries. 
 
The low-energy Hamiltonian in the presence of the two $\mathcal{A}^{\mcl{x}}_{1g}$ ($\mcl{x}$= I or II)  masses is given by Eq. \ref{eq:general term in Nambu basis}, with the form of the mass matrix given by Eq. \ref{Eq_superconducting_mass_matrix} with $d=1$.  The $\mcl{A}^{I}_{1g}$ singlet is particularly simple and arises from the direct product of the singlets in flavour and valley-subband sectors (see Eq.~\ref{eq_decomposition_in_antisym_1}) such that
\begin{align}
    {\rm Singlet-I}~:&~\begin{array}{l}
    m^{\mcl{A}^{I}_{1g}} = \Sigma_{13}\tau_1\sigma_0,\\
\Delta^{\mathcal{A}^{I}_{1g}}=\langle\chi^{T}m^{\mcl{A}^{I}_{1g}}\chi \rangle.
    \end{array}
    \label{eq_pair1310}
\end{align}

The $\mcl{A}^{II}_{1g}$ singlet, on the other hand, arises from the direct product of triplets in the flavour and valley-subband sectors as mentioned in Eq. \ref{eq_decomposition_in_sym_2} and this results in  
\begin{align}
    {\rm Singlet-II}~:&\begin{array}{l}
m^{\mcl{A}^{II}_{1g}} = \frac{\left(\Sigma_0\tau_0\sigma_2+\Sigma_{12}\tau_3\sigma_2-\Sigma_{23}\tau_2\sigma_0\right)}{\sqrt{3}}\\
\\
\Delta^{\mcl{A}^{II}_{1g}} = \langle \chi^T \,m^{\mcl{A}^{II}_{1g}}\,\chi \rangle
    \end{array}
    \label{eq_sc_amp_A1g_II}
\end{align}
Note that while both the masses are $\mcl{A}_{1g}$ singlets, they are composed of different combinations of the flavour and chiral sectors, which lead to different consequences, with the first being unitary, while the second one a rather rare example of a non-unitary singlet SC, {\it i.e.},
\begin{align}
    \left[\Delta^{\mathcal{A}^I_{1g}}m^{\mathcal{A}^I_{1g}}\right]\cdot \left[\Delta^{\mathcal{A}^I_{1g}}m^{\mathcal{A}^I_{1g}}\right]^\dagger&=|\Delta^{\mathcal{A}^I_{1g}}|^2\Sigma_0\tau_0\sigma_0,
    \label{eq_ma1g1}
\end{align}
and
\begin{align}    &\left[\Delta^{\mathcal{A}^{II}_{1g}}m^{\mathcal{A}^{II}_{1g}}\right]\cdot  \left[\Delta^{\mathcal{A}^{II}_{1g}}m^{\mathcal{A}^{II}_{1g}}\right]^\dagger\nonumber\\
&=|\Delta^{\mathcal{A}^{II}_{1g}}|^2\left[\Sigma_{0}\zeta_0+\frac{2}{3}(\Sigma_{12}\zeta_1+\Sigma_{13}\zeta_2-\Sigma_{23}\zeta_3)\right],
    \label{eq_ma1g2}
\end{align}
where $\zeta_0=\tau_0\sigma_0$ and $\zeta_i (i=1,2,3)$ are given in Eq. \ref{eq_chiralsu2}.  

\subsubsection{\texorpdfstring{$\mcl{A}^{I}_{1g}$}{} Singlet}
\label{sec_anti_singlet_section}

Diagonalising the mean field Hamiltonian (Eq. \ref{eq:general term in Nambu basis}) for singlet-I (Eq. \ref{eq_pair1310}) provides two bands with gapped dispersion for the Bogoliubov quasi-particles of the form
\begin{eqnarray}\label{antisymmetric singlet spectrum}
E^\pm(k_x,k_y,\Delta^{\mcl{A}^{I}_{1g}})=\pm \sqrt{k_x^2+k_y^2+\vert\Delta^{\mcl{{A}}^{I}_{1g}}\vert^2},
\end{eqnarray}
where, as usual,   $2\vert\Delta^{\mcl{A}^{I}_{1g}}\vert$ is the gap in the spectrum. Each band is 16-fold degenerate. This single gap is a direct outcome of the unitary pairing (Eq. \ref{eq_ma1g1}). Indeed, the SO(16) symmetry group (Eq. \ref{eq_so16}) is broken down to SO(8)$~\otimes~$SO(8) by the unitary pairing. The resultant SC vortex is featureless and is characterised by the usual homotopy classification $\pi_{1}(S^1)=Z$.

The above form of the pairing can be traced back to the lattice BCS Hamiltonian (in the global basis) of the form
\begin{align}\label{Lattice_ham_in_3/2_basis_2}
H^{global}_{\mathcal{A}_{I}} =&  H+ \sum_{\mbf{r}}\,\,{\left(\Delta_{lat}^{\mcl{A}^{I}_{1g}}\right)^*}\,\,
\Psi^{T}(\mbf{r})\,\Sigma_{13}\,\Psi(\mbf{r})+h.c.
\end{align}
where $H$ is given by Eq. \ref{eq_masaki_ham} and $\Psi(\mbf{r})$ corresponds to the $j=3/2$ orbitals. Note that the pairing is on-site, {\it i.e.},
\begin{align}
{\left(\Delta_{lat}^{\mcl{A}^{I}_{1g}}\right)^*}=\langle \Psi^{\dagger}(\mbf{r}) \Sigma_{13}(\Psi^{\dagger}(\mbf{r}))^T\rangle, 
\label{eq_aigunitarysinglet}
\end{align}
similar to the $s$-wave pairing.  In particular, given the explicit form of 
     \begin{eqnarray}
         \Sigma_{13}=-\frac{7}{3}\left(J_y-\frac{4}{7}J_y^3\right)=\begin{pmatrix}
             0& -i& 0& 0\\
             i& 0& 0& 0\\
             0& 0& 0& i\\
             0& 0& -i& 0\\
         \end{pmatrix},
     \end{eqnarray}
it is clear that the form is reminiscent of a {\it spin singlet} type pairing within the $\pm 1/2$ and $\pm 3/2$ orbitals, respectively, with a relative $\pi$-phase. Indeed, the relative $\pi$-phase can be understood as a pairing in the spin-$3/2$ $SU(2)$-singlet sector. {Note that in the local basis (Eq.~\ref{eq_phi_definition}),  all the pairings transform as $3/2\otimes 3/2$ representation~\cite{kim2018beyond} of the SU(2) generated by $(J_x,J_y,J_z)$~\cite{basusu8} that is a subgroup of the flavour SU(4).} Under spin $SU(2)$, the decomposition has the well-known form
\begin{align}\label{eq_SU_2decomp}
    \frac{3}{2}\otimes \frac{3}{2} = {\color{red} 0} \oplus 1 \oplus {\color{red} 2} \oplus 3.
\end{align}
Thus, the total spin angular momentum of the Cooper pair can be $J_T=0,1,2,3$, with the magnetic quantum number, $m_T=-J_T,\cdots, J_T$. The black (red) colour corresponds to symmetric (anti-symmetric) representations, such that only anti-symmetric (symmetric) representations are allowed by the Pauli exclusion principle when the real-space part of the wave function is symmetric (anti-symmetric) under spatial inversion.

For the above on-site pairing, the spin part of the Cooper-pair wave function is given by the $SU(2)$ singlet anti-symmetric wave-function~\cite{kim2018beyond}
\begin{align}
    |J_T=0, m_T=0\rangle=&\frac{1}{2}\left(|3/2,-3/2\rangle-|-3/2,3/2\rangle\right)\nonumber\\
    &-\frac{1}{2}\left(|1/2,-1/2\rangle-|-1/2,1/2\rangle\right)
    \label{eq_jtsinglet}
\end{align}
while its on-site spatial part is symmetric. The SU(2) singlet then demands the relative $\pi$ phase for the above on-site $s$-wave SC. The statement becomes manifest in the local basis, whence, on using Eq. \ref{eq_phi_definition}, the hopping term becomes diagonal while the pairing term remains unchanged, such that in the local basis there is pairing between the flavours (1,2) and (3,4), respectively, with a relative $\pi$-phase between the two sets of pairing terms. {Due to the inter-twinning of real and spin spaces, the change of the relative phase from $\pi$ leads to the mixing of the above singlet with the onsite $\mcl{E}_{u}$ doublet (corresponding to a part of the antisymmetric representation {\color{red}2} in Eq. \ref{eq_SU_2decomp}) studied in Section \ref{sec_eu_doublet1} resulting in breaking of inversion symmetry. This relative $\pi$-phase is therefore a direct outcome of the symmetry implementation. One can therefore conceive of fluctuations in relative phase that would correspond to an inversion-odd Leggett mode~\cite{leggett1966number}. Symmetry analysis shows that the gap of this Leggett mode appears to be driven by a quartic coupling with the $\mcl{E}_{u}$ pairing field, raising the possibility of an undamped Leggett mode investigated in superconducting Dirac materials~\cite{cuozzo2024leggett}. Clearly, such a Leggett mode is absent in the $j=1/2$ counterpart (Appendix \ref{sec_su2global}).}

\subsubsection{\texorpdfstring{$\mcl{A}^{II}_{1g}$}{}  Singlet}
\label{sec_a1g2}
\begin{figure}%[h!]
\includegraphics[scale=.38]{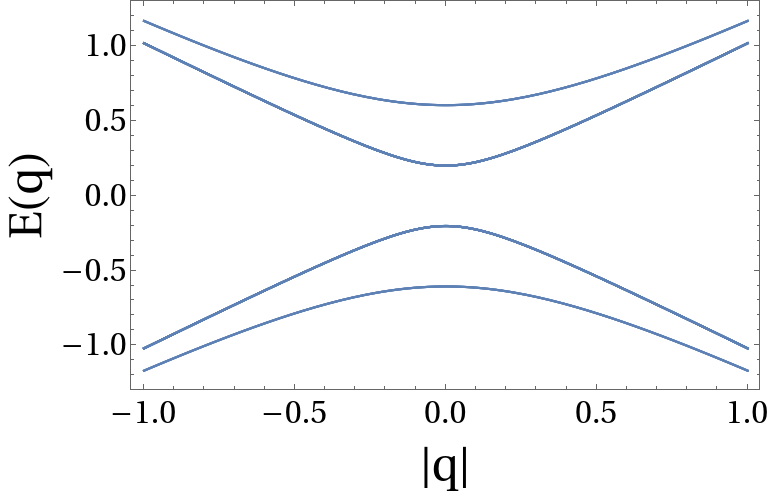}
\caption{Spectrum for the non-unitary $\mcl{A}_{1g}^{II}$ singlet (with $\Delta^{\mcl{A}^{II}_{1g}}=0.35$) around the Dirac points. Each of the uppermost and the lowermost bands is 4-fold degenerate, while each of the rest two is 12-fold degenerate. This degeneracy structure implies that the SO(16) symmetry of the free Hamiltonian breaks down into SO(2)$\otimes$SO(2)$\otimes$SO(6)$\otimes$SO(6). Here, $\vert \text{\bf q} \vert$ is measured from the position of Dirac cones.}
\label{symmetric gapped singlet spectrum}
\end{figure}

The $\mathcal{A}^{II}_{1g}$ singlet originates from the direct product of $\mcl{T}_{2g}$ lattice triplets in both the flavour and valley-subband sectors, as outlined in Eq. \ref{eq_decomposition_in_sym_2} and the corresponding pairing matrix is given in Eq.~\ref{eq_sc_amp_A1g_II}. Due to the non-unitary pairing (Eq. \ref{eq_ma1g2}), the spectrum (Fig. \ref{symmetric gapped singlet spectrum}) has a double gap structure, which can be checked by diagonalising the mean-field Hamiltonian.

\begin{figure}
    \centering
\includegraphics[width=.62\columnwidth]{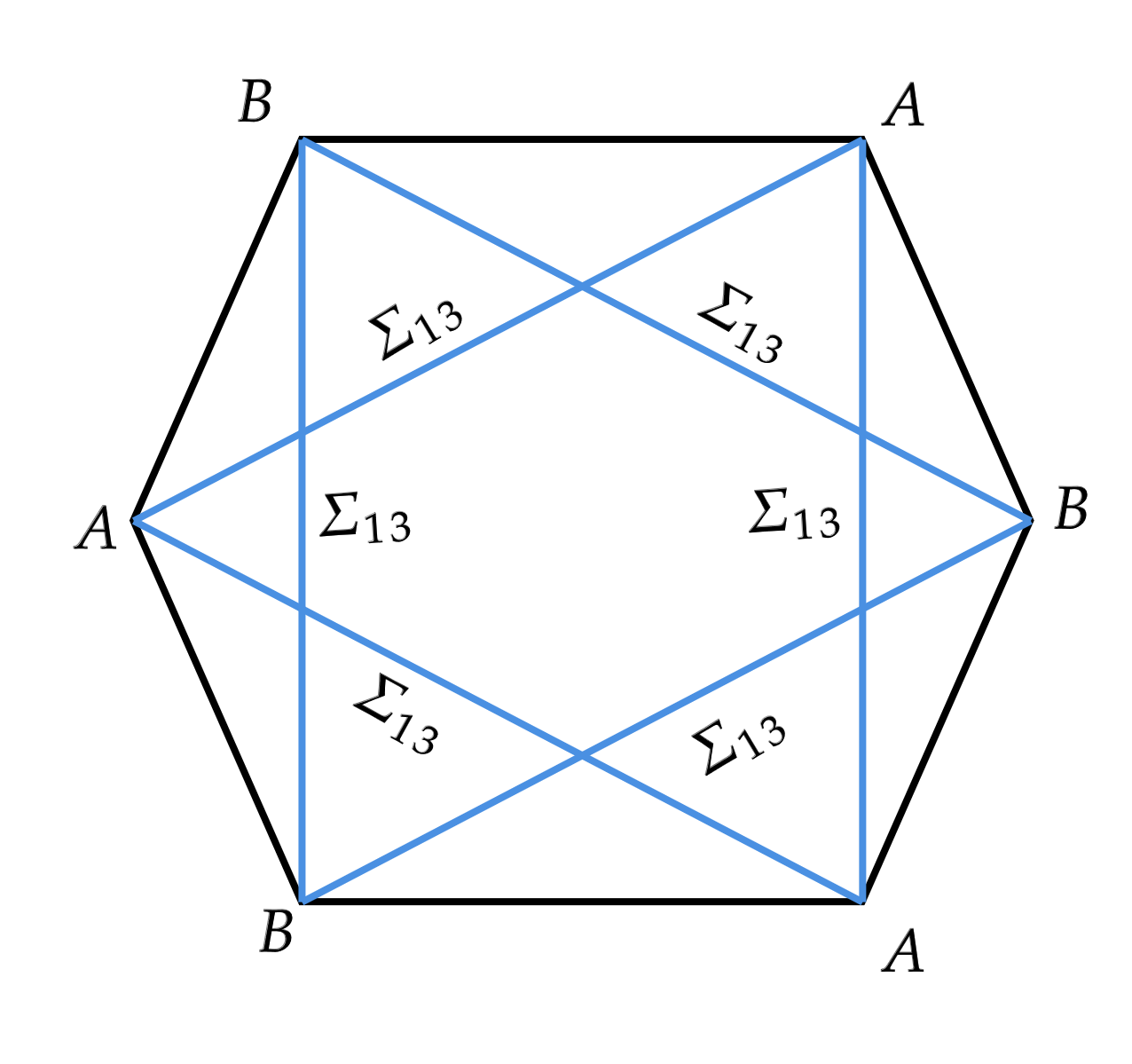}
    \caption{Schematic of the pairing in global basis for the $\mcl{A}^{II}_{1g}$ singlet on the lattice (Eq. \ref{Fig_lattice model_A_1g_II}). The blue bonds represent SC pairing amplitudes on the NNN bonds with the pairing matrix being $\Sigma_{13}$.}
    \label{Fig_lattice model_A_1g_II}
\end{figure}

Similar to the first singlet discussed above, the low-energy Hamiltonian in the presence of this superconducting mass can be interpreted as the low-energy limit of a lattice BCS Hamiltonian (in the global basis) of the form
\begin{align}\label{Lattice_ham_in_3/2_basis}
H^{global}_{\mathcal{A}_{II}} =&  H + \sum_{\braket{\braket{\mbf{rr'}}}}\,\,\left(\Delta_{lat}^{\mcl{A}^{II}_{1g}}\right)^*\,\,
\Psi^{T}(\mbf{r})\,\Sigma_{13}\,\Psi(\mbf{r'})+h.c.
\end{align}
where $H$ is the free fermion Hamiltonian (Eq. \ref{eq_masaki_ham}) and the  pairing matrix is given by $\Sigma_{13}$. However, unlike in the previous case (Eq. \ref{Lattice_ham_in_3/2_basis}),  the pairing is NNN, as pictorially shown in Fig.~\ref{Fig_lattice model_A_1g_II}. Thus, while the spin part of the Cooper pair wave function is still given by Eq. \ref{eq_jtsinglet}, the spatial part is symmetric, with weights on the NNN sites, resulting in an extended $s$-wave SC.

\subsubsection{Are the two \texorpdfstring{$\mcl{A}_{1g}$}{} superconductors distinct  ?}

Given that both the $\mcl{A}_{1g}$ SCs discussed above belong to the same Irrep, it is useful to understand the sense in which they represent distinct SCs, if at all. Comparing the lattice Hamiltonians in Eqs. \ref{Lattice_ham_in_3/2_basis_2} and \ref{Lattice_ham_in_3/2_basis}, it is clear that they respectively correspond to on-site $s-$wave and extended (NNN) $s$-wave SCs for the $j=3/2$ electrons. However, while  $\mcl{A}_{1g}^I$ is unitary, $\mcl{A}_{1g}^{II}$ is non-unitary as is evident from one and two gap spectra, respectively. This is because, in the global basis, the hopping Hamiltonian, $H$ (Eq. \ref{eq_masaki_ham}), does not have the spin-rotation symmetry. Therefore, a simple one-parameter adiabatic interpolation between them fails and is interrupted by a gapless point, which is in accordance with the fact that there is an intermediate phase transition, since they break different flavour symmetries (Eq. \ref{eq_pair1310} vs \ref{eq_sc_amp_A1g_II}). {Analyzing the pairing matrix in the local basis(Eq.~\ref{eq_phi_definition}) provides, $\mcl{A}^{I}_{1g}$ singlet belongs to SU(2) singlet sector (0 in Eq. \ref{eq_SU_2decomp}), while that of $\mcl{A}^{II}_{1g}$ belongs to the symmetric representation of the SU(2) (1 and 3 Eq. \ref{eq_SU_2decomp}). Based on this, we  conclude that the two $\mcl{A}_{1g}$ represent distinct SCs.}
%%%%%%%%%%%%%%%%

\subsection{The Gapless Singlets : \texorpdfstring{$\mcl{A}_{1u}$}{} and \texorpdfstring{$\mcl{A}_{2u}$}{}}\label{sec_gapless_singlets}
\begin{figure}
    \centering
\includegraphics[scale=.38]{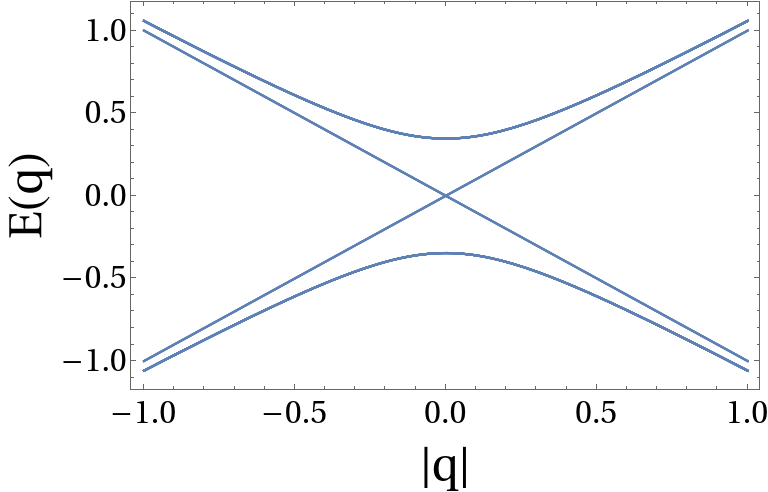}
    \caption{Dispersion for the $\mcl{A}_{1u}$ and the $\mcl{A}_{2u}$ Irreps { (for $\Delta^{\mcl{A}_{1u}}=0.30$)} around the Dirac points. The gapped bands are 12-fold degenerate and originate from the Dirac nodes at $M$-points in the global basis, while the two gapless bands are 4-fold degenerate and originate from the node at $\Gamma$-point in the global basis (Eq. \ref{Eq_singlet_in_global_basis}). }
    \label{Fig_Gapless_singlet}
\end{figure}

The $\mcl{A}_{2u}$ and the $\mcl{A}_{1u}$ singlets arise from the decomposition in Eq.~\ref{eq_decomposition_in_sym_3} and \ref{eq_decomposition_in_sym_4}, respectively, and  represent two distinct gapless SCs. Indeed, unlike the $\mcl{A}_{1g}$s, both the $\mcl{A}_{1u}$ and the $\mcl{A}_{2u}$ exhibit gapless Bogoliubov spectrum (see Fig~\ref{Fig_Gapless_singlet}). For both the singlets, the SO(16) symmetry of the free Hamiltonian is broken down to $SO(4)\otimes SO(6)\otimes SO(6)$. In the following, we first discuss the low-energy Hamiltonian along with a lattice realisation for these masses and then provide an understanding of these gapless modes by analysing these masses in the global basis. 

The low-energy Hamiltonian in the presence of these masses is given by Eq.~\ref{eq:general term in Nambu basis}. The corresponding mass matrix is given by Eq.~\ref{Eq_superconducting_mass_matrix} with $d=1$ for the $\mcl{A}_{1u}$ singlet with
\begin{eqnarray}
&& m^{\mcl{A}_{1u}} = \frac{1}{\sqrt{3}}(\frac{\sqrt{3}}{2}\Sigma_{35} +\frac{1}{2}\Sigma_{34})\tau_3\sigma_2\nonumber\\  
&& ~~~~~~~~+\frac{1}{\sqrt{3}}(\frac{\sqrt{3}}{2}\Sigma_{4}+\frac{1}{2}\Sigma_{5})\tau_0\sigma_2 +\frac{1}{\sqrt{3}}\Sigma_{14}\tau_2\sigma_0 ,
\end{eqnarray}
and 
\begin{eqnarray}\label{eq_sc_amp_A1u}
    \Delta^{\mcl{A}_{1u}} = \langle \chi^T \,m^{\mcl{A}_{1u}}\,\chi \rangle ,
\end{eqnarray}
such that the pairing is non-unitary, {\it i.e.},
\begin{align}
    &\left[\Delta^{\mcl{A}_{1u}}\,m^{\mcl{A}_{1u}}\right]\cdot\left[\Delta^{\mcl{A}_{1u}}\,m^{\mcl{A}_{1u}}\right]^\dagger\nonumber\\
    &=|\Delta^{\mcl{A}_{1u}}|^2\left[\Sigma_0\zeta_0+\frac{1}{3}\left(\Sigma_{23}\zeta_3-\Sigma_{12}\zeta_1-\Sigma_{13}\zeta_2\right)\right].
    \label{eq_nonunita1u}
\end{align}

For the $\mcl{A}_{2u}$ singlet, we similarly have an analogous mass matrix with 
\begin{eqnarray}
&& m^{\mcl{A}_{2u}} = \frac{1}{\sqrt{3}}(\frac{\sqrt{3}}{2}\Sigma_{34} -\frac{1}{2}\Sigma_{35})\tau_3\sigma_2\nonumber\\  &&~~~~~~~
+\frac{1}{\sqrt{3}}(-\frac{\sqrt{3}}{2}\Sigma_{5}+\frac{1}{2}\Sigma_{4})\tau_0\sigma_2 -\frac{1}{\sqrt{3}}\Sigma_{15}\tau_2\sigma_0,
\end{eqnarray}
and 
\begin{eqnarray}\label{eq_sc_amp_A2u}
    \Delta^{\mcl{A}_{2u}} = \langle \chi^T \,m^{\mcl{A}_{2u}}\,\chi \rangle.
\end{eqnarray}

Thus, the pairing in this case too is non-unitary, {\it i.e.}, $\left[\Delta^{\mcl{A}_{2u}}\,m^{\mcl{A}_{2u}}\right]\cdot\left[\Delta^{\mcl{A}_{2u}}\,m^{\mcl{A}_{2u}}\right]^\dagger$ has a form similar to Eq. \ref{eq_nonunita1u}. Thus, both the SCs have a double gap structure, with one of the gaps being exactly zero corresponding to four zero eigenvalues of the matrix on the right-hand-side of Eq. \ref{eq_nonunita1u} (the other twelve eigenvalues are non-zero equal and hence give rise to 12 fold degenerate gapped excitations) and is evident from Fig. \ref{Fig_Gapless_singlet}. This gapless structure becomes most transparent when the low-energy theory is written in the global basis (using the $\chi_g$ spinors defined in Eq.~\ref{eq:global basis}). Note that in the global basis~\cite{basusu8}, there are four valleys (see Fig. \ref{fig_BZ})-- one at $\Gamma$ point and three at the three inequivalent $M\equiv(M_1,M_2,M_3)$ points-- and each valley contributes a 4-component Dirac spinor as expressed in Eq. \ref{eq_globalspinor}. The corresponding Nambu spinor (in the global basis) is then given by 
\begin{equation}\label{Eq_Nambu_spinor_in_global_basis}
    \chi^{N}_{g} = {(\tilde{\chi}_{g\Gamma},{\tilde{\chi}_{g\Gamma}}^{\dagger},\tilde{\chi}_{gM_1},{\tilde{\chi}_{gM_1}}^{\dagger},\tilde{\chi}_{gM_2},{\tilde{\chi}_{gM_2}}^{\dagger},\tilde{\chi}_{gM_3},{\tilde{\chi}_{gM_3}}^{\dagger})}^T.
\end{equation}

In this basis the mass matrices (Eq. \ref{Eq_superconducting_mass_matrix}), $M^{\mcl{A}_u}_{global}$, corresponding to the $\mcl{A}_{1u}$ and the $\mcl{A}_{2u}$ singlets have the generic form of
\begin{equation}\label{Eq_singlet_in_global_basis}
\left(\begin{array}{cc|cc|cc|cc}
0_{4\times4}&0_{4\times4}&0_{4\times4}&0_{4\times4}&0_{4\times4}&0_{4\times4}&0_{4\times4}&0_{4\times4} \\
0_{4\times4}&0_{4\times4}&0_{4\times4}&0_{4\times4}&0_{4\times4}&0_{4\times4}&0_{4\times4}&0_{4\times4} \\  \hline
0_{4\times4}&0_{4\times4}&0_{4\times4}& \mcl{W_{1}}&0_{4\times4}&0_{4\times4}&0_{4\times4}&0_{4\times4} \\
0_{4\times4}&0_{4\times4}&\mcl{W_{1}}^{\dagger}&0_{4\times4}&0_{4\times4}&0_{4\times4}&0_{4\times4}&0_{4\times4} \\ \hline
0_{4\times4}&0_{4\times4}&0_{4\times4}&0_{4\times4}&0_{4\times4}&\mcl{W_{2}}&0_{4\times4}&0_{4\times4} \\
0_{4\times4}&0_{4\times4}&0_{4\times4}&0_{4\times4}&\mcl{W_{2}}^{\dagger}&0_{4\times4}&0_{4\times4}&0_{4\times4} \\ \hline
0_{4\times4}&0_{4\times4}&0_{4\times4}&0_{4\times4}&0_{4\times4}&0_{4\times4}&0_{4\times4}&\mcl{W_{3}} \\
0_{4\times4}&0_{4\times4}&0_{4\times4}&0_{4\times4}&0_{4\times4}&0_{4\times4}&\mcl{W_{3}}^{\dagger}&0_{4\times4} \\ 
\end{array}\right),
\end{equation}
where $\mcl{W_{i}}$ are $4\times 4$ matrices which are essentially the pairing matrices for the $\chi_{gM_i}$ spinor (for $i=1,2,3$). From the structure of the matrices, it is clear that the pairing is absent for the Dirac fermions $\chi_{g\Gamma}$, which then remains massless due to lattice symmetries as well as an emergent $SO(4)$, similar to the protection of the $\Gamma$-DSM in Ref. \cite{basusu8}, and is discussed in Appendix \ref{sec_symmetry_analysis_gapless_modes}. The $SO(4)$ can either be broken via inducing spin-octupole Hall order~\cite{basusu8} or admixing the above SC with the gapped $\mathcal{A}_{1g}$ singlet SC discussed above in Section \ref{sec_anti_singlet_section}.

\begin{figure}
  \centering
\includegraphics[scale=0.5]{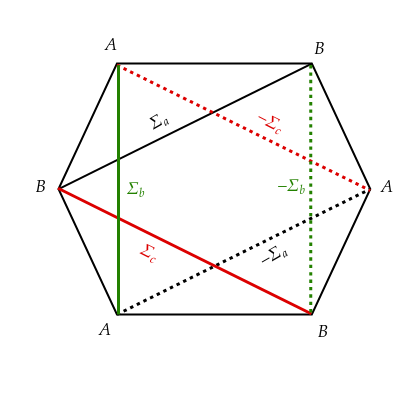}
    \caption{Schematic of the pairing for for $\mcl{A}_{1u}$ and $\mcl{A}_{2u}$ singlets the lattice BCS Model  in global basis (Eq. \ref{eq_lattice_ham_nnn}). The connecting lines indicate NNN pairing amplitudes with their colours representing the pairing matrix. Solid(dashed) lines indicate the sign of the pairing term is +ve(-ve) with the matrices $(\Sigma_a, \Sigma_b,\Sigma_c)$ described in Eqs. \ref{eq_pairingsinglet1} and \ref{eq_pairingsinglet2} respectively for the two singlets.}
    \label{Fig_lattice model_A_1u_2}
\end{figure}

The above low-energy forms of the two gapless SCs stem from a mean-field lattice Hamiltonian with NNN  pairing, which, in the global basis (similar to Eq. \ref{Lattice_ham_in_3/2_basis}), is given by 
\begin{align}\label{eq_lattice_ham_nnn}
    H^{global}_{\mcl{A}_{1u(2u)}} = H+ \sum_{\braket{\braket{\mbf{rr'}}}}&\left[\left(\Delta_{lat}^{\mcl{A}_{1u(2u)}}\right)^*~~
\Psi^{T}(\mbf{r})~\mcl{X}_{\mbf{rr'}}~\Psi(\mbf{r'})\right.\nonumber\\
&~~~~~~~~~~~~~~~~~~~~~~~~+\left.h.c.\right],
\end{align}
where $\mcl{X}_{\mbf{rr'}}$ is the pairing matrix between the NNN sites at $\mbf{r}$ and $\mbf{r'}$ that are pictorially shown in Fig.~\ref{Fig_lattice model_A_1u_2}, where $\Sigma_a, \Sigma_b$ and $\Sigma_c$ are the pairing matrices on the different NNN bonds. For the $\mcl{A}_{1u}$ singlet, these paring matrices respectively are
\begin{align}
-\frac{1}{2}(\Sigma_{24}+\sqrt{3}\Sigma_{25}),~~\Sigma_{24},~~\frac{1}{2}(\sqrt{3}\Sigma_{25}-\Sigma_{24}), 
\label{eq_pairingsinglet1}
\end{align}
while, for the $\mcl{A}_{2u}$ singlet, they are
\begin{align}
\frac{1}{2}(\Sigma_{25}-\sqrt{3}\Sigma_{24}),~~-\Sigma_{25},~~ \frac{1}{2}(\sqrt{3}\Sigma_{24}+\Sigma_{25}).
\label{eq_pairingsinglet2}
\end{align}

Each of the Cooper pair wave functions is symmetric in real space with weights at NNN sites, while the spin wave functions for the three bonds with $\Sigma_a, \Sigma_b, \Sigma_c$ (Fig. \ref{Fig_lattice model_A_1u_2}) are anti-symmetric and given by
\begin{align}
    |\Phi_a\rangle=&\frac{1}{2}|J_T=2, m_T=0\rangle+\sqrt{\frac{3}{8}}|J_T=2,m_T=2\rangle\nonumber\\
    &-\sqrt{\frac{3}{8}}|J_T=2,m_T=-2\rangle,\\
    |\Phi_b\rangle=&|J_T=2,m_T=0\rangle,
    \label{eq_phib1}\\
    |\Phi_c\rangle=&-\frac{1}{2}|J_T=2, m_T=0\rangle+\sqrt{\frac{3}{8}}|J_T=2,m_T=2\rangle\nonumber\\
    &-\sqrt{\frac{3}{8}}|J_T=2,m_T=-2\rangle,
\end{align}
for $\mcl{A}_{1u}$ and
\begin{align}
    |\Phi_a\rangle=&\frac{\sqrt{3}}{2}|J_T=2, m_T=0\rangle+\sqrt{\frac{1}{8}}|J_T=2,m_T=2\rangle,\nonumber\\
    &-\sqrt{\frac{1}{8}}|J_T=2,m_T=-2\rangle\\
    |\Phi_b\rangle=&\frac{1}{\sqrt{2}}\left(|J_T=2,m_T=2\rangle-|J_T=2,m_T=-2\rangle\right),
    \label{eq_phib2}\\
    |\Phi_c\rangle=&\frac{\sqrt{3}}{2}|J_T=2, m_T=0\rangle-\sqrt{\frac{1}{8}}|J_T=2,m_T=2\rangle\nonumber\\
    &+\sqrt{\frac{1}{8}}|J_T=2,m_T=-2\rangle
\end{align}
for $\mcl{A}_{2u}$.

Noticeably, while the spatial part of the wave-function is symmetric and as a result the spin-part is anti-symmetric, the structure of the latter depends on the actual direction of the bonds in the former (Fig. \ref{Fig_lattice model_A_1u_2}) in both the superconductors, due to the SOC, resulting in inversion odd SCs.
%%%%%%%%%%%%%%%%%%

\section{ Doublet Superconductors}
\label{sec_doubletsc}
There are four doublet bilinears, one with the irrep $\mcl{E}_g$ and three with $\mcl{E}_u$, given in Eq.~\ref{eq_decomposition_in_antisym_2}, \ref{eq_decomposition_in_sym_2}, \ref{eq_decomposition_in_sym_3} and \ref{eq_decomposition_in_sym_4} respectively. 
The three $\mcl{E}_u$ masses represent a single gapped SC, while the $\mcl{E}_g$ doublet represents a nodal SC. In the following, we first discuss the three $\mcl{E}_u$ doublets and then the $\mcl{E}_g$ doublet.
%%%%%%%%%%%%%

\subsection{The \texorpdfstring{3 $\mcl{E}_u$}{} Doublets}\label{sec_gapped_doublets}
The three $\mcl{E}_u$ doublets, denoted as $\mcl{E}_u^I, \mcl{E}_u^{II}$ and $\mcl{E}_u^{III}$, arise in Eq.~\ref{eq_decomposition_in_antisym_2}, \ref{eq_decomposition_in_sym_3} and \ref{eq_decomposition_in_sym_4} respectively. The low-energy Hamiltonian in the presence of $\mcl{E}_u^\mcl{x}$ (for $\mcl{x}=I,II,III$) doublet is given by Eq. \ref{eq:general term in Nambu basis} with the mass matrices, $M^{\mcl{E}^{\mcl{x}}_{u}}$ of the form given by Eq.~\ref{Eq_superconducting_mass_matrix} 
with the mass matrices $m_1^{\mcl{E}_u^{\mcl{x}}}$ and $m_2^{\mcl{E}_u^{\mcl{x}}}$  given by
\begin{align}
    \mcl{E}_u^I &:\left\{\begin{array}{l}
      m^{\mcl{E}^I_{u}}_1 =\Sigma_{24}\tau_1\sigma_0  \\
      m^{\mcl{E}^I_{u}}_2 =\Sigma_{25}\tau_1\sigma_0
    \end{array}\right.
    \end{align}
    \begin{align}
    \mcl{E}_u^{II} &:\left\{\begin{array}{l}
    m^{\mcl{E}^{II}_{u}}_1 =\frac{1}{\sqrt{6}}(\frac{1}{2}\Sigma_{35}-\frac{\sqrt{3}}{2}\Sigma_{34})\tau_3\sigma_2 -\sqrt{\frac{2}{3}}\Sigma_{15}\tau_2\sigma_0\\
    ~~~~~~~~+\frac{1}{\sqrt{6}}(\frac{\sqrt{3}}{2}\Sigma_5-\frac{1}{2}\Sigma_4)\tau_0\sigma_2\\
    m^{\mcl{E}^{II}_{u}}_2 =\frac{1}{\sqrt{2}}\frac{\Sigma_{35}-\sqrt{3}\Sigma_{34}}{2}\tau_3\sigma_2 -\frac{1}{\sqrt{2}}\frac{\sqrt{3}\Sigma_5-\Sigma_4}{2}\tau_0\sigma_2.
    \end{array}\right.
    \end{align}
    \begin{align}
    \mcl{E}_u^{III} &:\left\{\begin{array}{l}
    m^{\mcl{E}^{III}_{u}}_1 =-\frac{1}{\sqrt{6}}(\frac{1}{2}\Sigma_{34}+\frac{\sqrt{3}}{2}\Sigma_{35})\tau_3\sigma_2+\sqrt{\frac{2}{3}}\Sigma_{14}\tau_2\sigma_0  \nonumber \\ 
    ~~~~~~~~~ -\frac{1}{\sqrt{6}}(\frac{\sqrt{3}}{2}\Sigma_4+\frac{1}{2}\Sigma_5)\tau_0\sigma_2 \\
    m^{\mcl{E}^{III}_{u}}_2 =\frac{1}{\sqrt{2}}\frac{\Sigma_{34}+\sqrt{3}\Sigma_{35}}{2}\tau_3\sigma_2 -\frac{1}{\sqrt{2}}\frac{\sqrt{3}\Sigma_4+\Sigma_5}{2}\tau_0\sigma_2.
    \end{array}\right.
\end{align}

The corresponding pairing amplitudes, 
\begin{align}
    \Delta_{i}^{\mcl{E}^x_u}=\langle\chi^T m_i^{\mcl{E}^x_u}\chi\rangle~~~~~ (i=1,2),
\end{align}
can be parametrised as

\begin{align}\label{eq_sc_amp_E1u}
    (\Delta^{\mcl{E}^{\mcl{x}}_{u}}_1, \Delta^{\mcl{E}^{\mcl{x}}_{u}}_2)= \Delta^{\mcl{E}_u^{\mcl{x}}} e^{i\tilde{\phi}} (\cos \theta , \sin \theta e^{i\tilde{\gamma}}),
\end{align} 
where $\Delta^{\mcl{E}_u^{\mcl{x}}} = \sqrt{|\Delta_1^{\mcl{E}_u^{\mcl{x}}}|^2+|\Delta_2^{\mcl{E}_u^{\mcl{x}}}|^2}$, $\tilde{\phi},\tilde{\phi}+\tilde{\gamma} \in (0,2\pi]$ are the phases of the superconducting amplitudes and $\theta  = \tan^{-1}\left(|\Delta_2^{\mcl{E}_u^{\mcl{x}}}|/|\Delta_1^{\mcl{E}_u^{\mcl{x}}}|\right)$. For a general phase difference $(\tilde\gamma\neq 0)$, the SC breaks the microscopic time-reversal symmetry as explained in Sec.~\ref{sec_classification}. 

All three doublets give rise to non-unitary SCs, and the details of the Bogoliubov spectrum depend on the values of the various parameters in the superconducting amplitude (Eq. \ref{eq_sc_amp_E1u}). As mentioned earlier, these three $\mcl{E}_u$ doublets represent a single superconducting phase. This can be demonstrated through a similar analysis as done in Sec.~\ref{sec_a1g_singlets}. We can find an adiabatic path between any two of them without closing the quasi-particle energy gap and without breaking further microscopic symmetries. Thus, we will discuss the $\mcl{E}_u^I$ doublet in detail below, while the Irreps $\mcl{E}^{II}_{u}$ and $\mcl{E}^{III}_{u}$ are discussed in Appendix \ref{sec_E_u_doublet_II_III}.
%%%%%%%%%%%%%%%%%

\subsubsection{\texorpdfstring{$\mcl{E}_u^I$}{} Doublet}\label{sec_eu_doublet1}

We rewrite the mass matrix (Eq. \ref{Eq_superconducting_mass_matrix}) as
\begin{align}
    m^{\mcl{E}^I_{u}}({\bf d}) = \Delta^{\mcl{E}_u^1}\left({d}_1^{\mcl{E}_u^I}m^{\mcl{E}^I_u}_1+ {d}_2^{\mcl{E}_u^I}m^{\mcl{E}^I_u}_2\right) =|\Delta^{\mcl{E}^{I}_{u}}|~{\bf d}\cdot{\bf m^{\mcl{E}^I_{u}}},
\end{align}
where 
\begin{align}\label{eq_E_u_dublet_parametrization}
{\bf d}=~e^{i\tilde{\phi}}\left(\cos{\theta},~\sin{\theta}~e^{i\tilde{\gamma}}\right)
\end{align}
is a two-component complex vector and $|\Delta^{\mcl{E}^{I}_{u}}|$ is the magnitude of the pairing. Checking for unitarity, we have
\begin{align}
     &\left[\boldsymbol{\Delta}^{\mcl{E}^I_{u}}\cdot{\bf m^{\mcl{E}^I_{u}}}\right]\cdot \left[\boldsymbol{\Delta}^{\mcl{E}^I_{u}}\cdot{\bf m^{\mcl{E}^I_{u}}}\right]^{\dagger} =\nonumber  \\ 
&\quad|\boldsymbol{\Delta^{\mcl{E}^I_u}}|^2\left(\Sigma_0\tau_0\sigma_0 + \frac{i}{2}\sin(2\theta)\sin(\tilde\gamma) \left[m^{\mcl{E}^1_u}_1,m^{\mcl{E}^1_u}_2\right]
\right), \label{eq_eu1nonunitarity} \\
&\text{where}~
\frac{1}{2i}[m^{\mcl{E}^1_u}_1,m^{\mcl{E}^1_u}_2] =\Sigma_{45}\tau_0\sigma_0. \nonumber 
\end{align}
The eigenvalues of Eq. \ref{eq_eu1nonunitarity} are given by
\begin{align}
     \lambda^{\mcl{E}^{I}_u}_{\pm} = (\Delta^{\mcl{E}^I_u})^2\left( 1\pm\sin(\tilde{\gamma})\sin(2\theta)\right), %\nonumber   \\
% = |\boldsymbol{\Delta^{\mcl{E}^I_u}}|^2\left( 1\pm \frac{1}{i}\left({\bf d}\times {\bf d^{*}}\right)\right).
\end{align}
(with each being 8-fold degenerate) such that for the TR invariant (breaking) manifold, $\tilde\gamma= 0~(\theta,\tilde\gamma\neq 0)$, and we get a unitary (non-unitary) SC and consequently a single (double) gap structure. Interestingly, on the isolated points within the TR breaking manifold of SCs, when $\theta=\pm \pi/4, \pm 3\pi/4$ and $\tilde\gamma=\pm\pi/2$, corresponding to
\begin{align}
\sin(\tilde\gamma)\sin(2\theta) = \pm 1,
\label{eq_isonode}
\end{align}
when one of the eigenvalues vanishes, leading to the collapse of the smaller of the two pairing gaps, we have 8-fold gapless (nodal) Bogoliubov excitations.

\begin{figure}
\includegraphics[width=.55\columnwidth]{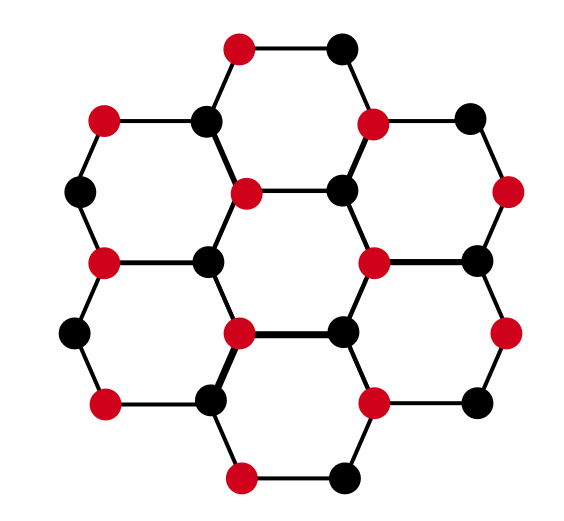}
\caption{Schematic figure for the pairing amplitude of the $\mathcal{E}^I_{u}$ on the lattice in the global basis (Eq. \ref{eq_doublet_onsite_y_r}). Black dots represents onsite pairing with pairing matrix $\Sigma_i\in (\Sigma_{24},\Sigma_{25})$ and Red dots represents onsite pairing with pairing matrix $-\Sigma_i$ with $\{\Sigma_{24}, \Sigma_{25}\}$ being the two members of the doublet. The relative negative sign of the pairings on the two sublattices ensures that the SC is odd under inversion.}
    \label{Fig_lattice_model_E_u}
\end{figure}

The above low-energy theory is obtained from a mean field lattice Hamiltonian in the global basis of the form similar to Eq.~\ref{Lattice_ham_in_3/2_basis_2} with onsite pairing between the $j=3/2$ electrons given by
\begin{align}\label{eq_doublet_onsite_y_r}
\sum_{\bf r}f(\textbf{r})&\left(\left(\Delta^{\mcl{E}^I_u}_{lat,1}\right)^*\Psi^T(\textbf{r})\,\Sigma_{24}\,\Psi(\textbf{r})\right.\nonumber\\
&\left.+\left(\Delta^{\mcl{E}^I_u}_{lat,2}\right)^*\Psi^T(\textbf{r})\,\Sigma_{25}\,\Psi(\textbf{r})\right). \nonumber \\
\end{align}
Here, \( f(r) \) (\(= \pm 1 \)) is a function defined on the lattice sites \( \mbf{r} \) (see Fig.~\ref{Fig_lattice_model_E_u}). This modulation \( f(\mbf{r}) \) breaks the point group symmetry while preserving the translational symmetry of the honeycomb lattice, and corresponds to staggered on-site pairing.
%where $f(\mbf{r})(=\pm 1)$ is a function of the lattice site $\mbf{r}$  (see Fig.~\ref{Fig_lattice_model_E_u}). The modulation $f({\bf r})$ breaks point group symmetry, but not the translations of the honeycomb lattice, and corresponds to staggered on-site pairing. 
The corresponding anti-symmetric spin wave function of the Cooper pair is given by Eqs. \ref{eq_phib1} and \ref{eq_phib2} respectively for the two components. Thus, the Cooper pairs for this doublet are made up of a linear combination of a subset of anti-symmetric $J_T=2 ~(m_T=0,\pm 2)$ spin multiplet states. 

Further insights about the staggered on-site pairing become apparent by going back to the continuum limit, whence using Eq. \ref{eq_E_u_dublet_parametrization} we get
\begin{align}\label{eq_E_g_doublet_secondary_OP}
    \hat{\bf n}\equiv(n_1,n_2,n_3)={\bf d}^\dagger\mathbf{\sigma}{\bf d}=(\cos\tilde{\gamma}\sin2\theta,\sin \tilde{\gamma}\sin2\theta,\cos2\theta)
\end{align}
such that $(n_1,n_3)$ transforms as $\mathcal{E}^e_{g}$, and $n_2$ transforms as $\mathcal{A}_{2g}^o$ under lattice symmetries. The TRI sub-manifold is then spanned by $n_2=0$, while the pure TR breaking manifold is given by $n_1=n_3=0$, with the latter being an isolated point corresponding to $\theta=\pm\pi/4, \tilde{\gamma}=\pm\pi/2$. Note that at these isolated points, the smaller of the two gaps vanishes, and we get a gapless TR breaking SC (Eq. \ref{eq_isonode}). The underlying microscopic symmetries allow for the leading order anisotropic term 
\begin{align}
  \lambda_1n_2^2 +  \lambda_2\left( n_1^2+n_3^2 \right)
  \label{eq_anisotropysecondaryeg}
\end{align}
in the Landau free energy, which dictates details of whether the TR invariant/breaking manifold is chosen, depending on the signs of the coupling constants $\lambda_1, \lambda_2$.

\paragraph*{Vortex structure:}It is interesting to look at the possible topological defects in the presence of this doublet mass. Apart from the usual U(1) vortex in the Nambu sector (texture in the real and imaginary part of superconducting amplitude), we find that there is also another stable vortex defect possible, which we discuss now. This is clear by noting that the order parameter (Eq. \ref{eq_E_u_dublet_parametrization}) lives in $(S^1\times S^2)/Z_2$ manifold, similar to the spinor bosons~~\cite{PhysRevLett.87.080401,Ueda_2014,subrotopaper} which is broken down to $(S^1\times S^1)/Z_2$~\cite{PhysRevLett.87.080401,subrotopaper} in the TRI subspace. Thus, the vortices generically are characterized by $\pi_1((S^1\times S^2)/Z_2)=Z$ and reduces to $\pi_1((S^1\times S^1)/Z_2)=Z\times Z$~\cite{PhysRevLett.87.080401,Ueda_2014,subrotopaper} in the TRI manifold. 
%%%%%%%%%%%%%%%%%%%%

\subsection{The Gapless Doublet Mass (\texorpdfstring{$\mcl{E}_g$}{}) }
\label{sec_doubleteg}

\begin{figure}%[h]
    \begin{subfigure}[]
    {\includegraphics[width=0.90\columnwidth]{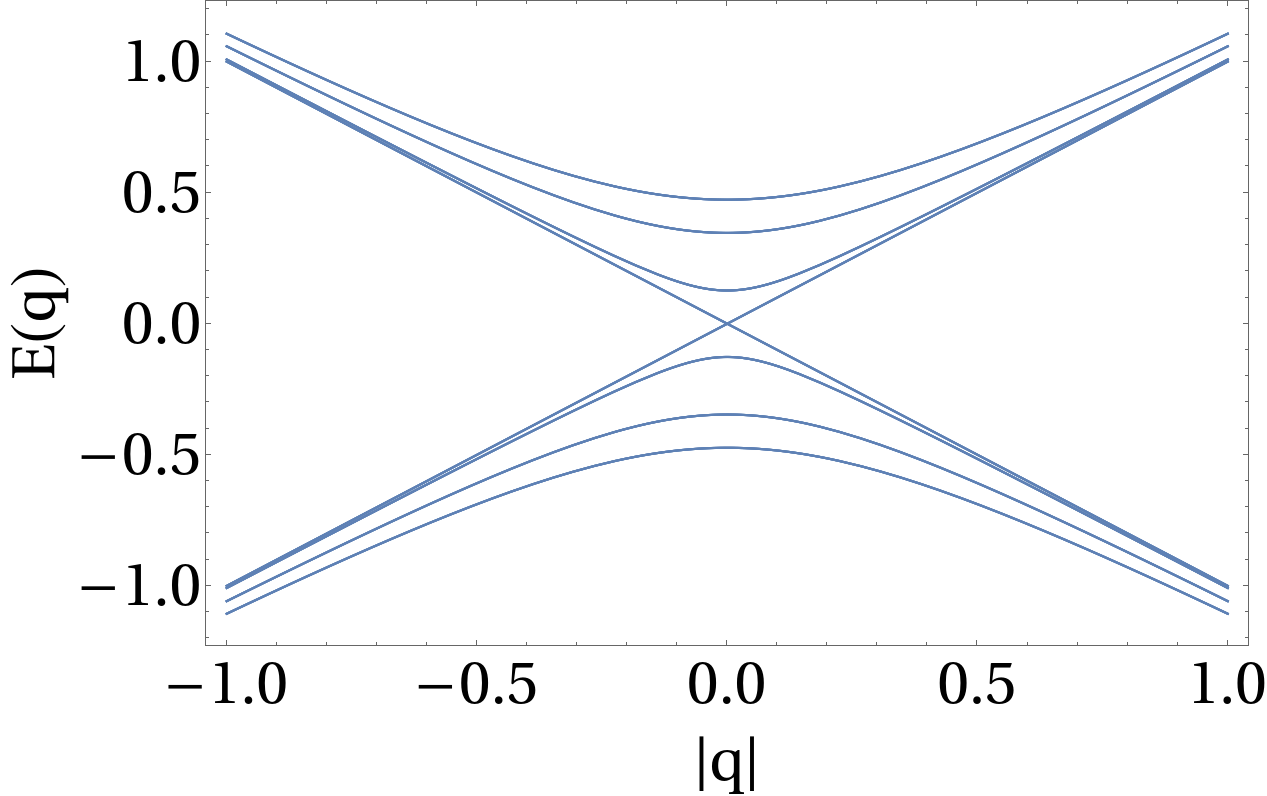}}
    \label{fig_Gapless Doublet_general_Spectrum}
    \end{subfigure}
    \begin{subfigure}[]{
         \includegraphics[width=0.90\columnwidth]{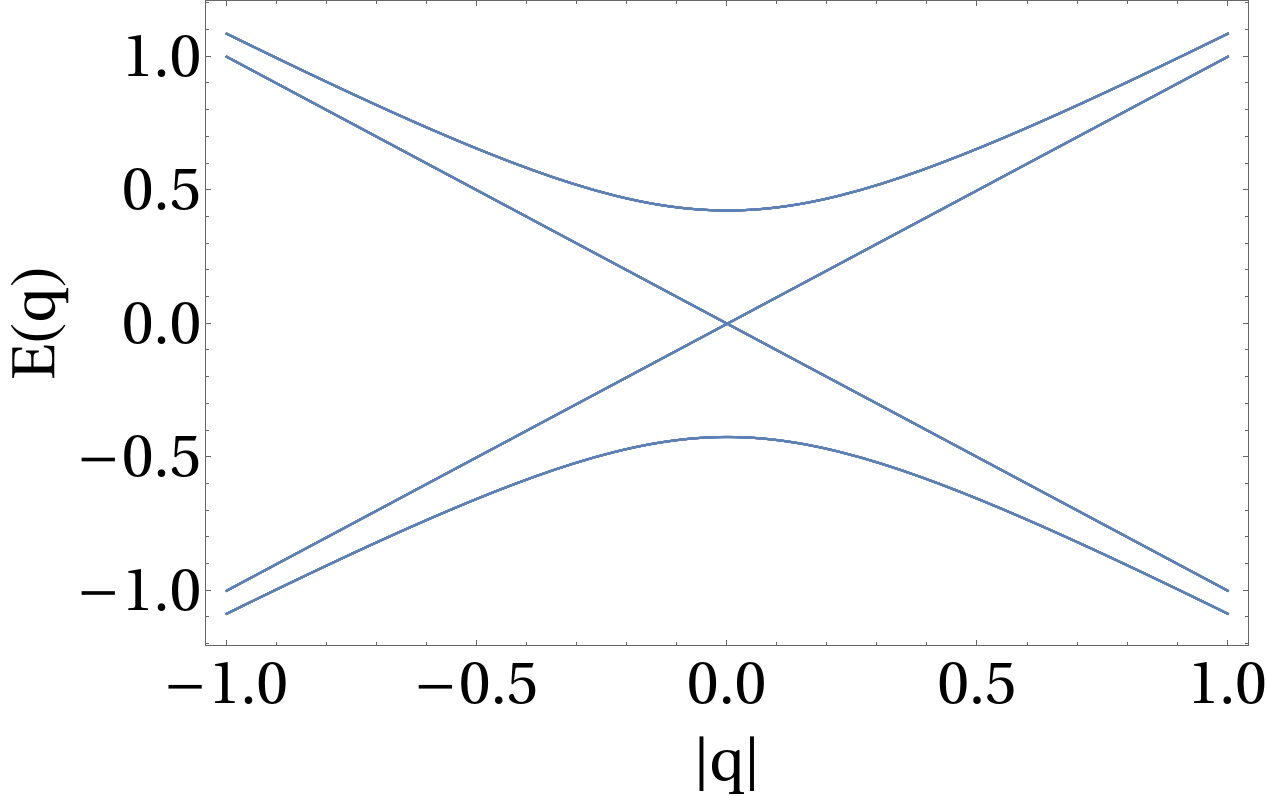}}
           \label{fig_Gapless Doublet_special_Spectrum}
    \end{subfigure}
    \caption{ Gapless Doublet Spectrum ({for $|\Delta^{\mcl{E}_{g}}|=0.30$}) (a) For general values of $\theta$ (here it is shown for $\theta=\pi/4$), there are 8 bands each of them is 8-fold degenerate. Out of these 8 bands, two bands touch linearly at $\bf q=0$, providing 8-gapless modes.  (b) For $\theta \in \{\pm\pi/3,\pm2\pi/3, 0\}$ (here it is shown for $\theta=\pi/3$), Apart from the usual two bands, extra bands are touching each other at $\bf q=0$, providing 16-gapless modes.}
    \label{fig_gapless_doublet_spectrum}
\end{figure}

The $\mcl{E}_g$ doublet masses arise from direct product of the $\mcl{T}_{2g}$ lattice Irrep in flavour and the $\mcl{T}_{2g}$ Irrep in valley-subband sector (see Eq.~\ref{eq_decomposition_in_sym_2}).
The low-energy Hamiltonian in the presence of this doublet mass is given by Eq. \ref{eq:general term in Nambu basis}, with the mass matrix, {$M^{\mcl{E}_{g}}$, given }by a form similar to Eq. \ref{Eq_superconducting_mass_matrix} with

\begin{eqnarray}\label{eq_E_g_doublet_matrices}
&& m^{\mcl{E}_{g}}_1 =\frac{1}{\sqrt{2}}(\Sigma_0\tau_0\sigma_2-\Sigma_{12}\tau_3\sigma_2), 
\\
&& m^{\mcl{E}_{g}}_2 =\frac{1}{\sqrt{6}}(\Sigma_0\tau_0\sigma_2+\Sigma_{12}\tau_3\sigma_2+2\Sigma_{23}\tau_2\sigma_0), 
\end{eqnarray}
and 
\begin{eqnarray}\label{eq_sc_amp_A1u_2}
    \Delta^{\mcl{E}_{g}}_1 = \langle \chi^T \,m^{\mcl{E}_{g}}_1\,\chi\rangle \,\,\, , \,\,\,\Delta^{\mcl{E}_{g}}_2 = \langle \chi^T \,m^{\mcl{E}_{g}}_2\,\chi  \rangle, 
\end{eqnarray}
which can be parametrised in a way similar to Eq. \ref{eq_sc_amp_E1u} and gives rise to a non-unitary SC which is TR invariant only for $\gamma=0,\pi$ and breaks it otherwise. 

\begin{figure}%[h]\label{Fig_lattice_model_E_u_12}
\includegraphics[width=1\columnwidth]{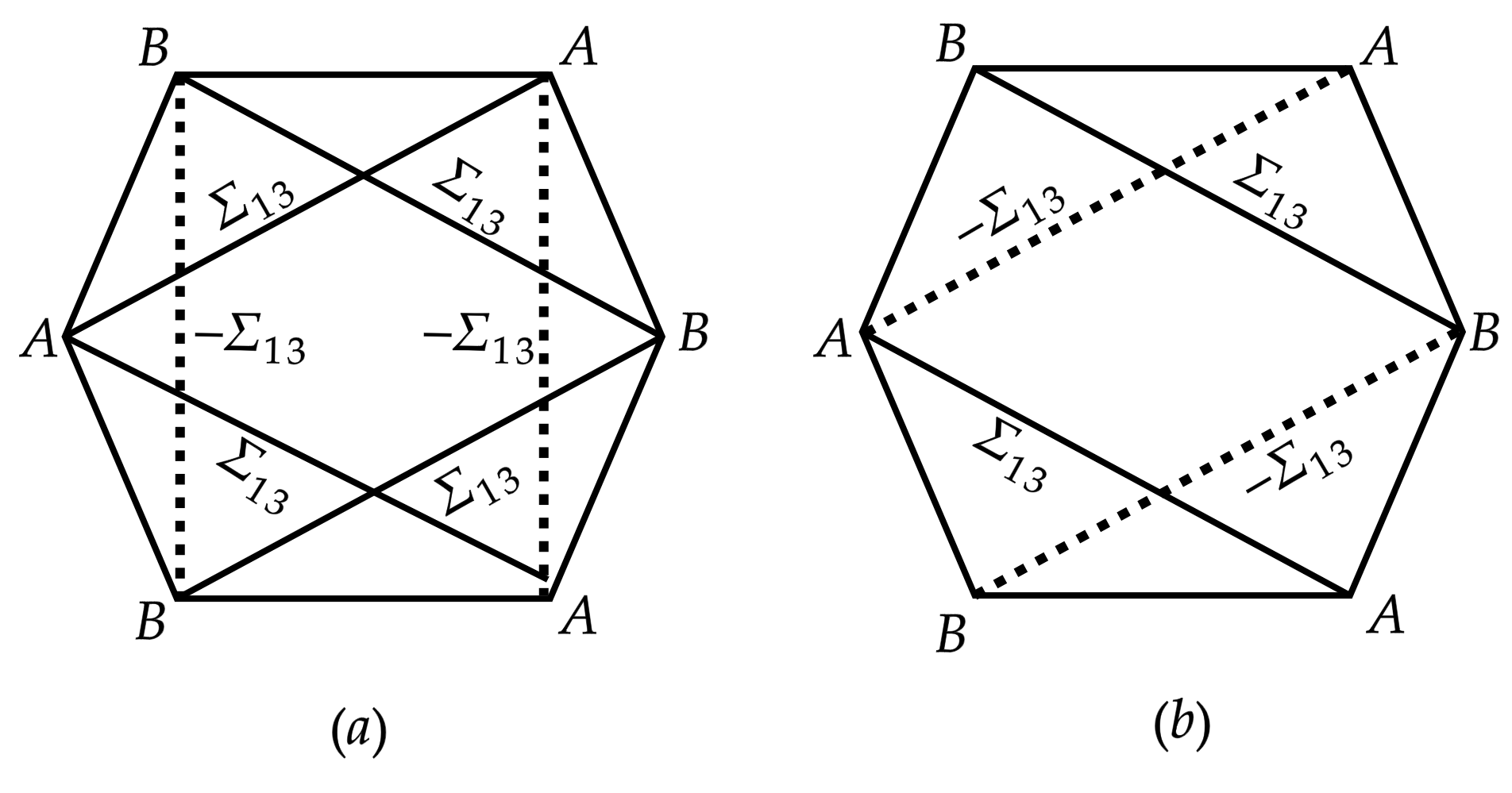}
% \label{Fig_lattice_model_E_u_2}
    \caption{Schematics of the pairing for the two components ($m^{\mathcal{E}_g}_2$, $m^{\mathcal{E}_g}_1$) of the $\mcl{E}_g$ doublet on the lattice in the global basis (Eq. \ref{eq_latticepairingeg}). The pairing amplitudes are on NNN bonds, and the solid (dashed) lines are related to each other by the change in sign of the pairing matrix as indicated.}
      \label{Fig_lattice_model_E_u_12}
  \end{figure}

Notably, however the eigenvalues of the matrix $m^{\mcl{E}_g}({\bf d}).(m^{\mcl{E}_g}({\bf d}))^{\dagger}$ is
\begin{align}
   &\Bigg(0,~ \frac{8}{3}(\Delta^{\mcl{E}_g})^2\sin^{2}(\theta),\nonumber \\
    &\frac{1}{3}(\Delta^{\mcl{E}_g})^2\left( 4+2\cos(2\theta)+2\sqrt{3}\cos(\tilde{\gamma})\sin(2\theta)
    \right),\nonumber \\
     &\frac{1}{3}(\Delta^{\mcl{E}_g})^2\left( 4+2\cos(2\theta)-2\sqrt{3}\cos(\tilde{\gamma})\sin(2\theta)
    \right)\Bigg),
\end{align}
with each being 4-fold degenerate. Thus, the spectrum of the Bogoliubov quasi-particles is generically gapless, with the degeneracy of the number of gapless nodes varying over the order parameter manifold. The resultant low-energy Bogoliubov spectrum is shown in Fig.  \ref{fig_gapless_doublet_spectrum}(a) whose origin can be traced back to the Dirac nodes at the $\Gamma$-point in the global basis -- similar to the case of $\mathcal{A}_{1u}$ and $\mathcal{A}_{2u}$ singlets discussed in Sec. \ref{sec_gapless_singlets} above (also Appendix \ref{sec_symmetry_analysis_gapless_modes}). Extra isolated gapless modes can appear on the TRI subspace ($\tilde\gamma=0,\pi$) for $\theta=0,\pi,\pm\pi/3,\pm 2\pi/3$ as shown in Fig. \ref{fig_gapless_doublet_spectrum}(b).

On the lattice, these doublet masses correspond to NNN pairing. The equivalent lattice Hamiltonian in this case is similar to that given in Eq.~\ref{eq_lattice_ham_nnn} with the pairing amplitudes on the NNN bonds (in the global basis) being given by
\begin{eqnarray}
    \mcl{X}_{\mbf{rr'}} = f(\mbf{r,r'})\Sigma_{13}={ -f({\bf r,r'})\frac{7}{3}\left(J_y-\frac{4}{7}J_y^3\right)}
    \label{eq_latticepairingeg}
\end{eqnarray}
Here, $f(\mbf{r,r'})(=\pm 1)$ is depends on the particular NNN bond. For $\Delta_2^{\mcl{E}_g}=0$, $f(\mbf{r,r'})$ is pictorially shown in Fig.~\ref{Fig_lattice_model_E_u_12}(a). Similarly, Fig.~\ref{Fig_lattice_model_E_u_12}(b) shows the pairing pattern for $\Delta_1^{\mcl{E}_g}=0$. The resultant Cooper-pair wave-function is an anti-symmetric spin singlet and is given by Eq. \ref{eq_jtsinglet}, while the spatial part is symmetric with modulation on the NNN bonds, as shown in the figures. Thus, these too correspond to nodal SCs with anisotropic NNN pairing.
%%%%%%%%%%%%%%%%%%%%%%%%%

\section{Triplet Superconductors}
\label{sec_tripletsc}
Finally, we turn to the triplet SCs. There are eight triplet Irreps (Eqns~\ref{eq_decomposition_in_sym_1} - \ref{eq_decomposition_in_sym_4}) : (3)$\mcl{T}_{1g}$, (2)$\mcl{T}_{1u}$, (2)$\mcl{T}_{2u}$, and $\mcl{T}_{2g}$. These irreps result in {\it six} distinct superconducting phases {since the two $\mcl{T}_{1u}$ give rise to a single SC, and so does the two $\mcl{T}_{2u}$}. Specifically, two of the $\mcl{T}_{1g}$, the $\mcl{T}_{1u}$, and the $\mcl{T}_{2u}$ are gapped SCs, while the remaining $\mcl{T}_{1g}$ and the $\mcl{T}_{2g}$ are SCs with gapless Dirac nodes. All these SCs correspond to non-unitary pairing generically as we describe below. 

%%%%%%%%%%%%%%%%%%%%
\subsection{The \texorpdfstring{$\mcl{T}_{1g}$}{} Triplet superconductors }
\label{sec_t_1g}

There are three $\mcl{T}_{1g}$ triplets corresponding to Eqs. \ref{eq_decomposition_in_antisym_3}, \ref{eq_decomposition_in_sym_1} and \ref{eq_decomposition_in_sym_2}. All three triplets generically break TR, but have extended TRI sub-space. While the first one is a flavour anti-symmetric, the other two are flavour symmetric. However, in the TRI sub-manifold, we find that the three triplets cannot be adiabatically connected without breaking further microscopic lattice symmetries or time reversal. Particularly once TR is broken, the three SCs can be adiabatically connected. Further, while two of them are generically gapped, the third one is a nodal SC. Due to this, we consider them as separate SCs which we denote by $\mcl{T}_{1g}^x$ with $x=I, II, III$. We discuss each of them now separately below.

The low-energy Hamiltonian for these SCs is given in Eqns.~\ref{eq:general term in Nambu basis} and \ref{Eq_superconducting_mass_matrix} for $\mcl{T}_{1g}^{\mcl{x}}$ (for $\mcl{x}=I,II, III$)  with mass matrices of the form
\begin{align}
    \mcl{T}_{1g}^I &:\left\{\begin{array}{l}
    m_{1}^{\mcl{T}^{I}_{1g}} =\Sigma_{1}\tau_1\sigma_0,\\ m_{2}^{\mcl{T}^{I}_{1g}}=\Sigma_3\tau_1\sigma_0,\\
    m_{3}^{\mcl{T}^{I}_{1g}}=i\Sigma_{45}\tau_{1}\sigma_0.
    \end{array}\right.
    \label{eq_t1g1mass}
    \end{align}
    \begin{align}
    \mcl{T}_{1g}^{II} &:\left\{\begin{array}{l}
    m_{1}^{\mcl{T}^{II}_{1g}} =\Sigma_{2}\tau_2\sigma_0,\\
    m_{2}^{\mcl{T}^{II}_{1g}}=\Sigma_2\tau_3\sigma_2,\\
    m_{3}^{\mcl{T}^{II}_{1g}}=i\Sigma_{2}\tau_{0}\sigma_2.
    \end{array}\right.
    \label{eq_t1g2mass}
\end{align}
and
\begin{align}
\mcl{T}_{1g}^{III}=\left\{\begin{array}{l}
m_{1}^{{\mcl{T}}_{1g}} =(\Sigma_0\tau_3\sigma_2-\Sigma_{12}\tau_0\sigma_2)/\sqrt{2}\\
m_{2}^{{\mcl{T}}_{1g}}=(-\Sigma_{23}\tau_0\sigma_2-\Sigma_{0}\tau_2\sigma_0)/\sqrt{2}\\   
m_{3}^{{\mcl{T}}_{1g}}= i(\Sigma_{12}\tau_2\sigma_0-\Sigma_{23}\tau_3\sigma_2)/\sqrt{2}.
\end{array}\right.
\label{eq_massmatt1ggapless}
\end{align}
The corresponding pairing amplitudes are given by
\begin{eqnarray}
    \Delta_i^{\mcl{T}_{1g}^{\mcl{x}}} = \braket{\chi^T m_i^{\mcl{T}_{1g}^{\mcl{x}}}\chi}~~~~(i=1,2,3).
\label{eq_deltat1g}
\end{eqnarray}

As discussed in Sec.~\ref{sec_classification}, that triplet breaks the microscopic time-reversal symmetry for a general value of the superconducting amplitudes, which, for each triplet, can be generically parametrised as 
\begin{align}
    \boldsymbol{\Delta}^{\mcl{T}^x_{1g}}=|\Delta^{\mcl{T}^{x}_{1g}}|~{\bf d}^x
    \label{eq_paramtrip}
\end{align}
where ${\bf d}^x\equiv(d^x_1, d^x_2, d^x_3)$ is a 3-component complex vector that spans the order parameter manifold $(S^1\times CP^2)/Z_2$~\cite{massey_1973,kuiper_1974} via
\begin{align}\label{eq_triplet_in _theta_phi_I}
(d_1^x, d_2^x, d_3^x) = e^{i\tilde{\phi}}(\cos \theta, e^{i\tilde{\gamma}_1}\sin \theta \cos \phi, e^{i \tilde{\gamma}_2}\sin \theta \sin \phi).
\end{align}
where, $\tilde\phi$ is the superconducting phase and $\theta, \tilde\gamma_1$ and $\tilde\gamma_2$ specify the direction in the triplet space. Note that the TRI sub-manifold is given by $\tilde\gamma_1=\tilde\gamma_2=0$ or $\theta=0,\pi$ {whence the order parameter manifold reduces to $(S^1\times S^2)/Z_2$ and corresponding vortices are characterized by $\pi_1((S^1\times S^2)/Z_2)= Z$.} 
%%%%%%%%%%%%%%%%%%%%%%%%

\subsubsection{\texorpdfstring{$\mcl{T}_{1g}^I$}{} Triplet}

A general mass term for this triplet is written as,
\begin{align}
    m^{\mcl{T}^I_{1g}}({\bf d}) = |\Delta^{\mcl{T}_{1g}^I}|\left({d}_1^{I}m^{\mcl{T}^I_{1g}}_1+ {d}_2^{I}m^{\mcl{T}^I_{1g}}_2+{d}_3^{I}m^{\mcl{T}^I_{1g}}_3\right) \nonumber \\=\boldsymbol{\Delta}^{\mcl{T}^I_{1g}}\cdot{\bf m}^{\mcl{T}^I_{1g}},
\end{align}
such that the unitarity condition leads to
\begin{align}\label{Eq_t1g_unitary}
 \left[ \boldsymbol{\Delta}^{\mcl{T}^I_{1g}}\cdot{\bf m}^{\mcl{T}^I_{1g}}\right] \cdot  \left[ \boldsymbol{\Delta}^{\mcl{T}^I_{1g}}\cdot{\bf m}^{\mcl{T}^I_{1g}}\right]^{\dagger} &= |\Delta^{\mcl{T}^{I}_{1g}}|^2  \times \nonumber \\ \Bigg(\Sigma_0\tau_0\sigma_0  + \frac{1}{2} \left({\bf d}^I \times {{\bf d}^I}^{*}\right)&\cdot \left({\bf m}^{\mcl{T}^{I}_{1g}} \times {{\bf m}^{\mcl{T}^{I}_{1g}}}^{\dagger}\right) \Bigg),
 \end{align}
which implies it is unitary provided $\bf d^I\times {d^I}^*=0$, {\it i.e.}, on the TRI manifold. On the other hand, on the TRB manifold, there are two gaps (as can be obtained by diagonalizing Eq. \ref{Eq_t1g_unitary}) generically except at when $\vert{\bf d}^I\times {{\bf d}^I}^* \vert  =\pm 1$ whence the smaller of the two gaps collapse leading to a nodal SC. However, unlike the doublets ({\it e.g.}, Eq. \ref{eq_isonode}), the above condition of obtaining a node can be satisfied on extended sub-spaces of the order-parameter manifold. 

\begin{figure}%[t]
\includegraphics[width=.55\columnwidth]{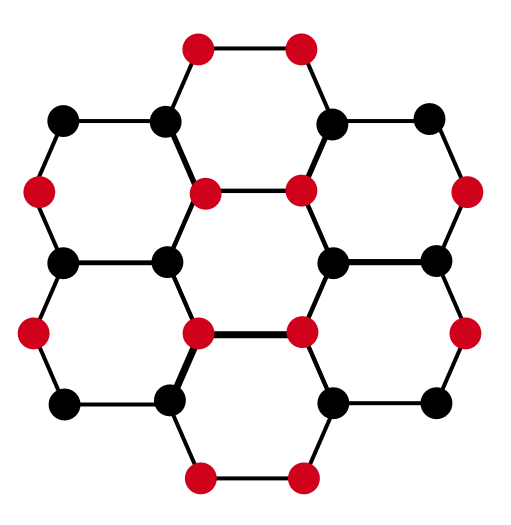}
    \caption{Schematics of the pairing for $m_1^{\mcl{T}^I_{1g}}$ (Eq. \ref{eq_t1g1mass}) bilinear belonging to the ${\mcl{T}^I_{1g}}$ triplet in global basis (Eq. \ref{eq_tiglatticeham}). Here, Black (Red) dots represent on-site pairing in the flavour sector with pairing matrix $\Sigma_1(-\Sigma_1)$. {Pairing amplitudes oscillates at momenta corresponding to $M_2$ point in the BZ(Fig.~\ref{fig_BZ}).} Lattice model for the other two components of the triplets can be generated by acting with ${\bf S_6}$ (Table.~\ref {tab_triplet_irrep}).}
\label{fig_Triplet_lattice_model_T_1G_onsite}
     \end{figure}

The corresponding lattice Hamiltonian (for mass matrix $m_{1}^{\mcl{T}^I_{1g}}$) has a form similar to Eq.~\ref{Lattice_ham_in_3/2_basis_2} with on-site pairing term ($\mcl{Y}_{\mbf{r}}$) at the lattice site $\mbf{r}$ is given by
\begin{eqnarray}
\mcl{Y}_{\mbf{r}}=&\left(\Delta^{\mcl{T}^I_{1g}}_{lat,1}\right)^*\left(f(\mbf{r})\Psi^T(\mbf{r})\Sigma_1\Psi(\mbf{r})\right), 
\label{eq_tiglatticeham}
\end{eqnarray}
where $f(\mbf{r}) (= \pm 1)$ is a function that creates a vertical stripy pattern of pairing on the honeycomb lattice, as illustrated in Fig.~\ref{fig_Triplet_lattice_model_T_1G_onsite}. As is evident from the figure, the pairing breaks translation symmetry in addition to point group symmetries. The pairing amplitude is at a finite momentum corresponding to the $M_2$ point in the BZ(Fig.~\ref{fig_BZ}). The spin-wave function of the Cooper pair is anti-symmetric and is given by
\begin{align}
|\Phi_1\rangle=&\frac{1}{\sqrt{2}}\left(|J_T=2,m_T=2\rangle+|J_T=2,m_T=-2\rangle\right)
\end{align}
which should be contrasted with Eq. \ref{eq_phib2}.

The lattice Hamiltonian for the other two masses in this triplet can be obtained by applying $C_3$ rotations and  correspond to the other two stripy patterns on the honeycomb lattice, with {$\braket{\Psi^T\Sigma_3\Psi}$} and $\braket{\Psi^T\Sigma_{45}\Psi}$ pairings, with the corresponding spin-wave function for the Cooper pair being
  \begin{align}
      |\Phi_3\rangle=\frac{1}{\sqrt{2}}\left(|J_T=2,m_T=1\rangle-|J_T=2,m_T=-1\rangle\right),
  \end{align}
  and
  \begin{align}
      |\Phi_{45}\rangle=\frac{1}{\sqrt{2}}\left(|J_T=2,m_T=1\rangle+|J_T=2,m_T=-1\rangle\right).
  \end{align}
  
  Therefore, these correspond to pair density wave (PDW) SCs~\cite{PhysRevLett.88.117001,PhysRevLett.99.127003,agterberg2020physics} (with or without TR) where the pairing amplitude oscillates at finite momentum. The nature of these PDWs and the splitting of their degeneracies can be studied using methods similar to Eq. \ref{eq_E_g_doublet_secondary_OP}, albeit using SU(3) matrices (see Appendix \ref{appen_su3secondary}).

%%%%%%%%%%%%%%%%%%%%%%%

\subsubsection{\texorpdfstring{$\mcl{T}_{1g}^{II}$}{} triplet}

The analysis of this triplet (spectrum and unitarity condition) is very similar to $\mcl{T}_{1g}^I$ discussed above, leading to a PDW SC with a two-gap structure on the TRB manifold, which reduces to a single gap on the TRI subspace, as before. 

\begin{figure}%[h!]
\includegraphics[width=0.42\textwidth, height=.28\textwidth]{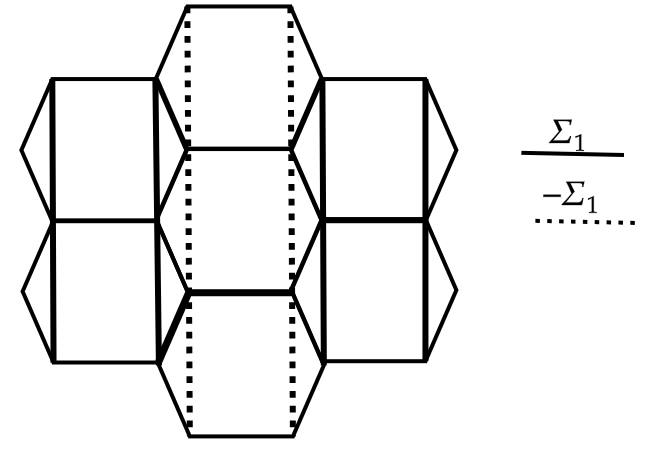}
    \caption{Schematic of NNN pairing for the bilinear $m_1^{\mcl{T}^{II}_{1g}}$ (Eq. \ref{eq_t1g2mass}) component of $(\mcl{T}^{II}_{1g})$ in global basis (Eq. \ref{eq_t1g2nnnpairing}). Here dotted (Solid) line represents pairing in the flavour sector with Pairing matrix $\Sigma_1(-\Sigma_1)$. Pairing amplitudes oscillate at momenta corresponding to the $M_2$ point in the BZ(Fig.~\ref{fig_BZ}).Pairings for the two other components of the triplet can be obtained by acting with ${\bf S_6}$ on the present one (Table.~\ref{tab_triplet_irrep}).}
\label{fig_Triplet_lattice_model_T_1G_2_onsite}
     \end{figure}

The lattice Hamiltonian is similar to that given in Eq.~\ref{eq_lattice_ham_nnn}. For the $m_1^{\mcl{T}^{II}_{1g}}$  mass of this triplet, the pairing amplitude on the NNN bond connecting the lattice sites at $\mbf{r}$ and $\mbf{r'}$ is given by
\begin{eqnarray}
    \mcl{X}_{\mbf{rr'}} = f(\mbf{r,r'})\Sigma_1.
\label{eq_t1g2nnnpairing}
\end{eqnarray}
Here, $f(\mbf{r,r'})(=\pm 1)$ is such that it forms a vertical stripy pattern on the honeycomb lattice (see Fig.~\ref{fig_Triplet_lattice_model_T_1G_2_onsite}). The lattice model for other masses of the triplet can be obtained by action of ${\bf S}_6$, which leads to NNN pairing with stripy pattern along $M_1$ and $M_3$ momenta (Fig. \ref{fig_BZ}). Thus, this is a NNN version of the $\mcl{T}^{I}_{1g}$ triplet.
 
%%%%%%%%%%%%%%%%
\subsubsection{\texorpdfstring{$\mcl{T}_{1g}^{III}$}{} triplet}\label{sec_t1g_III}

\begin{figure}%[h!]
   \begin{subfigure}[]{
    \includegraphics[width=.92\columnwidth]{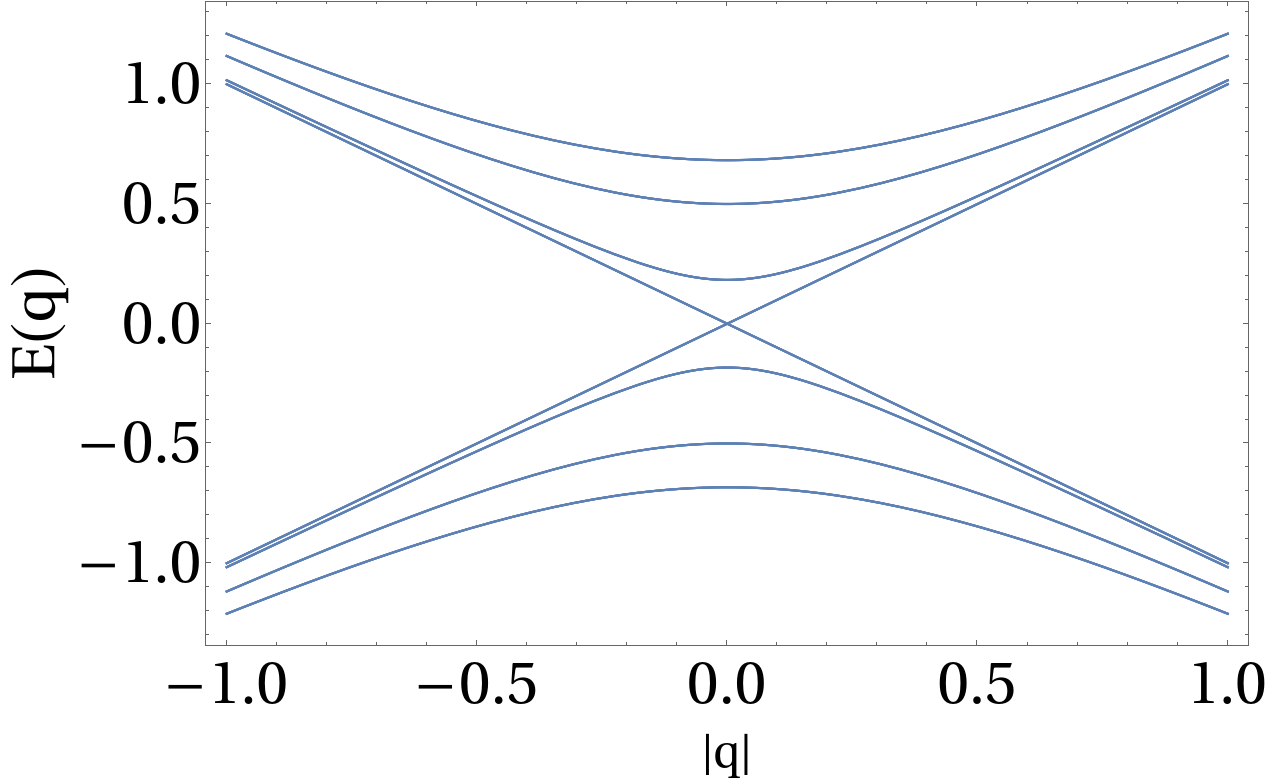}}    \label{fig_Gapless triplet_general_Spectrum}
    \end{subfigure}
    \begin{subfigure}[]{
        \includegraphics[width=.80\columnwidth]{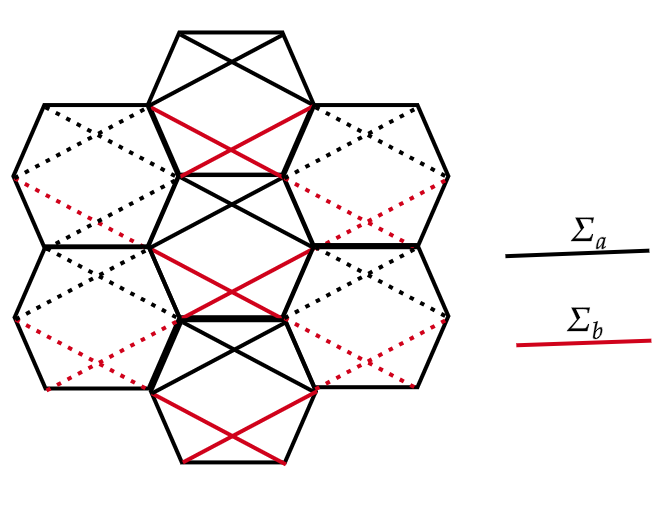}}\end{subfigure}
    \caption{(a) Spectrum for gapless triplet $\mcl{T}_{1g}$  on TRI manifold, for general ($\theta,\phi$) and { (for $|\Delta^{\mcl{T}_{1g}}|=0.31$}). There are two bands which touch each other linearly at $\mbf{q}=0$; each is four-fold degenerate. For values of $(\theta, \phi)$ on the Great circle (Fig.~\ref{triplet parameter space}), the number of bands touching each other at $\bf q=0$ increases and the two bands touching linearly at $\bf q=0$, each one is 8-fold degenerate. (b) Lattice model for mass $m^{\mcl{T}_{1g}}_1$ of the same triplet with NNN pairing indicated by connecting lines with the colors of lines red (black) representing the pairing matrices $\Sigma_a(\Sigma_b)$  whose forms are mentioned in Eq.~\ref{eq_latt_mode_T_1g}. {  Momenta at which the pairing amplitudes oscillates corresponds to $M_2$ point in the BZ(Fig.~\ref{fig_BZ}).}}
\label{fig_Triplet_lattice_model_T_1g}
\end{figure}

Finally, we turn to the $\mcl{T}_{1g}^{III}$ triplet that arises from the direct product of the flavour and valley sub-band triplets (Eq. \ref{eq_decomposition_in_sym_2}). This results in a nodal PDW SC with excitation spectrum given in Fig. \ref{fig_Triplet_lattice_model_T_1g}(a). 

The structure of the Bogoliubov spectrum is best understood by writing the generic mass matrix (using Eqs. \ref{eq_deltat1g} and \ref{eq_massmatt1ggapless}) in the global basis (see Eq.~\ref{Eq_Nambu_spinor_in_global_basis}), whence its structure reduces to a form similar to Eq.~\ref{Eq_singlet_in_global_basis}. Therefore, like the $\mcl{A}_u$ singlets discussed above and the $\Gamma$-DSM of Ref. \cite{basusu8}, the mass matrix here too has a zero block corresponding to the Dirac fermions at the BZ centre, leading to the nodal PDW SC.

\begin{figure}%[h!]
 \includegraphics[width=.75\columnwidth]{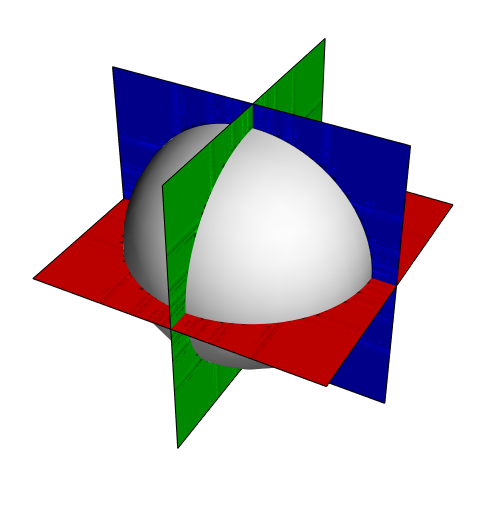}
\caption{Schematic diagram for parameter space of Triplets on TRI manifold represented by Sphere. Circles obtained from the intersection of sphere with the blue-, green- and red-plane (representing $d_1=0$ ,$d_2=0$ and $d_3=0$ planes respectively) correspond to the Great circle discussed for $\mcl{T}_{1g}$ triplet.}
    \label{triplet parameter space}
\end{figure}

Interestingly, on the TRI manifold~($\bf d \times d^* =0$), there are extra gapless modes in the spectrum for some special linear combinations of the masses.  We find that in the parameter space of TRI points, on the great circles (Fig.~\ref{triplet parameter space}), two of the gapped bands in Fig.~\ref{fig_Triplet_lattice_model_T_1g}(a)
touch each other at $\mbf{q}=0$ and the degeneracy of the gapless bands becomes 16 (also see the related discussion in Ref.~\cite{basusu8}).

The masses in this triplet correspond to the lattice Hamiltonian of the form given in Eq.~\ref{eq_lattice_ham_nnn}. For the first mass in this triplet, the pairing matrix $\mcl{X}_{\mbf{rr'}}$ (in the global basis) on the NNN bonds is shown pictorially in Fig.~\ref{fig_Triplet_lattice_model_T_1g}(b) with
\begin{eqnarray}\label{eq_latt_mode_T_1g}
    &&\Sigma_a = i\Sigma_{23} = i{\frac{7}{3}\left(J_{x}-\frac{4}{7}J^3_{x} \right)},\\
    &&\Sigma_b = -i\Sigma_{23}.
\end{eqnarray}

This leads to a symmetric spin wave function for the Cooper pairs 
\begin{align}
    |\Phi_{23}\rangle=\frac{1}{\sqrt{5}}\left(2|J_T=3,m_T=0\rangle+|J_T=1,m_T=0\rangle\right),
    \label{eq_cooper23}
\end{align}
with an antisymmetric real-space part. The other two masses can be obtained by acting with ${\bf S}_6$, and the resultant mass matrices are $\Sigma_{12}$ and $\Sigma_0$ respectively.

%%%%%%%%%%%%%%%%%%%%%

\subsection{The \texorpdfstring{$\mcl{T}_{2g}$}{} triplet superconductor}
\label{sec_t2g}

The single $\mcl{T}_{2g}$ SC arises from the direct product of triplets in both the flavour and the valley sub-band sectors  (Eq.~\ref{eq_decomposition_in_sym_2}) and correspond to nodal SC.  

\begin{figure}%[h]
  \begin{subfigure}[]{
    \includegraphics[width=.92\columnwidth]{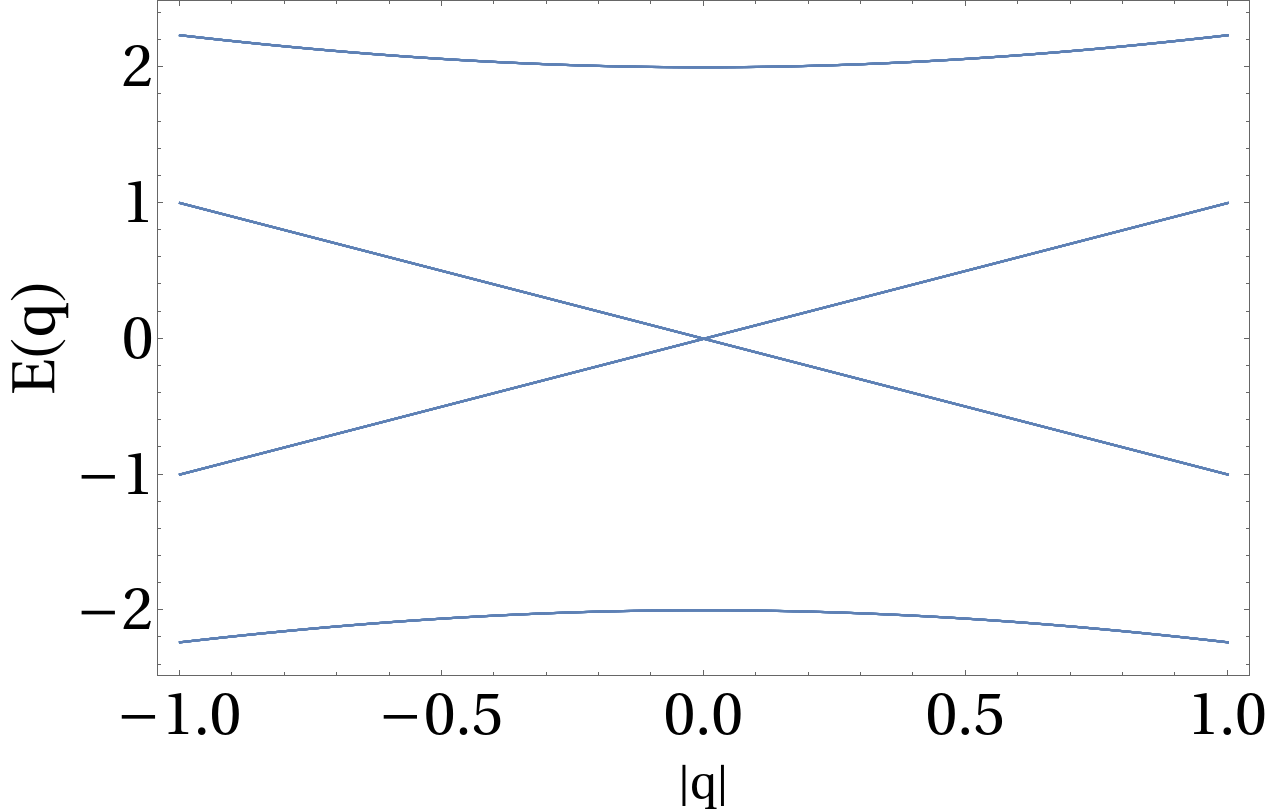}}  \label{fig_Gapless triplet_t_2g_special_Spectrum}
    \end{subfigure}
    \begin{subfigure}[]{\includegraphics[width=.90\columnwidth]{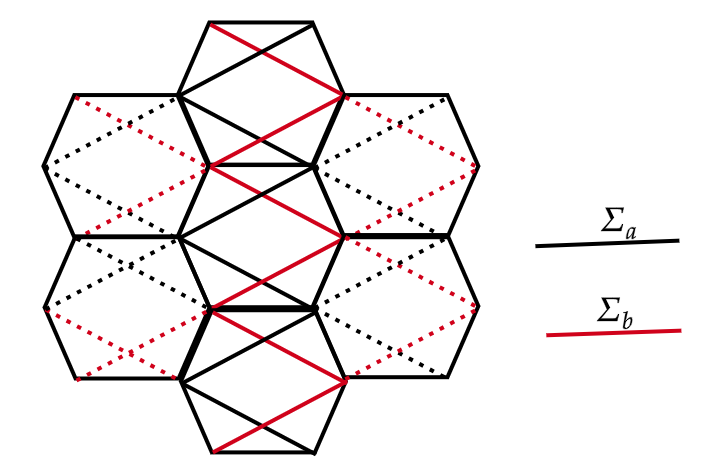}}
     
    \end{subfigure}
        \caption{(a) Spectrum for gapless Triplet $\mcl{T}_{2g}$ on TRI manifold $\forall (\theta,\phi$) on TRI manifold (for  $|\Delta^{\mcl{T}_{2g}}|=1$).There are four bands, each of which is 8-fold degenerate. Two of the bands touch at $\mbf{q}=0$. Here on TRI points, the number of gapless modes remains the same for all parameters. (b) Lattice model for mass $m^{\mcl{T}_{2g}}_1$ of the same triplet with NNN pairing indicated by connecting lines with the colors of lines red (black) representing the pairing matrices $\Sigma_a(\Sigma_b)$ form of which are mentioned in Eq.~\ref{eq_latt_mode_T_1g}. The pairing amplitudes oscillates at momenta corresponding to $M_2$ point in the BZ(Fig.~\ref{fig_BZ}).}
  \label{fig_Triplet_lattice_model_T_2g}
\end{figure}

The pairing amplitude is similar to Eq. \ref{eq_deltat1g}, which can be parametrised using Eq. \ref{eq_paramtrip}, and the mass matrices are given by
\begin{eqnarray}
& m_{1}^{\mcl{T}_{2g}} &=\frac{1}{\sqrt{2}}(\Sigma_0\tau_3\sigma_2+\Sigma_{12}\tau_0\sigma_2), \nonumber  \\ 
& m_{2}^{\mcl{T}_{2g}}&=\frac{1}{\sqrt{2}}(-\Sigma_{23}\tau_0\sigma_2+\Sigma_{0}\tau_2\sigma_0), \nonumber  \\   
& m_{3}^{\mcl{T}_{2g}}&= \frac{i}{\sqrt{2}}(\Sigma_{12}\tau_2\sigma_0+\Sigma_{23}\tau_3\sigma_2),
\label{eq_massmatt2ggapless}
\end{eqnarray}
such that the mass matrix (Eq. \ref{Eq_superconducting_mass_matrix} in the global basis (Eq. \ref{Eq_Nambu_spinor_in_global_basis}), $M^{\mcl{T}_{2g}}_{global}$, has the generic form
\begin{equation}\label{Eq_triplet_in_global_basis} \left(\begin{array}{cc|cc|cc|cc}
0_{4\times4}&0_{4\times4}&0_{4\times4}&\mcl{R}_{A}&0_{4\times4}&\mcl{R}_{B}&0_{4\times4}&\mcl{R}_{C} \\
0_{4\times4}&0_{4\times4}&\mcl{R}^{\dagger}_{A}&0_{4\times4}&\mcl{R}^{\dagger}_{B}&0_{4\times4}&\mcl{R}^{\dagger}_{C} & 0_{4\times4}\\  \hline
0_{4\times4}&\mcl{R}_{A}&0_{4\times4}& 0_{4\times4}&0_{4\times4}&0_{4\times4}&0_{4\times4}&0_{4\times4} \\
\mcl{R}^{\dagger}_{A}&0_{4\times4}&0_{4\times4}&0_{4\times4}&0_{4\times4}&0_{4\times4}&0_{4\times4}&0_{4\times4} \\ \hline
0_{4\times4}&\mcl{R}_{B}&0_{4\times4}&0_{4\times4}&0_{4\times4}&0_{4\times4}&0_{4\times4}&0_{4\times4} \\
\mcl{R}^{\dagger}_{B}&0_{4\times4}&0_{4\times4}&0_{4\times4}&0_{4\times4}&0_{4\times4}&0_{4\times4}&0_{4\times4} \\ \hline
0_{4\times4}&\mcl{R}_{C}&0_{4\times4}&0_{4\times4}&0_{4\times4}&0_{4\times4}&0_{4\times4}&0_{4\times4} \\
\mcl{R}^{\dagger}_{C}&0_{4\times4}&0_{4\times4}&0_{4\times4}&0_{4\times4}&0_{4\times4}&0_{4\times4}&0_{4\times4} \\ 
\end{array}\right)
\end{equation}
such that in the global basis it corresponds to inter-valley pairing between the three $M$-valleys with the $\Gamma$-valley. The structure of the mass matrix is similar to the M-DSM phase of Ref~\cite{basusu8} and yields 8-fold degenerate gapless nodes as can be seen from the eigenvalues of $ \left[ \boldsymbol{\Delta}^{\mcl{T}_{2g}}\cdot{\bf m}^{\mcl{T}_{2g}}\right] \cdot  \left[\boldsymbol{\Delta}^{\mcl{T}_{2g}}\cdot{\bf m}^{\mcl{T}_{2g}}\right]^{\dagger}$. The resultant spectrum is shown in Fig.~\ref{fig_Triplet_lattice_model_T_2g}(a). The gapless manifold has an effective $SU(4)$ symmetry at low energy. It would be interesting to understand the nature of phases that can be obtained by breaking this emergent $SU(4)$~\cite{Ueda_superconductivity}. 

On the lattice, in the global basis, the first mass of this triplet corresponds to NNN pairing of the form given in Eq.~\ref{eq_lattice_ham_nnn}, where the NNN pairing matrices ($\mcl{X}_{\mbf{rr'}}$)  given by Eq. \ref{eq_latt_mode_T_1g}
, albeit with a different hopping structure as shown in Fig.~\ref{fig_Triplet_lattice_model_T_2g}(b), and corresponds to a finite momentum ordering with the spin wave function of the Cooper pair given by Eq. \ref{eq_cooper23}. Thus, this corresponds to a non-unitary nodal PDW SC. 
%%%%%%%%%%%%%%%%%%%

\subsection{The \texorpdfstring{$\mcl{T}_{2u}$}{} Triplet superconductors}
\label{sec_t2u}

These two ($\mcl{T}_{2u}^I$ and $\mcl{T}_{2u}^{II}$)  SCs arise from flavour symmetric spaces (Eqs. \ref{eq_decomposition_in_sym_3} and \ref{eq_decomposition_in_sym_4}). An analysis similar to that performed in Sec.~\ref{sec_a1g_singlets} for the $\mcl{A}_{1g}$ singlets demonstrate that these two irreducible representations give rise to the same superconducting phase as both Irreps can be adiabatically connected via one-parameter interpolation without closing spectrum gap and also not breaking any further microscopic symmetries. 

\begin{figure}%[h]
    \begin{subfigure}[]{
        \includegraphics[width=.95\columnwidth]{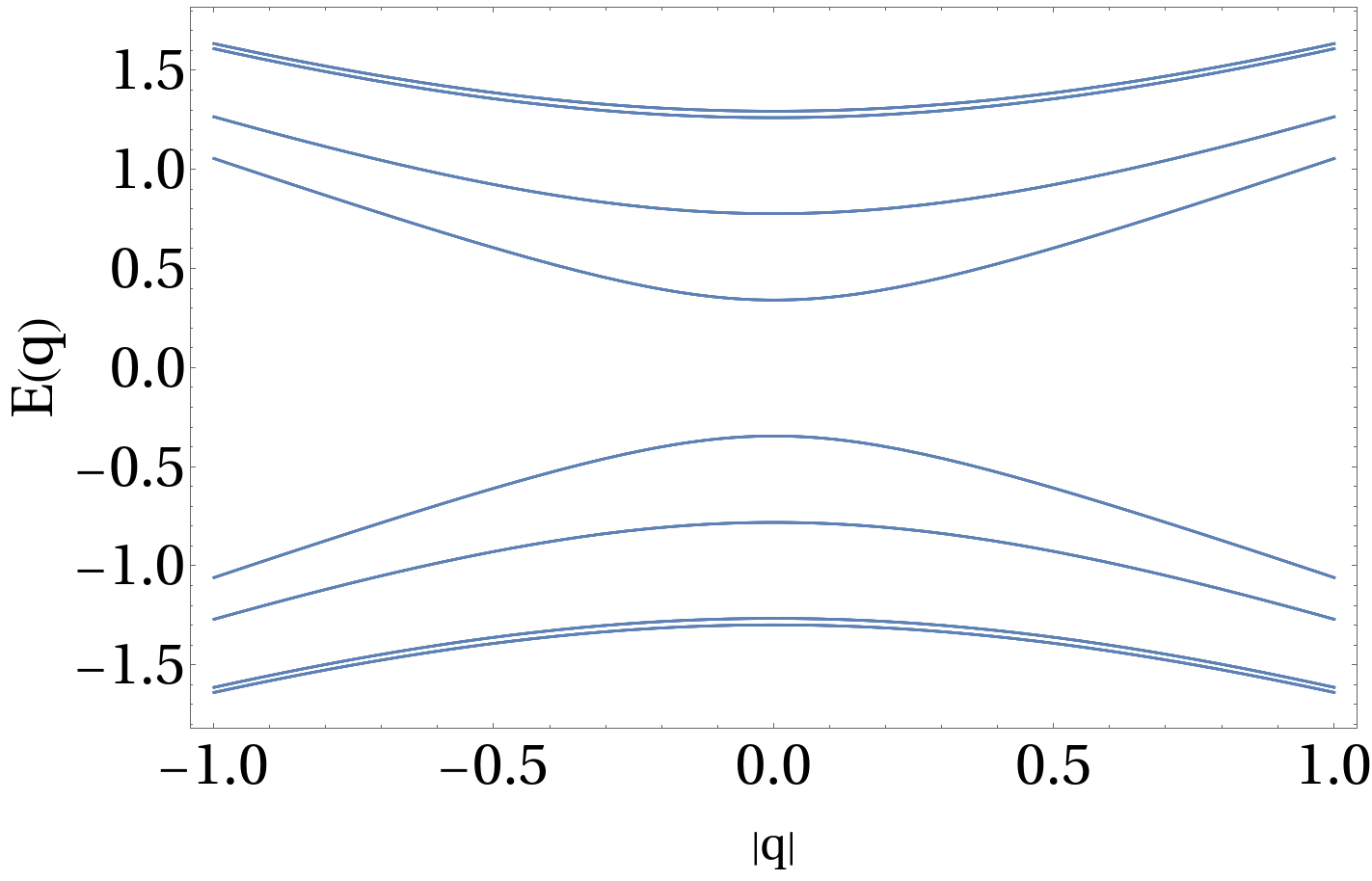}}
    \end{subfigure}
      \begin{subfigure}[]
 {   \includegraphics[width=.95\columnwidth]{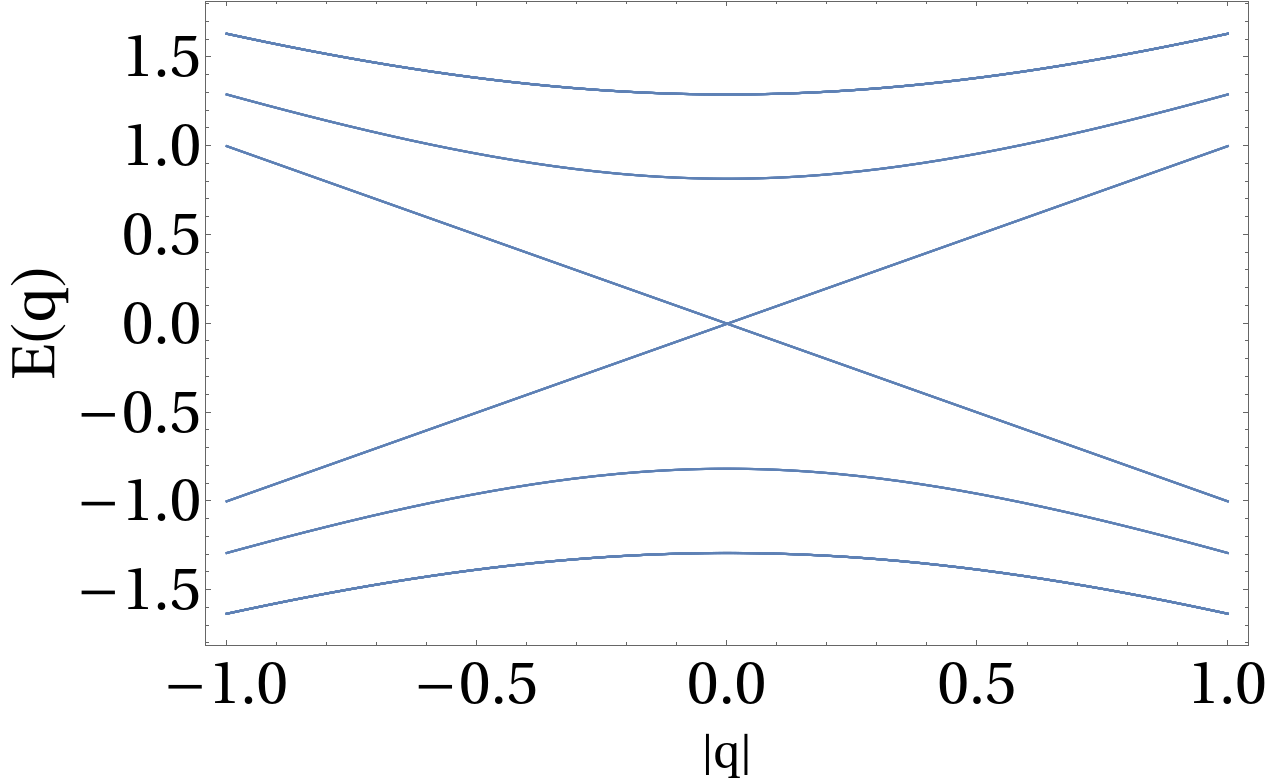}}
    \end{subfigure}
     \caption{Spectrum for $\mcl{T}_{1u}$ and $\mcl{T}_{2u}$ Irreps on TRI manifold (for { ($|\Delta^{\mcl{T}_{1u}}|=1$} or $ |\Delta^{\mcl{T}_{2u}}|=1$): (a) For general ${\bf d}$ on TRI manifold the spectrum is gapped. Here spectrum is shown for ${\bf d} = (1/\sqrt{6},1/\sqrt{6},2/\sqrt{6})$. Eigenvalues depend on ${\bf d}$. (b) There are gapless modes at special points $\left({\bf d} = \left(\pm\frac{1}{\sqrt{3}},\pm\frac{1}{\sqrt{3}},\pm\frac{1}{\sqrt{3}}\right)\right)$ on the TRI manifold (which is $S^2$ in this case).}
    \label{fig_T_2u_special_Spectrum}
\end{figure}

The mass matrices for the $\mcl{T}_{2u}^{\mcl{x}}$ (for $\mcl{x}=I,II$) triplets are given by Eq.~\ref{Eq_superconducting_mass_matrix} with $d=3$, with
\begin{eqnarray}
    \Delta_{i}^{\mcl{T}^{\mcl{x}}_{2u}} = \Braket{\chi^Tm_{i}^{\mcl{T}^{\mcl{x}}_{2u}}\chi},~~~~~(i=1,2,3)
\end{eqnarray}
where
\begin{eqnarray}
&& m_{1}^{\mcl{T}^{I}_{2u}} =\frac{1}{\sqrt{2}}\left[(\frac{\sqrt{3}\Sigma_{34}}{2}-\frac{\Sigma_{35}}{2})\tau_0\sigma_2-(\frac{\Sigma_4}{2}-\frac{\sqrt{3}\Sigma_5}{2})\tau_3\sigma_2\right], \nonumber  \\ 
&& m_{2}^{\mcl{T}^{I}_{2u}}=\frac{1}{\sqrt{2}}(\frac{1}{2}\Sigma_{4}-\frac{\sqrt{3}}{2}\Sigma_{5})\tau_2\sigma_0+\frac{1}{\sqrt{2}}\Sigma_{15}\tau_0\sigma_2, \nonumber  \\   
&&m_{3}^{\mcl{T}^{I}_{2u}}= \frac{i}{\sqrt{2}}(\frac{\sqrt{3}}{2}\Sigma_{34}-\frac{1}{2}\Sigma_{35})\tau_2\sigma_0-\frac{i}{\sqrt{2}}\Sigma_{15}\tau_3\sigma_2, 
\end{eqnarray}
for $\mcl{T}_{2u}^I$, and 
\begin{eqnarray}
&& m_{1}^{\mcl{T}^{II}_{2u}} =\frac{1}{\sqrt{2}}\left[(\frac{\sqrt{3}\Sigma_{35}+\Sigma_{34}}{2})\tau_0\sigma_2+(\frac{\Sigma_{5}+\sqrt{3}\Sigma_4}{2})\tau_3\sigma_2\right], \nonumber  \\ 
&& m_{2}^{\mcl{T}^{II}_{2u}}=\frac{1}{\sqrt{2}}(\frac{1}{2}\Sigma_{5}+\frac{\sqrt{3}}{2}\Sigma_{4})\tau_2\sigma_0+\frac{1}{\sqrt{2}}\Sigma_{14}\tau_0\sigma_2, \nonumber  \\   
&& m_{3}^{\mcl{T}^{II}_{2u}}= -\frac{i}{\sqrt{2}}(\frac{\sqrt{3}}{2}\Sigma_{35}+\frac{1}{2}\Sigma_{34})\tau_2\sigma_0+\frac{i}{\sqrt{2}}\Sigma_{14}\tau_3\sigma_2, 
\end{eqnarray}
for $\mcl{T}_{2u}^{II}$. Using the parametrization similar to Eq.~\ref{eq_triplet_in _theta_phi_I}, it is straightforward to show that both the triplets ($\mcl{T}^x_{2u}~~, x\in (I,II))$ correspond to non-unitary PDW SCs that break TR, provided ${\bf d}^x \times {{\bf d}^x}^*\neq 0$. However, unlike the triplets discussed above, the present ones are non-unitary even within the TRI subspace and exhibit a multi-gap structure, with the gap magnitudes dependent on ${\bf d}^x$. 
In fact, at isolated points on the TRI sub-space ${\bf d}^x= \left( \pm \frac{1}{\sqrt{3}},\pm \frac{1}{\sqrt{3}},\pm \frac{1}{\sqrt{3}} \right) $, the smaller of the two gaps go to zero to yield a nodal SC (Fig.~\ref{fig_T_2u_special_Spectrum}) with the gapless nodes arising from the mixing of the $\Gamma$ and the $M$ valleys allowed by finite momentum pairing. On moving away from the TRI sub-space, the above isolated gapless points appear to bifurcate, but the full analysis of the fate of these isolated nodal SCs needs to be explored further. 

 \begin{figure}%[t]
\includegraphics[width=1\columnwidth]{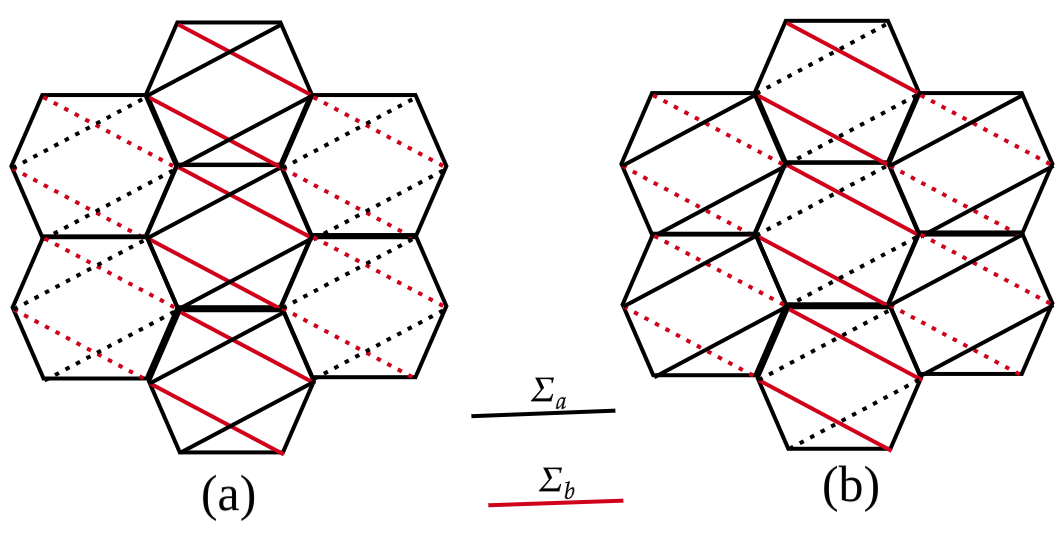}
    \caption{(a) Schematic for the NNN pairing for $m_{1}^{\mcl{T}^{I}_{2u}} $ ($m_{1}^{\mcl{T}^{II}_{2u}} $) bilinears of the Triplets $\mcl{T}^{I}_{2u}$ $(\mcl{T}^{II}_{2u}$) in global basis. The black (red) colour of connecting lines represents the pairing matrix $\Sigma_{a}$ ($\Sigma_{b}$) mentioned in Eq.~\ref{Eq_Pairing_matrix_T_2u_1}~(Eq.~\ref{Eq_Pairing_matrix_T_2u_2}) while dashed lines correspond to -ve of the matrices. (b) Schematic for the NNN pairing for $m_{1}^{\mcl{T}^{I}_{1u}} $ ($m_{1}^{\mcl{T}^{II}_{1u}} $) bilinears of the Triplets $\mcl{T}^{I}_{1u}$ $(\mcl{T}^{II}_{1u}$) in global basis with the notation and form of matrices $(\Sigma_a, \Sigma_b$) being same as used in (a) subfigure. For both lattice models, momenta at which the pairing amplitudes oscillate corresponding to $M_2$ point in the BZ(Fig.~\ref{fig_BZ}).} 
\label{fig_Triplet_lattice_model}
     \end{figure}

The mean field lattice Hamiltonian has the form similar to Eq.~\ref{eq_lattice_ham_nnn}, with NNN pairing as shown in Fig.~\ref{fig_Triplet_lattice_model}, where the pairing matrices, $\mcl{X}_{\mbf{rr'}}$,   corresponding to $m_{1}^{\mcl{T}^{I}_{2u}}$ and $m_{1}^{\mcl{T}^{II}_{2u}}$ are given by (with reference to Fig. \ref{fig_Triplet_lattice_model})
\begin{align}\label{Eq_Pairing_matrix_T_2u_1}
    \Sigma_b = i \left(\frac{\sqrt{3}}{2}\Sigma_{14}-\frac{1}{2}\Sigma_{15}\right),~~\Sigma_a =i\left(\frac{\sqrt{3}}{2}\Sigma_{14}+\frac{1}{2}\Sigma_{15}\right),
\end{align}
and
\begin{align}\label{Eq_Pairing_matrix_T_2u_2}
    \Sigma_b = i \left(\frac{\sqrt{3}}{2}\Sigma_{15}+\frac{1}{2}\Sigma_{14} \right),~~
    \Sigma_a =-i\left(\frac{\sqrt{3}}{2}\Sigma_{15}-\frac{1}{2}\Sigma_{14}\right),
\end{align}
respectively. The corresponding spin-wave function for the Cooper pair is symmetric and given by
\begin{align}
    |\Phi_b\rangle=&\sqrt{\frac{3}{5}}|J_T=1,m_T=0\rangle-\sqrt{\frac{3}{20}}|J_T=3,m_T=0\rangle\nonumber\\
    &+\frac{1}{2\sqrt{2}}\left(|J_T=3,m_T=2\rangle+|J_T=3, m_T=-2\rangle\right),\\
    |\Phi_a\rangle=&\sqrt{\frac{3}{5}}|J_T=1,m_T=0\rangle-\sqrt{\frac{3}{20}}|J_T=3,m_T=0\rangle\nonumber\\
    &-\frac{1}{2\sqrt{2}}\left(|J_T=3,m_T=2\rangle+|J_T=3, m_T=-2\rangle\right),
\end{align}
for $\mcl{T}_{2u}^I$, and
\begin{align}
    |\Phi_b\rangle=&\frac{2\sqrt{2}}{5}|J_T=1,m_T=0\rangle-\frac{\sqrt{2}}{5}|J_T=3,m_T=0\rangle\nonumber\\
    &-\sqrt{\frac{3}{5}}\left(|J_T=3,m_T=2\rangle+|J_T=3, m_T=-2\rangle\right),\\
    |\Phi_a\rangle=&\frac{2\sqrt{2}}{5}|J_T=1,m_T=0\rangle-\frac{\sqrt{2}}{5}|J_T=3,m_T=0\rangle\nonumber\\
    &+\sqrt{\frac{3}{5}}\left(|J_T=3,m_T=2\rangle+|J_T=3, m_T=-2\rangle\right),
\end{align}
for $\mcl{T}_{2u}^{II}$. The other two components are obtained by symmetry transformations given in Table \ref{tab_triplet_irrep}.

%%%%%%%%%%%%%%%%%%%%%%

\subsection{The \texorpdfstring{$\mcl{T}_{1u}$}{} triplet superconductors} \label{sec_t1u}

There are two $\mcl{T}_{1u}$ ($\mcl{T}_{1u}^I$ and $\mcl{T}_{1u}^{II}$) triplets that arise in Eq.~\ref{eq_decomposition_in_sym_3} and  \ref{eq_decomposition_in_sym_4},  which can be adiabatically connected without breaking further symmetries, and hence represent the same SC. The pairing amplitudes are given by expressions similar to Eq. \ref{eq_deltat1g} with mass matrices
\begin{eqnarray}
& &m_{1}^{\mcl{T}^{I}_{1u}} =\frac{1}{\sqrt{2}}\left[(\frac{\sqrt{3}\Sigma_{34}}{2}-\frac{\Sigma_{35}}{2})\tau_0\sigma_2+(\frac{\Sigma_{4}}{2}-\frac{\sqrt{3}\Sigma_5}{2})\tau_3\sigma_2\right], \nonumber  \\ 
& &m_{2}^{\mcl{T}^{I}_{1u}}=\frac{1}{\sqrt{2}}(\frac{1}{2}\Sigma_{4}-\frac{\sqrt{3}}{2}\Sigma_{5})\tau_2\sigma_0 -\frac{1}{\sqrt{2}}\Sigma_{15}\tau_0\sigma_2, \nonumber  \\   
& &m_{3}^{\mcl{T}^{I}_{1u}}= \frac{i}{\sqrt{2}}(\frac{\sqrt{3}}{2}\Sigma_{34}-\frac{1}{2}\Sigma_{35})\tau_2\sigma_0 +\frac{i}{\sqrt{2}}\Sigma_{15}\tau_3\sigma_2. \nonumber \\ 
\end{eqnarray}
for $\mcl{T}_{1u}^I$, and
\begin{eqnarray}
    && m_{1}^{\mcl{T}^{II}_{1u}} =\frac{1}{\sqrt{2}}\left[(\frac{\sqrt{3}\Sigma_{35}}{2}+\frac{\Sigma_{34}}{2})\tau_0\sigma_2-(\frac{\Sigma_5}{2}+\frac{\sqrt{3}\Sigma_4}{2})\tau_3\sigma_2\right], \nonumber  \\ 
&& m_{2}^{\mcl{T}^{II}_{1u}}=\frac{1}{\sqrt{2}}(\frac{1}{2}\Sigma_{5}+\frac{\sqrt{3}}{2}\Sigma_{4})\tau_2\sigma_0-\frac{1}{\sqrt{2}}\Sigma_{14}\tau_0\sigma_2,\nonumber  \\   
& &m_{3}^{\mcl{T}^{II}_{1u}}= -\frac{i}{\sqrt{2}}(\frac{\sqrt{3}}{2}\Sigma_{35}+\frac{1}{2}\Sigma_{34})\tau_2\sigma_0-\frac{i}{\sqrt{2}}\Sigma_{14}\tau_3\sigma_2, \nonumber \\ 
\end{eqnarray}
for $\mcl{T}_{1u}^{II}$. {
The spectrum (on TRI manifold) for this triplet is the same as that of the $\mcl{T}_{2u}$ triplets in the sense that the number of bands and the gap structure is the same. It should be noticed that for general points in the parameter space, each of the $\mcl{T}_{u}$ SC has 8 bands, each band is 4-fold degenerate. Out of them, 8 of the bands become gapless on the special points $\left({\bf d} = \left(\pm\frac{1}{\sqrt{3}},\pm\frac{1}{\sqrt{3}},\pm\frac{1}{\sqrt{3}}\right)\right)$ on TRI manifold. This similarity in the spectrum is also extended to the TRB manifold. The resultant PDW SCs are very similar to the $\mcl{T}_{2u}$ ones just discussed, albeit with a different lattice symmetry-breaking pattern as indicated by the Irrep. It is evident from the lattice model for $m^{\mcl{T}^I_{1u}}_1$($m^{\mcl{T}^{II}_{1u}}_1$) shown in Fig.~\ref{fig_Triplet_lattice_model}(b).

%%%%%%%%%%%%%%%%%%%%%%%%%%%%%%%%%%%%%%%%%%%%
\section{Summary and Outlook}
\label{sec_summary}

In this work, we have presented the superconducting phases (Tables \ref{tab_singlets}, \ref{tab_doublets} and \ref{tab_triplets}) that are naturally proximate the SU(8) DSM that may be realised in honeycomb lattice materials with strong SOC at $1/4$th filling. The resultant unconventional SCs differ from the usually studied ones in two aspects-- (1) the larger spin representation stemming from the $j=3/2$ orbitals, and (2) non-trivial implementation of the microscopic symmetries due to the SOC-induced mixing of spin and real spaces. As a result, the different SC phases proximate to the SU(8) DSM are very different from that of graphene~\cite{chamon2012masses} even though the microscopic symmetries and the lattice structures are the same. Indeed, the larger spin representation facilitates substantially generalizes (compared to $S=1/2$~\cite{scalapino2012common,Ueda_superconductivity}, {and the discussion of $j=1/2$ spin-orbit coupled fermions presented in Appendix.~\ref{sec_su2global}}) the scope of the interplay between the {\it spin} and superconducting pairing channels to realise various unconventional SCs. While superconductivity in higher spin-representations, arising due to strong SOC, is well known in several rare earth and actinide compounds~\cite{RevModPhys.81.1551,RevModPhys.74.235,PhysRevB.94.174513,Ueda_superconductivity,volovik1985superconducting,PhysRevLett.116.177001}, the SOC concomitantly is responsible for lowering the symmetry in such materials distinguishing them from the present study. 

One of the central outcomes of the above ingredients is multiple-gap superconductors with the possibility of tuning the gap magnitudes with the direction of the SC order-parameter, providing a way to tune a gapped SC to a nodal one. This leads to several interesting unconventional SCs, including a particularly interesting case is that of an unconventional even-parity non-unitary superconductor. Further, the tuning of pairing gaps and associated relative phases opens interesting questions for the associated Leggett modes~\cite{leggett1966number,cuozzo2024leggett} for future studies.

The catalogue of the unconventional SCs presented here raises several interesting questions. Given the plethora of unconventional SCs discussed here, it would be interesting to understand the topological properties~\cite{chiu2016classification,schnyder2009classification,PhysRevB.65.212510,PhysRevB.70.054502,PhysRevB.79.094504,kim2018beyond} of the SCs and their implications in various tunnel junctions~\cite{PhysRevResearch.3.013051} involving these SCs~\cite{zhang2024finite}. A different question pertains to the nature of unconventional quantum phase transitions~\cite{Grover_2008} involving these SCs and the normal phases, which are intimately related to each other at low energies due to the emergent SO(16) symmetry. We hope that our results would fuel interest in experimental studies of candidate materials~\cite{PhysRevResearch.5.043219} that will provide concrete context to study such issues in future.  Finally, a similar analysis would apply to the even number of surface Dirac cones for weak TI, which are also stabilised by strong SOC. However, the symmetry of the low-energy Dirac theory will differ from SU(8).
%%%%%%%%%%%%%%%%%%%%%%

\begin{acknowledgments}
    The authors thank T. Saha Dasgupta, K. Sengupta, H. R. Krishnamurthy and R. Sensarma for discussion. We acknowledge the use of the open-source software \textit{Groups, Algorithms, Programming - a System for Computational Discrete Algebra} (GAP). SB acknowledges funding by the Swarna Jayanti fellowship of SERB-DST (India) Grant No. SB/SJF/2021-22/12. SB and VBS acknowledge funding from DST, Government of India (Nano mission), under Project No. DST/NM/TUE/QM-10/2019 (C)/7. AC, BM and SB acknowledge support of the Department of Atomic Energy, Government of India, under project no. RTI4001. BM acknowledges the support of Early Career Scheme of Hong Kong Research Grant Council with grant No.~26309524.
\end{acknowledgments}

%%%%%%%%%%%%%%%%%%%%%%%
\appendix

\section{Summary of the low-energy Dirac theory in the global basis}
\label{appen_global}

The low-energy Hamiltonian in the global basis (Fig. \ref{fig_BZ}) is obtained by performing a unitary transformation on the Dirac spinors in the local basis, $\boldsymbol{\chi}$ (Eq. \ref{Eq_Dirac_spinor_Xi}),
\begin{eqnarray}\label{eq:global basis}
     \boldsymbol{\chi}_g = U_{global}~~\boldsymbol{\chi}.
\end{eqnarray}
where  $U_{global}$ is a $16\times 16$ unitary matrix whose explicit form is given in Ref. \cite{basusu8} and $\boldsymbol{\chi}_g$ is the spinor in the global basis, {\it i.e.},
\begin{align}
\boldsymbol{\chi}_g=(\chi_{g\Gamma},\chi_{gM_1},\chi_{gM_2},\chi_{gM_3})^T
\label{eq_globalspinor}
\end{align}
where each $\chi_{gv}$   corresponds to a $4$-component spinor corresponding to the Dirac fermions at the four valleys $v=(\Gamma, M_1, M_2, M_3)$ in the global basis (Fig. \ref{fig_BZ}). The low-energy Hamiltonian (Eq. \ref{eq_dirac_intro}) in the global basis is then given by
 \begin{align}
H_D  &=
 i v_F \int d^2{\bf x}~ \boldsymbol{\chi}_g^{\dagger}\left[\left(\Sigma_0\tilde\Sigma_{23}\right)\partial_1 +\left(-\Sigma_0\tilde\Sigma_{24}\right)\partial_2 \right]\boldsymbol{\chi}_g.
 \label{eq_freediracchitilde}
\end{align}
 Here, $\Sigma_0$ represents the four Dirac valleys in the global basis. The $\tilde \Sigma_i$ are $4\times 4$ Hermitian matrices which are of the same form as the $\Sigma_i$ matrices, but, unlike the $\Sigma_i$ matrices, they do not exclusively act on the flavour space. 
%%%%%%%%%%%%%%%%%%%%%%%%%%%%%%%

\section{The BdG Hamiltonian and the pairing amplitudes for the Dirac superconductors}
\label{appen_bdg}

The BdG Hamiltonian for the SU(8) Dirac fermions is given by~\cite{PhysRevLett.97.067007,PhysRevLett.97.217001}
\begin{align}
    H_{MF}=\int d^2{\bf q}~\tilde\chi^\dagger_N({\bf q})~\mathcal{H}_{BdG}({\bf q})~\tilde\chi_N({\bf q})
    \label{eq_bdg}
\end{align}
where
\begin{align}
    \chi_N({\bf q})=[\chi({\bf q}), \left(\mathcal{T}\chi({\bf q})\right)^*]^T
    \label{eq_nambuspinor}
\end{align}
is the Nambu spinor with $\chi$ given by Eq. \ref{Eq_Dirac_spinor_Xi} and $\mathcal{T}$ is the TR symmetry operator such that~\cite{basusu8}
\begin{align}
    \mathcal{T}\chi({\bf q})&=i\Sigma_{13}\tau_1\sigma_0\mathcal{K}\chi(-{\bf q})\nonumber\\
    &=i\Sigma_{13}\mathcal{K}[\tau_1\chi_1,\tau_1\chi_2,\tau_1\chi_3,\tau_1\chi_4]^T
\end{align}
with $\mathcal{K}$ being the complex conjugation operator and 
\begin{align}
\tau_1\chi_f=\left(\chi_{f1-},\chi_{f2-},\chi_{f1+},\chi_{f2+}\right)
\end{align}
where $\chi_f$ is given by Eq. \ref{Eq_Dirac_spinor_Xi_f}. Putting everything together, Eq. \ref{eq_nambuspinor} becomes
\begin{align}
    \tilde\chi_N({\bf q})&=\left(\begin{array}{cc}
    \mathbb{I}_{16}& 0 \\
    0 & i\Sigma_{13}\tau_1\sigma_0\mathcal{K}\\
    \end{array}\right)\left(\begin{array}{c}
    \boldsymbol{\chi}({\bf q})\\
    \boldsymbol{\chi}^*({\bf -q})\\
    \end{array}\right)\equiv\mathcal{Y}\cdot\chi_{N}
\end{align}
where $\chi_N$ is defined in Eq. \ref{eq:Nambu spinor in momentum space}. Finally,
\begin{align}
    \mathcal{H}_{BdG}({\bf q})=\left(\begin{array}{cc}
    \hat{\varepsilon}_{\bf q} & \hat{\Delta}_{\bf q}\\
    \hat{\Delta}^\dagger_{\bf q} & -\hat{\varepsilon}_{\bf q}\\
    \end{array}\right)
\end{align}
with $\hat{\varepsilon}_{\bf q}=\frac{v_F}{2}\Sigma_0(\alpha_1 q_x+\alpha_2 q_y)$ being the free Dirac Hamiltonian and $\hat\Delta_{\bf q}$ being the pairing matrix.

The relation between the form of $H_{MF}$ in Eq. \ref{eq_bdg} and that of Eq. \ref{eq:general term in Nambu basis} is obtained via the transformation
\begin{align}
    \mathcal{Y}^{-1}\cdot\mathcal{H}_{BdG}\cdot\mathcal{Y}=\frac{v_{F}}{2}(\tilde\alpha_1 q_x+\tilde\alpha_2 q_y)+M^{SC}
\end{align}
where the two terms in the RHS are given by Eqs. \ref{eq:Hamiltonian in nambu basis} and \ref{Eq_superconducting_mass_matrix} respectively. Note that under the above transformation, $\hat\Delta\cdot\hat\Delta^\dagger$ remains invariant.

Eq. \ref{eq_bdg} is  diagonalized~\cite{Ueda_superconductivity} via Bogoliubov transformations 
\begin{align}
    \tilde\chi_{N}({\bf q})=\mathcal{W}_{\bf q}\Gamma({\bf q})
\end{align}
where $\Gamma({\bf q})$ is the 32-component Bogoliubov fermions while $\mathcal{W}_{\bf q}$ is a $32\times32$ unitary matrix ({\it i.e.} $\mathcal{W}_{\bf q}\cdot\mathcal{W}_{\bf q}^\dagger=1$) of the form
\begin{align}
\mathcal{W}_{\bf q}=\left(\begin{array}{cc}
        u_{\bf q} & v_{\bf q}\\
        v^*_{-\bf q} & u^*_{-\bf q}\\
        \end{array}\right)
\end{align}
with each of $u$ and $v$ being $16\times16$ matrix.  The energy spectrum is then obtained from
\begin{align}
    E_{\bf q}=\mathcal{W}_{\bf q}^\dagger\cdot \mathcal{H}_{BdG}\cdot \mathcal{W}_{\bf q}~~~~~\Rightarrow~~~~\mathcal{W}_{\bf q}\cdot E_{\bf q}=\mathcal{H}_{BdG}\cdot \mathcal{W}_{\bf q}
    \label{eq_bdgdiag}
\end{align}
where $E_{\bf q}$ is the diagonalized form having the structure
\begin{align}
E_{\bf q}=\left(\begin{array}{cc}
\mathcal{E}_{{\bf q}} & 0\\
0 & -\mathcal{E}_{\bf q}\\
\end{array}\right)
\end{align}
with $\mathcal{E}_{\bf q}$ being a $16\times16$ diagonal matrix with entries $\mathcal{E}_{{\bf q},a}~(a=1, 2,\cdots, 16)$. The second form of Eq. \ref{eq_bdgdiag} is expanded to get
\begin{align}
    u_{\bf q}\mathcal{E}_{\bf q}&=\hat{\varepsilon}_{\bf q}u_{\bf q}+\hat\Delta_{\bf q}v^*_{-{\bf q}}\\
    v^*_{-{\bf q}}\mathcal{E}_{\bf q}&=\hat\Delta^\dagger_{\bf q}u_{\bf q}-\hat{\varepsilon}_{{\bf q}}v^*_{-{\bf q}}
\end{align}
For Dirac dispersion, since $\hat{\varepsilon}^2_{\bf q}\propto \mathbb{I}$, generalizing the methods outlined in Ref. \cite{Ueda_superconductivity}, we get
\begin{align}
   u_{\bf q}\left(\mathcal{E}^2_{\bf q}-\hat{\varepsilon}_{\bf q}^2\right)=\hat\Delta_{\bf q}\hat\Delta^\dagger_{\bf q} u_{\bf q}
   \label{eq_dispersionbdg}
\end{align}

The solution then depends on whether the pairing is unitary ($\hat{\Delta}\cdot\hat{\Delta}^\dagger =|\Delta|^2\mathbb{I}_{16\times16}$) or non-unitary ($\hat{\Delta}\cdot\hat{\Delta}^\dagger=\sum_{\alpha\beta\gamma}A_{\alpha\beta\gamma}\Sigma_\alpha\tau_\beta\sigma_\gamma$). 

Note that a similar analysis is applicable in principle to the lattice theory. This may be slightly easier in the local basis where the $SU(4)$ flavour symmetry is manifest, however, the actual solutions of the eigenvectors are substantially complicated in practice. 

Knowing the transformation from the basis in Eq. \ref{eq:Nambu spinor in real space} to that in Eq.~\ref{eq_nambuspinor}, the general pairing matrix $M^{Sc}$ (Eq.~\ref{Eq_superconducting_mass_matrix}) transforms as
\begin{align}
    M^{Sc} \to \mcl{Y}M^{Sc}\mcl{Y}^\dagger
\end{align}

\section{The Majorana representation}\label{appen_majorana}

As a first step to obtain the Majorana representation, we perform a unitary transformation on the spinors, $\boldsymbol{\chi}$ as
\begin{eqnarray}\label{eq_majorana_basis_rot}
\tilde{\boldsymbol{\chi}}=\tilde U~\boldsymbol{\chi}.
\end{eqnarray}
Here, $\tilde U=\tilde{U}_2\tilde{U}_1$ is a product of 16-dimensional unitary matrices which are given by
\begin{equation}\label{eq_def_of_U_1}
\tilde U _1=  \frac{1}{2}\Big(\Sigma_0  (\tau_0+\tau_3) \sigma_0 + \Sigma_0 (\tau_0-\tau_3)\sigma_2\Big).
\end{equation}
and
\begin{equation}
   \tilde U_2= I_{8\times8}\otimes \exp\left[{\rm i} \pi \sigma_1^{maj}/4\right]
\end{equation}
where $\sigma^{maj}_i$ (for $i=0,\cdots,3$) are the Pauli matrices acting in the mixed valley-subband sector. The free Hamiltonian in $\tilde{\boldsymbol{\chi}}$ basis has the following form,
 \begin{align}
     H_D=-iv_F\int d^2\mathbf{x} ~\tilde{\boldsymbol{\chi}}^{\dagger}(\mathbf{x})(\alpha'_1\partial_x+\alpha'_2\partial_y)\tilde{\boldsymbol{\chi}}(\mbf{x})
     \label{eq_Hamiltonian before Majorana}
 \end{align}
with 
\begin{align}
 &   \alpha_1'=\tilde U. \alpha_x.\tilde U^{\dagger}=I_{8\times8}\otimes \sigma^{maj}_1,
\label{eq_alphax_prime} \\
&   \alpha_2'=\tilde U.\alpha_y.\tilde U^{\dagger}=I_{8\times8}\otimes \sigma^{maj}_3.
\label{eq_alphay_prime}
\end{align}
Performing the basis rotation mentioned in Eq.~\ref{eq_majorana_basis_rot} mixes the
valley and subband sectors. Now we define the Majorana fermions, $\boldsymbol{\eta}$, as
\begin{equation}\label{eq_def_of_eta}
\boldsymbol{\eta} =\Big(\boldsymbol{\eta}_r,\boldsymbol{\eta}_I\Big)^T= \Bigg(\frac{1}{2}\big(\tilde{\boldsymbol{\chi}}+\tilde{\boldsymbol{\chi}}^{\dagger}\big),\frac{1}{2i}\big(\tilde{\boldsymbol{\chi}}-\tilde{\boldsymbol{\chi}}^{\dagger}\big)\Bigg)^T.
\end{equation}

The Hamiltonian in Eq.~\ref{eq_Hamiltonian before Majorana} has the following form in the Majorana basis,
\begin{equation}\label{eq_free_H_in_Majorana}
    H_{D}=-i\frac{v_F}{2}\int d^2\textbf{x}\,\boldsymbol{\eta}^T \big(\alpha^m_1\partial_1  + \alpha^m_2\partial_2\big)~\boldsymbol{\eta},
\end{equation}
where
\begin{align}
&&\alpha^m_1=\begin{pmatrix}
\alpha'_1&0\\ 0& \alpha'_1
\end{pmatrix}=
I_{16\times16}\otimes\sigma^{maj}_1, 
\nonumber\\
&&\alpha^m_2=\begin{pmatrix}
\alpha'_2&0\\ 0& \alpha'_2
\end{pmatrix}=
I_{16\times16}\otimes\sigma^{maj}_3. 
\label{Eq_alpha_maj_matrices}
\end{align}
The form of the Hamiltonian in Eq.~\ref{eq_free_H_in_Majorana} has manifest SO(16) symmetry, where generators of the $SO(16)$ are of the form 
\begin{align}
    g_{majorana}\otimes \sigma^{maj}_0
    \label{eq_so16}
\end{align}
where $g_{majorana}$ are 16-dimensional real anti-symmetric matrices, i.e., $g_{majorana}^T=-g_{majorana}$, and are formed out of the 4 flavours, 2 Majorana and 2 mixed valley-subband components.

 As shown in Eq.~\ref{eq_free_H_in_Majorana}, the matrices \{$\alpha_1^m, \alpha_2^m$\}, when written in the notation introduced in Eq.~\ref{form_of_mass_term}, are of the form \{$X_{0001}, X_{0003}$\}. Thus, the matrices that anticommute with $\{\alpha_1^m, \alpha_2^m\}$ has the form $X_{\alpha \beta \gamma 2}$, for which there are 256 possibilities. Furthermore, in the Majorana basis, the particle-hole constraint (Eq. \ref{eq:constraint on Nambu hamiltonian}) is equivalent to the condition that $X_{\alpha\beta\gamma2}$ is anti-symmetric. This requires that 
\begin{equation}
(\mu_{\alpha}\Sigma_{\beta}\tau_{\gamma})^T= \mu_{\alpha}\Sigma_{\beta}\tau_{\gamma}.
\label{eq_scm}
\end{equation}
{\it i.e.}, all $16\times16$ symmetric matrices which are 136 in number. Note that $136=1\oplus135$, where $1$ is the SO(16) singlet and $135$ is the irreducible representation of SO(16) made up of the rank-2 traceless symmetric SO(16) tensor.

\section{Transformation of Irreps under lattice.}\label{app:C}

The symmetry transformations of the different Irreps~\cite{basusu8} are summarized in the following tables~\ref{tab_singlet_irrep}, \ref{tab_doublet_irrep_22} and \ref{tab_triplet_irrep} for completeness.
 \begin{table}
\begin{tabular}{|c|c|c|c|c|c|c|c|c|}\hline
Irrep & mass & ${\bf T}_1$ & ${\bf T}_2$ & ${\bf C}_2'$ & ${\bf C}_3$ & ${\bf S}_6$ &{\bf I} & ${\bf \sigma}_d$ \\\hline
  $\mcl{A}_{1g}$ &
   M & M & M &M & M & M & M & M  \\
    $\mcl{A}_{\text{2g}}$ &
   M & M & M & -M & M & M & M & M  \\
$\mcl{A}_{\text{1u}}$ &M&
   M &M & M &M & -M & -M& -M\\
 $\mcl{A}_{\text{2u}}$ &
   M & M& M & -M &M & -M & -M& M \\\hline
\end{tabular}
\caption{Table for one-dimensional Irreps of microscopic symmetries}
\label{tab_singlet_irrep}
\end{table}

%\begin{widetext}
\begin{center}%\label{tab_doublet_irrep_22}
\begin{table*}
\begin{tabular}{|c|c|c|c|c|c|c|c|c|}\hline
\text{Irrep} & \text{mass} & ${\bf T_1}$ & ${\bf T_2}$ & ${\bf C_2'}$ & ${\bf C_3}$ & ${\bf S_6}$ &${\bf I}$ & ${\bf \sigma_d}$ \\\hline
$\mcl{E}_{\text{u}}$ & $\text{M}_{1}$ & $\text{M}_{1}$ & $\text{M}_{1}$ & $\text{M}_{1}$  & $\frac{1}{2}$
   $\left(-\text{M}_{1}-\sqrt{3} \text{M}_{2}\right)$ & $\frac{1}{2} \left(\text{M}_{1}-\sqrt{3} \text{M}_{2}\right)$ & $-\text{M}_{1}$ &
  $ -\text{M}_{1}$  \\
 & $\text{M}_{2} $& $\text{M}_{2}$ & $\text{M}_{2}$ & $-\text{M}_{2}$ & $\frac{1}{2} \left(\sqrt{3}
   \text{M}_{1}-\text{M}_{2}\right)$ & $\frac{1}{2} \left(\sqrt{3} \text{M}_{1}+\text{M}_{2}\right)$ & $-\text{M}_{2}$ & $\text{M}_{2}$ \\\hline
  $ \mcl{E}_{\text{g}}$ & $\text{M}_{1}$ & $\text{M}_{1}$ & $\text{M}_{1}$ &$ -\text{M}_{1}$  & $\frac{1}{2} \left(\sqrt{3}
   \text{M}_{2}-\text{M}_{1}\right)$ & $\frac{1}{2} \left(-\text{M}_{1}-\sqrt{3} \text{M}_{2}\right)$ & $\text{M}_{1} $& $-\text{M}_{1}$  \\
 & $\text{M}_{2}$ & $\text{M}_{2}$ & $\text{M}_{2}$ & $\text{M}_{2}$ & $\frac{1}{2} \left(-\sqrt{3}
   \text{M}_{1}-\text{M}_{2}\right)$ & $\frac{1}{2} \left(\sqrt{3} \text{M}_{1}-\text{M}_{2}\right)$ & $\text{M}_{2}$ & $\text{M}_{2}$ \\\hline
\end{tabular}
\caption{Table for  two-dimensional Irreps of microscopic symmetries}
\label{tab_doublet_irrep_22}
\end{table*}
\end{center}
%\end{widetext}

\begin{table}
\begin{tabular}{|c|c|c|c|c|c|c|c|c|}\hline
 \text{Irrep} & \text{mass} & ${\bf T}_1$ & ${\bf T}_2$ & ${\bf C}_2'$ & ${\bf C}_3$ & ${\bf S}_6$ &{\bf I} & ${\bf \sigma}_d$ \\\hline
 $\mcl{T}_{\text{1g}}$  & $\text{M}_{1}$ & -$\text{M}_{1}$ & -$\text{M}_{1}$ & $\text{M}_{1}$  & $\text{M}_{3}$ & $\text{M}_{2}$ &
   $\text{M}_{1}$ & $\text{M}_{1}$ \\
 & $\text{M}_{2}$ & $\text{M}_{2}$ & -$\text{M}_{2}$ & $\text{M}_{3}$  & $\text{M}_{1}$ & $\text{M}_{3}$ &
   $\text{M}_{2}$ & $\text{M}_{3}$  \\
 & $\text{M}_{3}$ & -$\text{M}_{3}$ & $\text{M}_{3}$ & $\text{M}_{2}$  & $\text{M}_{2}$ & $\text{M}_{1}$ &
   $\text{M}_{3}$ & $\text{M}_{2}$  \\\hline
    $\mcl{T}_{\text{2g}}$ &  $\text{M}_{1}$ & -$\text{M}_{1}$ & -$\text{M}_{1}$ & -$ \text{M}_{1}$ &  $\text{M}_{3}$ &  $\text{M}_{2}$ &
    $\text{M}_{1}$ & - $\text{M}_{1}$  \\
 &  $\text{M}_{2}$ &  $\text{M}_{2}$ & - $\text{M}_{2}$ & - $\text{M}_{3}$ &  $\text{M}_{1}$ &  $\text{M}_{3}$ &
    $\text{M}_{2}$ & -$\text{M}_{3}$  \\
  &  $\text{M}_{3}$ & -$\text{M}_{3}$ &  $\text{M}_{3}$ & -$\text{M}_{2}$ &  $\text{M}_{2}$ &  $\text{M}_{1}$&
    $\text{M}_{3}$ & -$\text{M}_{2}$ \\\hline
  $\mcl{T}_{\text{1u}}$ &  $\text{M}_{1}$ & -$\text{M}_{1}$ & -$\text{M}_{1}$ &  $\text{M}_{1}$  &  $\text{M}_{3}$ & -$\text{M}_{2}$ &
   -$\text{M}_{1}$ & -$\text{M}_{1}$\\
  &  $\text{M}_{2}$ &  $\text{M}_{2}$ & - $\text{M}_{2}$ &  $\text{M}_{3}$  &  $\text{M}_{1}$ & - $\text{M}_{3}$ &
   - $\text{M}_{2}$ & - $\text{M}_{3}$ \\
  &  $\text{M}_{3}$ & - $\text{M}_{3}$ &  $\text{M}_{3}$ &  $\text{M}_{2}$ &  $\text{M}_{2}$ & - $\text{M}_{1}$ &
   - $\text{M}_{3}$ & - $\text{M}_{2}$  \\\hline
   $\mcl{T}_{\text{2u}}$ &  $\text{M}_{1}$ & -$ \text{M}_{1}$ & - $\text{M}_{1}$ & - $\text{M}_{1}$  &  $\text{M}_{3}$ & - $\text{M}_{2}$ &
   - $\text{M}_{1}$ &  $\text{M}_{1}$  \\
 &  $\text{M}_{2} $&  $\text{M}_{2}$ & - $\text{M}_{2}$ & - $\text{M}_{3}$  &  $\text{M}_{1}$ & - $\text{M}_{3}$ &
   - $\text{M}_{2}$ &  $\text{M}_{3}$  \\
  &  $\text{M}_{3}$ & - $\text{M}_{3}$ &  $\text{M}_{3}$ & - $\text{M}_{2}$  &  $\text{M}_{2}$ & - $\text{M}_{1}$ &
   - $\text{M}_{3}$ &  $\text{M}_{2}$  \\\hline  
\end{tabular}
\caption{Table for  three-dimensional Irreps of microscopic symmetries}
\label{tab_triplet_irrep}
\end{table}

%%%%%%%%%%%%%%%%%%%%%

\section{ Gapless Protection: Symmetry Analysis for \texorpdfstring{$\Gamma$}{}-DSM}\label{sec_symmetry_analysis_gapless_modes}

The Bogoliubov spectrum for the two gapless non-unitary singlet SCs in Sec.~\ref{sec_gapless_singlets} has a two-gap structure with one gap exactly zero, as explained in the main text. The gapless sector in both cases is four-fold degenerate with the Hamiltonian for the gapless sector 
 (in the local basis)  given by
\begin{eqnarray} 
&&    H_{Gapless} = \int d^2\textbf{q} \left(\chi_{G}^{\dagger}\,(\tilde{\zeta}_1\,q_1 +\tilde{\zeta}_2\,q_2) \chi_{G} \right)
    \label{Eq_projecting in gapless sector} 
\end{eqnarray}
where (not to be confused with Eq. \ref{eq_chiralsu2})
\begin{eqnarray}
&&  \tilde{\zeta}_1=\begin{pmatrix}
\tilde{\beta}_3\tilde{\gamma}_1 & 0_{4\times4}\\
0_{4\times4}&\tilde{\beta}_3\tilde{\gamma}_1
\end{pmatrix} \, ,\, \tilde{\zeta}_2  
=\begin{pmatrix}
-\tilde{\beta}_0\tilde{\gamma}_2 & 0_{4\times4}\\
0_{4\times4}&\tilde{\beta}_0\tilde{\gamma}_2
\end{pmatrix}
\label{Eq_form_of_gapless_zeta} 
\end{eqnarray}
with $\tilde{\beta}_i$ and $\tilde{\gamma}_i$ are the 2-dimensional Identity and the Pauli matrices and $\chi_G$ is a $8$-component Nambu spinor obtained from an unitary transformation of Eq. \ref{eq:Nambu spinor in momentum space} and is given by
\begin{align}
    \chi_G=(\eta_{G}, \left[\eta_{G}^\dagger\right]^T)
\end{align}
where $\eta_{G}$ is a 4-component Dirac spinor capturing the gapless sector of Fig. \ref{Fig_Gapless_singlet} that is given by
\begin{align}
    \eta_{G}=\left(\begin{array}{c}
\frac{\text{$\chi
   $(q)}_{31-}}{\sqrt{2}}+\frac{\text{$\chi $(q)}_{42+}}{\sqrt{2}} \\ 
   \frac{\text{$\chi $(q)}_{41+}}{\sqrt{2}}-\frac{\text{$\chi
   $(q)}_{32-}}{\sqrt{2}} \\
   \frac{\text{$\chi $(q)}_{22-}}{\sqrt{2}}-\frac{\text{$\chi $(q)}_{11+}}{\sqrt{2}} \\ \frac{\text{$\chi
   $(q)}_{21-}}{\sqrt{2}}+\frac{\text{$\chi $(q)}_{12+}}{\sqrt{2}} \\
    \end{array}\right)
\end{align}
The Hamiltonian in Eq. \ref{Eq_projecting in gapless sector} has a SO(4) symmetry which becomes manifest in terms of a Majorana representation.

Notably, these gapless fermions can be gapped out via a mass term similar to that discussed in the main text (Eq. \ref{eq_sc_mean_field}), albeit with an $8$-dimensional mass-matrix of the form
\begin{equation}\label{eq_label_8_dim_matrix}
\tilde{\mcl{m}}_{abc}= \mu_{a}\tilde{\beta}_b\tilde{\gamma}_c~~~~~ a,b,c=0,1,2,3  
\end{equation}
where $\mu$ still acts in the Nambu space. The lattice symmetry transformations of 10 different allowed mass terms are given in Table \ref{tab_gapless protecting table}, and it shows that most of them are not allowed without further lattice symmetry breaking. However, the three masses given by the matrices
$\{\tilde{\mcl{m}}_{021}, \tilde{\mcl{m}}_{202}, \tilde{\mcl{m}}_{102}\}$ 
which does not break any lattice symmetries, but are forbidden by the emergent $SO(4)$ symmetry of Eq. \ref{Eq_projecting in gapless sector}. The mass term $\tilde{m}_{021}$ is the projection of bilinear  $\chi^{\dagger}\Sigma_{45}\tau_3\sigma_3\chi$
which is quantum Spin-octupole Hall insulator($\mcl{A}^{e}_{1g}$). while the two components of superconducting mass ($\tilde{m}_{202}$, $\tilde{m}_{102}$) are respectively the projections of the singlet ($\mcl{A}_{1g}$) SC $\chi^T \Sigma_{13}\tau_1\sigma_0 \chi$ discussed in the main text (Sec. \ref{sec_anti_singlet_section}).

\begin{table}
\begin{tabular}{|c|ccccc|} \hline
 Mass & ${\bf T_1}$ & ${\bf T_2}$ & ${\bf C_2'}$ & ${\bf C_3}$ & {\bf TR} \\ \hline
 $\tilde{\mcl{m}}_{021}$ & \text{Yes} & \text{Yes} & \text{Yes} & \text{Yes} & \text{Yes} \\
 $\tilde{\mcl{m}}_{102}$ & \text{Yes} & \text{Yes} & \text{Yes} & \text{Yes} & \text{No} \\
 $\tilde{\mcl{m}}_{120}$ & \text{Yes} & \text{Yes} & \text{No} & \text{No} & \text{No} \\
 $\tilde{\mcl{m}}_{132}$ & \text{Yes} & \text{Yes} & \text{Yes} & \text{No} & \text{No} \\
 $\tilde{\mcl{m}}_{202}$ & \text{Yes} & \text{Yes} & \text{Yes} & \text{Yes} & \text{Yes} \\
 $\tilde{\mcl{m}}_{220}$ & \text{Yes} & \text{Yes} & \text{No} & \text{No} & \text{Yes} \\
 $\tilde{\mcl{m}}_{232}$ & \text{Yes} & \text{Yes} & \text{Yes} & \text{No} & \text{Yes} \\
 $\tilde{\mcl{m}}_{303}$ & \text{Yes} & \text{Yes} & \text{No} & \text{No} & \text{No} \\
 $\tilde{\mcl{m}}_{311}$ & \text{Yes} & \text{Yes} & \text{Yes} & \text{No} & \text{No} \\
 $\tilde{\mcl{m}}_{333}$ & \text{Yes} & \text{Yes} & \text{No} & \text{Yes} & \text{No} \\ \hline
\end{tabular}
\caption{Table of mass terms. The notation used is mentioned in Eq.~\ref{eq_label_8_dim_matrix}. \textbf{Yes(No)} implies corresponding symmetry is \textbf{Not broken(broken)} by mass term.}
\label{tab_gapless protecting table}
\end{table}
%%%%%%%%%%%%%%%%%%%%%%

\section{Summary of the two other gapped doublets  \texorpdfstring{$\mcl{E}_{u}$}{}}\label{sec_E_u_doublet_II_III}

While, $\mcl{E}^{I}_{u}$ is discussed in detail in the main text (Sec. \ref{sec_gapped_doublets}), here we summarize the other two ($\mcl{E}^{II}_{u}$ and $\mcl{E}^{III}_{u}$) which are adiabatically connected to the first one.  Using the parametrization in Eq.~\ref{eq_sc_amp_E1u},  spectrum of $ \left[\boldsymbol{\Delta}^{\mcl{E}^x_{u}}\cdot{\bf m}^{{\mcl{E}}^x_{u}}\right]\cdot \left[\boldsymbol{\Delta}^{\mcl{E}^x_{u}}\cdot{\bf m}^{\mcl{E}^x_{u}}\right]^{\dagger}$ for $x\in\left(I,II\right)$, has the form,
\begin{align}
   |\Delta^{\mcl{E}^{x}_{u}}|^2 \Bigg(\frac{1\pm\sin (\tilde{\gamma} ) \sin (2\theta )}{2}~, \nonumber \\~\frac{1}{6} \left(5\pm3 \sin (\tilde{\gamma} ) \sin (2 \theta )-4 \cos (2 \theta )\right)\Bigg)
\end{align}
such that there are 4 distinct eigenvalues and each one is 4-fold degenerate. As it can be concluded from the spectrum even on TRI ($\tilde{\gamma}=0~\text{ or }\pi$), these SC are always gapped but not unitary. However, they develop nodes at isolated points on the TRB manifold. 

As mentioned in the main text, these two doublets correspond to the same phase as $\mcl{E}^{I}_u$ discussed before. It can be shown by considering the following deformation from $\mcl{E}^{I}_{u}$ to $\mcl{E}^{II}_u$ doublets, carried through parameter p
\begin{align}
        \tilde{M}_1(p) = \text{p}~\mu_1 \otimes m_{1}^{\mcl{E}^I_u} + (1-\text{p}) \mu_1\otimes (m_{1}^{\mcl{E}^{II}_u}+m_{2}^{\mcl{E}^{II}_u}) 
\end{align}
and further deformation to $\mcl{E}^{III}_u$ in the following way,
 \begin{align}
        \tilde{M}_2(p) = \text{p}~\mu_1 \otimes m_{1}^{\mcl{E}^{II}_u}+ (1-\text{p}) \mu_1\otimes m_{1}^{\mcl{E}^{III}_u}.
    \end{align}
Spectrum of $\tilde{M}_1$ and $\tilde{M}_2$ shows that none of the eigenvalues go to zero during deformation, and also no extra microscopic symmetries were broken. 

 The lattice model for $m_{1}^{\mcl{E}_u^{II}}$ and $m_{1}^{\mcl{E}_u^{III}}$ have NNN pairing as  shown in Fig.~\ref{Fig_lattice_model_E_u_NNN}, with the corresponding pairing matrices, \\
For $m_1^{\mcl{E}_{u}^{II}}$,
\begin{align}\label{eq_m_1_Eu_II_matirx}
    \Sigma_a=-\Sigma_{25}~,~\Sigma_b = \frac{\sqrt{3}\Sigma_{24}+\Sigma_{25}}{4}~,~\Sigma_c = \frac{\sqrt{3}\Sigma_{24}-\Sigma_{25}}{4}
\end{align}
and for $m_1^{\mcl{E}_{u}^{III}}$,
\begin{align}\label{eq_m_1_Eu_III_matirx}
    \Sigma_a=\Sigma_{24}~,~\Sigma_b = \frac{\sqrt{3}\Sigma_{25}-\Sigma_{24}}{4}~,~\Sigma_c = \frac{\sqrt{3}\Sigma_{25}+\Sigma_{24}}{4}
\end{align}

%%%%%%%%%%%%%

\begin{figure}%[h]\label{Fig_lattice_model_E_u_12}
\includegraphics[width=.55\columnwidth]{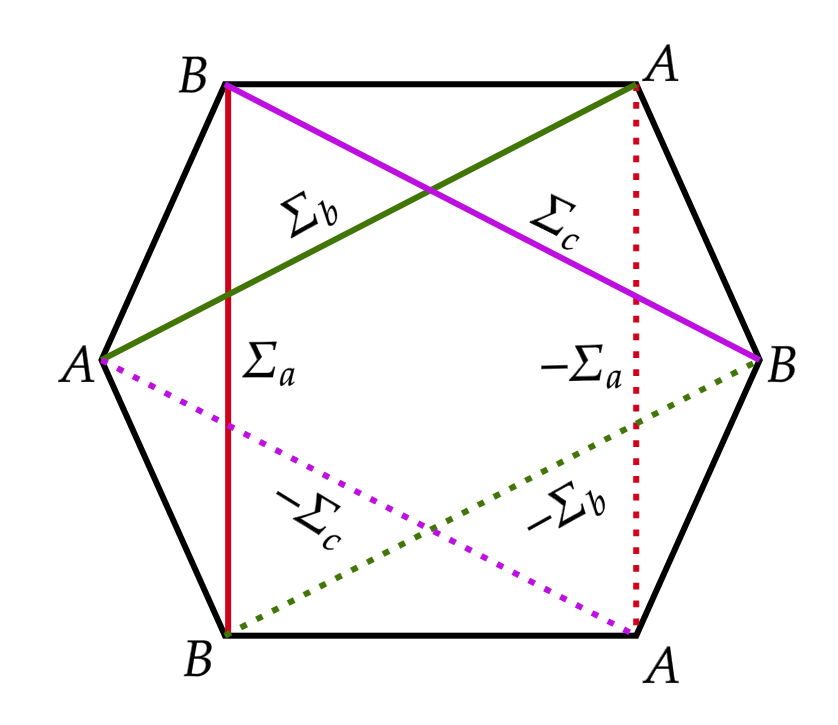}
% \label{Fig_lattice_model_E_u_2}
    \caption{Schematics of the pairing for the two components $m^{\mathcal{E}_u}_{II}$ and $m^{\mathcal{E}_u}_{III}$ with the corresponding pairing matrices mentioned in Eq.~\ref{eq_m_1_Eu_II_matirx} and ~\ref{eq_m_1_Eu_III_matirx} respectively. The pairing amplitudes are on NNN bonds, and the solid (dashed) lines are related to each other by the change in sign of the pairing matrix as indicated.}
      \label{Fig_lattice_model_E_u_NNN}
  \end{figure}
%%%%%%%%%%%%%%%%%%%

\section{The analysis of Triplet PDW}
\label{appen_su3secondary}
To understand the symmetry-breaking pattern for the triplets $\mcl{T}_{1g}$, here we will be looking at the secondary order parameter in the same spirit as done in Eq.~\ref{eq_E_g_doublet_secondary_OP}, but using SU(3) Gell-Mann matrices~\cite{wilson2024discretemodelgellmannmatrices}. There are 8 possible independent secondary order parameters defined as, 
\begin{align}
    {\bf \Lambda}_i = {\bf d}^{\dagger} \Lambda_i {\bf d} \text{ where }~i\in(1,..,8)
\end{align}
where $\Lambda_i$ are 3-dimensional Gell-Mann matrices \cite{wilson2024discretemodelgellmannmatrices}. In terms of  parameters $(\theta, \phi,\tilde{\gamma}_1, \tilde{\gamma_2})$, ${\bf \Lambda}$ has the following form:
\begin{align}
    {\bf \Lambda} = \begin{pmatrix}  {\bf \Lambda_1} \\ {\bf \Lambda_2} \\ {\bf \Lambda_3} \\ {\bf \Lambda_4} \\ {\bf \Lambda_5 } \\ {\bf \Lambda_6} \\
{\bf \Lambda_7} \\ {\bf \Lambda_8}   \end{pmatrix} =\begin{pmatrix} \cos \left(\tilde{\gamma} _1\right) \sin (2 \theta ) \cos (\phi ) \\ \sin \left(\tilde{\gamma}_1\right) \sin (2 \theta ) \cos (\phi ) \\ \frac{1}{4} \left(1-2 \sin ^2(\theta ) \cos (2 \phi
   )+3 \cos (2 \theta )\right)\\\cos \left(\tilde{\gamma}_2\right) \sin (2 \theta ) \sin (\phi ) \\ \sin \left(\tilde{\gamma}_2\right) \sin (2 \theta ) \sin (\phi ) \\ \cos \left(\tilde{\gamma}
   _1-\tilde{\gamma} _2\right) \sin ^2(\theta ) \sin (2 \phi )\\ -\sin \left(\tilde{\gamma} _1-\tilde{\gamma} _2\right)\sin ^2(\theta )\sin (2 \phi ) \\ \frac{1}{4} \left(6 \sin ^2(\theta
   ) \cos (2 \phi )+3 \cos (2 \theta )+1\right)\end{pmatrix}
\end{align}
These 8 breaks into $3\oplus3\oplus2$ under lattice symmetries, such that $({\bf \Lambda_1},{\bf \Lambda_4},{\bf \Lambda_6})$ and  $({\bf \Lambda_2},{\bf \Lambda_5},{\bf \Lambda_7})$ corresponds to $\mcl{T}^{e}_{1g}$ and $\mcl{T}^{o}_{2g}$ respectively  and $({\bf \Lambda_3},{\bf \Lambda_8})$ corresponds to $\mcl{E}^e_{g}$. As mentioned before,  $|{\bf d}\times {\bf d}^*|=0$ is the condition for TRI, 
which implies $\mcl{T}^o_{2g}$ should be zero.
The leading order anisotropic term in the free energy, which is allowed by lattice symmetries, is of the form,
\begin{align}
  \lambda_1 ({\bf \Lambda_1}^2+{\bf \Lambda_4}^2+{\bf \Lambda_6}^2)+&  \lambda_2 ({\bf \Lambda_2}^2+{\bf \Lambda_5}^2+{\bf \Lambda_7}^2) + \nonumber \\ &  \lambda_3\left( {\bf \Lambda_3}^2+{\bf \Lambda_8}^2 \right)
\end{align}
where signs of $\lambda_1,\lambda_2$ and $\lambda_3$ dictates symmetry breaking pattern of the SC.
Also, this analysis is the same for other triplets discussed in the main text.
%%%%%%%%%%%%

%%%%%%%%%%%%%%%%%%%%%%%%%%%%%%%%%

\section{Superconductivity in system with \texorpdfstring{$j=1/2$}{} orbitals}\label{sec_su2global}

In this appendix, we present our results for the same problem as studied in the main text, but for two orbitals {\it i.e.}, a $j=1/2$ doublet per site (in contrast with four orbitals per site for $j=3/2$ (Eq. \ref{eq_j3/2orb}) discussed in the main text) of the honeycomb lattice (fig.~\ref{fig_transformation}).%study the superconducting phases in a spin-orbit coupled system with two $j=1/2$ orbitals at each site (in contrast with four orbitals per site for $j=3/2$ discussed in the main text) of the honeycomb lattice (fig.~\ref{fig_transformation}). 
The Hamiltonian of the system is the same as Eq.~\ref{eq_masaki_ham} with $\mcl{U}^{global}_{\mbf{rr'}}$ are $2\times2$ Pauli matrices acting on the $j=1/2$ orbitals %{\color{red}\bf [Since we do not use $\tau_1,\tau_2, \tau_3$ in the rest, I have just removed them to avoid confusion with the use of $\tau$ in the main text.]} 
whose directed product around the hexagon is equal to $-\mathbb{I}_{2\times 2}$, as in the main text, indicating the $\pi$-flux. The system has (fig.~\ref{fig_transformation}) lattice translations ${\bf T}_1$ and ${\bf T}_2$, ${\bf C}_3$ rotations, $\boldsymbol{\sigma}_d$ dihedral reflection and time reversal (TR) $\mathbb{T}$, with, as in the main text, $\mathbb{T}^2 = -1$. 

Working in the global basis (see main text), this system has four non-degenerate bands. At quarter filling, two bands touch ``linearly'' at four points $\Gamma,M_1,M_2, M_3$ in the BZ leading to a low-energy description in terms of 2-component Dirac spinors $\chi_\nu$ where $\nu \in \{\Gamma,M_1,M_2, M_3\}$ at the four valleys. The low-energy theory, analogous to Eq.~\ref{eq_dirac_intro}, for this case is
\begin{equation}\label{eqn:su2normaldirac}
    H = -i v_F \int \textup{d}{^2 x} \chi^\dagger(x) (\alpha_1 \partial_1 + \alpha_2 \partial_2) \chi(x),
\end{equation}
a Dirac theory where $\chi(x)$  is now an 8-component Dirac spinor made up by stacking two-component Dirac spinors, $\chi_\nu$, from the four valleys. %the column vector of the Dirac spinors at the $4$ valleys, 
The $8\times8$ matrices $\alpha_i$ can be chosen as
$$
\alpha_i = \sigma_0 \sigma_0 \sigma_i
$$
Where $\sigma$s are Pauli matrices, and $v_F$ is a microscopic velocity scale (Note that for this appendix, we use $\sigma$ to denote the spin, valley and the band spaces, unlike in the main text). To prepare for the analysis of the superconducting masses, we recast this Hamiltonian in Nambu formulation (analogous to  Eq.~\ref{eq:Hamiltonian in nambu basis}) as
\begin{equation}
H_N = - i \frac{v_F}{2} \int \textup{d}{^2 x} \, \chi^\dagger_N(x) \left(\tilde{\alpha}_1 \partial_1 + \tilde{\alpha}_2 \partial_2 \right) \chi_N(x)
\end{equation}
where $\chi^\dagger_{N}(x) = ( \chi^\dagger(x) \;\;\; \chi^{T}(x))$. Here,
$$
\tilde{\alpha}_1 = M_{0001}, \;\;\; \tilde{\alpha}_2 = M_{3002}
$$
Where we have introduced a convenient notation 
$$
M_{\alpha \beta \gamma \delta} = \sigma_\alpha \sigma_\beta \sigma_\gamma \sigma_{\delta}
$$
For the set of all 256 $16 \times 16$ matrices obtained by Kronecker products of Pauli matrices which act on the Nambu, spin, valley and band spaces.

Matrices of the type $M_{\alpha \beta \gamma \delta}$ that anticommute with $\tilde{\alpha}_1$ and $\tilde{\alpha}_2$, and satisfy the conjugacy condition (analogous to Eq.~\ref{eq:constraint on Nambu hamiltonian})  
$$
M_{\alpha \beta \gamma \delta} + M_{1000} M_{\alpha \beta \gamma \delta}^T M_{1000} = 0$$ are the allowed masses. Among these, masses of the from $M_{0\beta \gamma \delta}$ and $M_{3 \beta \gamma \delta}$ are normal (non-superconducting) masses, while those of the type $M_{1\beta \gamma \delta}$ and $M_{2 \beta \gamma \delta}$ are superconducting masses describing pairing fields between the fermions. There are 16 matrices of the normal type, which result in seven distinct phases as shown in \cite{basusu8}. Here we focus on the superconducting masses of which 20 matrices, 10 each of the type $M_{1\beta \gamma \delta}$ and $M_{2 \beta \gamma \delta}$ which can be interpreted and the matrices that couple respectively to real and imaginary parts of the pairing amplitudes. 

The superconducting masses form an adjoint representation of the group generated by the symmetry elements of the system discussed above. This representation is broken down into irreducible components. We find that there are five superconducting phases, two of which are singlets, one doublet, and two triplets.  We will only show the $M_{2 \beta \gamma \delta}$ terms in the discussion below.

\subsection{\texorpdfstring{${\cal A}_1$}{} Unitary Singlet Superconductor}
This phase breaks the $\boldsymbol{\sigma}_d$ symmetry, with a mass term

\medskip
\centerline{
\begin{tabular}{c|c} \hline \hline
${\cal A}^{I}_1$& \\ \hline \hline 
\multirow{ 1}{*}{1}    & $ \left(1
\right)M_{2002}
 $ \\ 
 \hline \hline \end{tabular}
}
\medskip

This fully gapped superconductor has 8-fold-degenerate Bogoliubov quasi-particle bands and corresponds to the regular $s$-wave SC for $j=1/2$ electrons. While this phase is analogous to the ${\cal A}^I_{1g}$ phase in the system with $j=3/2$ discussed in the main text (see sec.~\ref{sec_anti_singlet_section}), the crucial difference is the absence of the extra pairing in the $j=3/2$ sector with a $\pi$-phase. This state is characterised by a uniform pairing (same at all sites) between the time-reversed $j=1/2$ fermions. 

\subsection{\texorpdfstring{${\cal A}^{II}_1$}{} Non-unitary Double-Gapped Singlet Superconductor}

This double gapped superconductor breaks the $\boldsymbol{\sigma}_d$ symmetry with a mass term

\medskip
\centerline{
\begin{tabular}{c|c} \hline \hline
${\cal A}_1^{II}$& \\ \hline \hline 
\multirow{ 1}{*}{1}    & $ \left(\frac{1}{\sqrt{3}}
\right)M_{2032}
+\left(\frac{1}{\sqrt{3}}
\right)M_{2302}
+\left(\frac{1}{\sqrt{3}}
\right)M_{2332}
 $ \\ 
 \hline \hline \end{tabular}
}
\medskip

The pairing matrix is non-unitary. The Bogoliubov bands break up into 2-fold degenerate bands around the $\Gamma$ point and $ 6$-fold degenerate bands around the $M$ points. The pairing is an extended $s$-wave type similar to the ${\cal A}^{II}_{1g}$ phase found in the system discussed in the main text (see sec.~\ref{sec_a1g2}). 

\subsection{\texorpdfstring{${\cal E}$}{} Non-unitary Gapless Doublet Superconductor}

This superconducting phase breaks ${\bf C}_3$ and $\boldsymbol{\sigma}_d$ symmetries. The doublet mass components are

\medskip
\begin{tabular}{c|c} \hline \hline
${\cal E}$& \\ \hline \hline 
\multirow{ 1}{*}{1}    & $ \left(\frac{1}{\sqrt{2}}
\right)M_{2032}
+\left(-\frac{1}{\sqrt{2}}
\right)M_{2332}
 $ \\ 
  \hline 
\multirow{ 1}{*}{2}    & $ \left(\frac{1}{\sqrt{6}}
\right)M_{2032}
+\left(-\sqrt{\frac{2}{3}}
\right)M_{2302}
+\left(\frac{1}{\sqrt{6}}
\right)M_{2332}
 $ \\ 
 \hline \hline \end{tabular}
\medskip

This {\color{red}nodal} superconductor has a quasi-particle band (doubly degenerate) at the $\Gamma$ point and three sets of doubly degenerate gapped quasi-particle bands. The gapped bands undergo crossings depending on the values of the components of the pairing amplitudes, multiplying the two masses. The pairing pattern includes anisotropic next-neighbour pairing (thus breaking the ${\bf C}_3$ symmetry), similar to the ${\cal E}_g$ doublet mass of sec.~\ref{sec_doubleteg}.
The order parameter manifold is also $(S^1 \times S^2)/Z_2$.

\subsection{\texorpdfstring{${\cal T}_2$}{} Non-unitary Gapless Triplet Superconductor}

This non-unitary superconductor breaks all the lattice symmetries, with mass components: 
\medskip
\centerline{
\begin{tabular}{c|c} \hline \hline
${\cal T}_2$& \\ \hline \hline 
\multirow{ 1}{*}{1}    & $ \left(\frac{1}{\sqrt{2}}
\right)M_{2112}
+\left(-\frac{1}{\sqrt{2}}
\right)M_{2222}
 $ \\ 
  \hline 
\multirow{ 1}{*}{2}    & $ \left(-\frac{1}{\sqrt{2}}
\right)M_{2102}
+\left(-\frac{1}{\sqrt{2}}
\right)M_{2132}
 $ \\ 
  \hline 
\multirow{ 1}{*}{3}    & $ \left(-\frac{1}{\sqrt{2}}
\right)M_{2012}
+\left(-\frac{1}{\sqrt{2}}
\right)M_{2312}
 $ \\ 
 \hline \hline \end{tabular}
}
\medskip

There is a four-fold degenerate Dirac cone quasi-particle band at the $\Gamma$ point, and a four-fold degenerate gapped quasi-particle band. Each of the mass components represents a pairing between the $\Gamma$-point states and $M_i$ point states, denoting a finite momentum pairing corresponding to the ${\bf K}_{M_i}$ wave-vectors of the BZ. This phase is analogous to the ${\cal T}_{2g}$ phase found in sec.~\ref{sec_t2g}. The order parameter manifold is $(S^1 \times CP^2)/Z_2$. The quasi-particle dispersion does not change its structure when the order parameter is varied on this manifold. Note, however, that the wavefunctions of the Dirac quasi-particles do change upon changing the order parameter on the said manifold.

\subsection{\texorpdfstring{${{\cal T}_1}$}{} Non-unitary Gapless Triplet Superconductor}

All lattice symmetries are broken in this superconducting phase, which carries a distinct three-dimensional representation (compared to the triplet discussed in the previous section) of the symmetry group of the system. The mass components are:

 \medskip
 \centerline{
 \begin{tabular}{c|c} \hline \hline
${\cal T}_1$& \\ \hline \hline 
\multirow{ 1}{*}{1}    & $ \left(\frac{1}{\sqrt{2}}
\right)M_{2112}
+\left(\frac{1}{\sqrt{2}}
\right)M_{2222}
 $ \\ 
  \hline 
\multirow{ 1}{*}{2}    & $ \left(-\frac{1}{\sqrt{2}}
\right)M_{2102}
+\left(\frac{1}{\sqrt{2}}
\right)M_{2132}
 $ \\ 
  \hline 
\multirow{ 1}{*}{3}    & $ \left(-\frac{1}{\sqrt{2}}
\right)M_{2012}
+\left(\frac{1}{\sqrt{2}}
\right)M_{2312}
 $ \\ 
 \hline \hline \end{tabular}
}
\medskip

For a generic set of pairing amplitudes, there is always a two-fold degenerate Dirac code quasi-particle band and three other gapped quasi-particle bands that are each two-fold degenerate. Here the pairing occurs between the states near the $M$ points in the BZ, {\it i.e.}, the triplet corresponds to pairing between $M_1$-$M_2$, $M_2$-$M_3$ and $M_3$-$M_3$, which gives the same vectors for the pair density as in the triplet phase discussed in the previous section. This phase is similar to the ${\cal T}_{1g}^{III}$ phase discussed in sec.~\ref{sec_t1g_III}. The order parameter manifold is again  $(S^2 \times CP^2)/Z_2$, and just as in sec.~\ref{sec_t1g_III}, there are points on the order parameter manifold where there are additional gapless modes. Interestingly, the states in the proximity of $\Gamma$ points are always gapless. Thus, the nature of gapless states at two generic points on the order parameter manifold remains unchanged; this is to be contrasted with the physics seen in the triplet superconducting phase discussed in the previous section.
%%%%%%%

\bibliography{ref}

\begin{thebibliography}{55}%
\makeatletter
\providecommand \@ifxundefined [1]{%
 \@ifx{#1\undefined}
}%
\providecommand \@ifnum [1]{%
 \ifnum #1\expandafter \@firstoftwo
 \else \expandafter \@secondoftwo
 \fi
}%
\providecommand \@ifx [1]{%
 \ifx #1\expandafter \@firstoftwo
 \else \expandafter \@secondoftwo
 \fi
}%
\providecommand \natexlab [1]{#1}%
\providecommand \enquote  [1]{``#1''}%
\providecommand \bibnamefont  [1]{#1}%
\providecommand \bibfnamefont [1]{#1}%
\providecommand \citenamefont [1]{#1}%
\providecommand \href@noop [0]{\@secondoftwo}%
\providecommand \href [0]{\begingroup \@sanitize@url \@href}%
\providecommand \@href[1]{\@@startlink{#1}\@@href}%
\providecommand \@@href[1]{\endgroup#1\@@endlink}%
\providecommand \@sanitize@url [0]{\catcode `\\12\catcode `\$12\catcode
  `\&12\catcode `\#12\catcode `\^12\catcode `\_12\catcode `\%12\relax}%
\providecommand \@@startlink[1]{}%
\providecommand \@@endlink[0]{}%
\providecommand \url  [0]{\begingroup\@sanitize@url \@url }%
\providecommand \@url [1]{\endgroup\@href {#1}{\urlprefix }}%
\providecommand \urlprefix  [0]{URL }%
\providecommand \Eprint [0]{\href }%
\providecommand \doibase [0]{https://doi.org/}%
\providecommand \selectlanguage [0]{\@gobble}%
\providecommand \bibinfo  [0]{\@secondoftwo}%
\providecommand \bibfield  [0]{\@secondoftwo}%
\providecommand \translation [1]{[#1]}%
\providecommand \BibitemOpen [0]{}%
\providecommand \bibitemStop [0]{}%
\providecommand \bibitemNoStop [0]{.\EOS\space}%
\providecommand \EOS [0]{\spacefactor3000\relax}%
\providecommand \BibitemShut  [1]{\csname bibitem#1\endcsname}%
\let\auto@bib@innerbib\@empty
%</preamble>
\bibitem [{\citenamefont {Boyack}\ \emph {et~al.}(2021)\citenamefont {Boyack},
  \citenamefont {Yerzhakov},\ and\ \citenamefont
  {Maciejko}}]{boyack2021quantum}%
  \BibitemOpen
  \bibfield  {author} {\bibinfo {author} {\bibfnamefont {R.}~\bibnamefont
  {Boyack}}, \bibinfo {author} {\bibfnamefont {H.}~\bibnamefont {Yerzhakov}},\
  and\ \bibinfo {author} {\bibfnamefont {J.}~\bibnamefont {Maciejko}},\
  }\bibfield  {title} {\bibinfo {title} {Quantum phase transitions in dirac
  fermion systems},\ }\href {http://dx.doi.org/10.1140/epjs/s11734-021-00069-1}
  {\bibfield  {journal} {\bibinfo  {journal} {The European Physical Journal
  Special Topics}\ }\textbf {\bibinfo {volume} {230}},\ \bibinfo {pages} {979}
  (\bibinfo {year} {2021})}\BibitemShut {NoStop}%
\bibitem [{\citenamefont {Kopnin}\ and\ \citenamefont
  {Sonin}(2008)}]{PhysRevLett.100.246808}%
  \BibitemOpen
  \bibfield  {author} {\bibinfo {author} {\bibfnamefont {N.~B.}\ \bibnamefont
  {Kopnin}}\ and\ \bibinfo {author} {\bibfnamefont {E.~B.}\ \bibnamefont
  {Sonin}},\ }\bibfield  {title} {\bibinfo {title} {Bcs superconductivity of
  dirac electrons in graphene layers},\ }\href
  {https://doi.org/10.1103/PhysRevLett.100.246808} {\bibfield  {journal}
  {\bibinfo  {journal} {Phys. Rev. Lett.}\ }\textbf {\bibinfo {volume} {100}},\
  \bibinfo {pages} {246808} (\bibinfo {year} {2008})}\BibitemShut {NoStop}%
\bibitem [{\citenamefont {Uchoa}\ \emph {et~al.}(2005)\citenamefont {Uchoa},
  \citenamefont {Cabrera},\ and\ \citenamefont
  {Castro~Neto}}]{PhysRevB.71.184509}%
  \BibitemOpen
  \bibfield  {author} {\bibinfo {author} {\bibfnamefont {B.}~\bibnamefont
  {Uchoa}}, \bibinfo {author} {\bibfnamefont {G.~G.}\ \bibnamefont {Cabrera}},\
  and\ \bibinfo {author} {\bibfnamefont {A.~H.}\ \bibnamefont {Castro~Neto}},\
  }\bibfield  {title} {\bibinfo {title} {Nodal liquid and $s$-wave
  superconductivity in transition metal dichalcogenides},\ }\href
  {https://doi.org/10.1103/PhysRevB.71.184509} {\bibfield  {journal} {\bibinfo
  {journal} {Phys. Rev. B}\ }\textbf {\bibinfo {volume} {71}},\ \bibinfo
  {pages} {184509} (\bibinfo {year} {2005})}\BibitemShut {NoStop}%
\bibitem [{\citenamefont {Castro~Neto}(2001)}]{PhysRevLett.86.4382}%
  \BibitemOpen
  \bibfield  {author} {\bibinfo {author} {\bibfnamefont {A.~H.}\ \bibnamefont
  {Castro~Neto}},\ }\bibfield  {title} {\bibinfo {title} {Charge density wave,
  superconductivity, and anomalous metallic behavior in 2d transition metal
  dichalcogenides},\ }\href {https://doi.org/10.1103/PhysRevLett.86.4382}
  {\bibfield  {journal} {\bibinfo  {journal} {Phys. Rev. Lett.}\ }\textbf
  {\bibinfo {volume} {86}},\ \bibinfo {pages} {4382} (\bibinfo {year}
  {2001})}\BibitemShut {NoStop}%
\bibitem [{\citenamefont {Uchoa}\ and\ \citenamefont
  {Neto}(2009)}]{PhysRevLett.102.109701}%
  \BibitemOpen
  \bibfield  {author} {\bibinfo {author} {\bibfnamefont {B.}~\bibnamefont
  {Uchoa}}\ and\ \bibinfo {author} {\bibfnamefont {A.~H.~C.}\ \bibnamefont
  {Neto}},\ }\bibfield  {title} {\bibinfo {title} {Comment on ``bcs
  superconductivity of dirac electrons in graphene layers''},\ }\href
  {https://doi.org/10.1103/PhysRevLett.102.109701} {\bibfield  {journal}
  {\bibinfo  {journal} {Phys. Rev. Lett.}\ }\textbf {\bibinfo {volume} {102}},\
  \bibinfo {pages} {109701} (\bibinfo {year} {2009})}\BibitemShut {NoStop}%
\bibitem [{\citenamefont {Herbut}(2006)}]{PhysRevLett.97.146401}%
  \BibitemOpen
  \bibfield  {author} {\bibinfo {author} {\bibfnamefont {I.~F.}\ \bibnamefont
  {Herbut}},\ }\bibfield  {title} {\bibinfo {title} {Interactions and phase
  transitions on graphene's honeycomb lattice},\ }\href
  {https://doi.org/10.1103/PhysRevLett.97.146401} {\bibfield  {journal}
  {\bibinfo  {journal} {Phys. Rev. Lett.}\ }\textbf {\bibinfo {volume} {97}},\
  \bibinfo {pages} {146401} (\bibinfo {year} {2006})}\BibitemShut {NoStop}%
\bibitem [{\citenamefont {Roy}\ and\ \citenamefont
  {Herbut}(2010)}]{PhysRevB.82.035429}%
  \BibitemOpen
  \bibfield  {author} {\bibinfo {author} {\bibfnamefont {B.}~\bibnamefont
  {Roy}}\ and\ \bibinfo {author} {\bibfnamefont {I.~F.}\ \bibnamefont
  {Herbut}},\ }\bibfield  {title} {\bibinfo {title} {Unconventional
  superconductivity on honeycomb lattice: Theory of kekule order parameter},\
  }\href {https://doi.org/10.1103/PhysRevB.82.035429} {\bibfield  {journal}
  {\bibinfo  {journal} {Phys. Rev. B}\ }\textbf {\bibinfo {volume} {82}},\
  \bibinfo {pages} {035429} (\bibinfo {year} {2010})}\BibitemShut {NoStop}%
\bibitem [{\citenamefont {Tinkham}(2004)}]{tinkham2004introduction}%
  \BibitemOpen
  \bibfield  {author} {\bibinfo {author} {\bibfnamefont {M.}~\bibnamefont
  {Tinkham}},\ }\href@noop {} {\emph {\bibinfo {title} {Introduction to
  superconductivity}}}\ (\bibinfo  {publisher} {Courier Corporation},\ \bibinfo
  {year} {2004})\BibitemShut {NoStop}%
\bibitem [{\citenamefont
  {Wilson}(2024)}]{wilson2024discretemodelgellmannmatrices}%
  \BibitemOpen
  \bibfield  {author} {\bibinfo {author} {\bibfnamefont {R.~A.}\ \bibnamefont
  {Wilson}},\ }\href {https://arxiv.org/abs/2401.13000} {\bibinfo {title} {A
  discrete model for gell-mann matrices}} (\bibinfo {year} {2024}),\ \Eprint
  {https://arxiv.org/abs/2401.13000} {arXiv:2401.13000 [math.GR]} \BibitemShut
  {NoStop}%
\bibitem [{\citenamefont {Beenakker}(2006)}]{PhysRevLett.97.067007}%
  \BibitemOpen
  \bibfield  {author} {\bibinfo {author} {\bibfnamefont {C.~W.~J.}\
  \bibnamefont {Beenakker}},\ }\bibfield  {title} {\bibinfo {title} {Specular
  andreev reflection in graphene},\ }\href
  {https://doi.org/10.1103/PhysRevLett.97.067007} {\bibfield  {journal}
  {\bibinfo  {journal} {Phys. Rev. Lett.}\ }\textbf {\bibinfo {volume} {97}},\
  \bibinfo {pages} {067007} (\bibinfo {year} {2006})}\BibitemShut {NoStop}%
\bibitem [{\citenamefont {Bhattacharjee}\ \emph {et~al.}(2007)\citenamefont
  {Bhattacharjee}, \citenamefont {Maiti},\ and\ \citenamefont
  {Sengupta}}]{PhysRevB.76.184514}%
  \BibitemOpen
  \bibfield  {author} {\bibinfo {author} {\bibfnamefont {S.}~\bibnamefont
  {Bhattacharjee}}, \bibinfo {author} {\bibfnamefont {M.}~\bibnamefont
  {Maiti}},\ and\ \bibinfo {author} {\bibfnamefont {K.}~\bibnamefont
  {Sengupta}},\ }\bibfield  {title} {\bibinfo {title} {Theory of tunneling
  conductance of graphene normal metal-insulator-superconductor junctions},\
  }\href {https://doi.org/10.1103/PhysRevB.76.184514} {\bibfield  {journal}
  {\bibinfo  {journal} {Phys. Rev. B}\ }\textbf {\bibinfo {volume} {76}},\
  \bibinfo {pages} {184514} (\bibinfo {year} {2007})}\BibitemShut {NoStop}%
\bibitem [{\citenamefont {Bhattacharjee}\ and\ \citenamefont
  {Sengupta}(2006)}]{PhysRevLett.97.217001}%
  \BibitemOpen
  \bibfield  {author} {\bibinfo {author} {\bibfnamefont {S.}~\bibnamefont
  {Bhattacharjee}}\ and\ \bibinfo {author} {\bibfnamefont {K.}~\bibnamefont
  {Sengupta}},\ }\bibfield  {title} {\bibinfo {title} {Tunneling conductance of
  graphene nis junctions},\ }\href
  {https://doi.org/10.1103/PhysRevLett.97.217001} {\bibfield  {journal}
  {\bibinfo  {journal} {Phys. Rev. Lett.}\ }\textbf {\bibinfo {volume} {97}},\
  \bibinfo {pages} {217001} (\bibinfo {year} {2006})}\BibitemShut {NoStop}%
\bibitem [{\citenamefont {Huang}\ \emph {et~al.}(2019)\citenamefont {Huang},
  \citenamefont {Zhou}, \citenamefont {Zhang}, \citenamefont {Yang},
  \citenamefont {Liu}, \citenamefont {Wang}, \citenamefont {Wan}, \citenamefont
  {Huang}, \citenamefont {Liao}, \citenamefont {Zhang} \emph
  {et~al.}}]{huang2019proximity}%
  \BibitemOpen
  \bibfield  {author} {\bibinfo {author} {\bibfnamefont {C.}~\bibnamefont
  {Huang}}, \bibinfo {author} {\bibfnamefont {B.~T.}\ \bibnamefont {Zhou}},
  \bibinfo {author} {\bibfnamefont {H.}~\bibnamefont {Zhang}}, \bibinfo
  {author} {\bibfnamefont {B.}~\bibnamefont {Yang}}, \bibinfo {author}
  {\bibfnamefont {R.}~\bibnamefont {Liu}}, \bibinfo {author} {\bibfnamefont
  {H.}~\bibnamefont {Wang}}, \bibinfo {author} {\bibfnamefont {Y.}~\bibnamefont
  {Wan}}, \bibinfo {author} {\bibfnamefont {K.}~\bibnamefont {Huang}}, \bibinfo
  {author} {\bibfnamefont {Z.}~\bibnamefont {Liao}}, \bibinfo {author}
  {\bibfnamefont {E.}~\bibnamefont {Zhang}}, \emph {et~al.},\ }\bibfield
  {title} {\bibinfo {title} {Proximity-induced surface superconductivity in
  dirac semimetal cd3as2},\ }\href
  {https://www.nature.com/articles/s41467-019-10233-w} {\bibfield  {journal}
  {\bibinfo  {journal} {Nature communications}\ }\textbf {\bibinfo {volume}
  {10}},\ \bibinfo {pages} {2217} (\bibinfo {year} {2019})}\BibitemShut
  {NoStop}%
\bibitem [{\citenamefont {Santos}\ \emph {et~al.}(2010)\citenamefont {Santos},
  \citenamefont {Neupert}, \citenamefont {Chamon},\ and\ \citenamefont
  {Mudry}}]{PhysRevB.81.184502}%
  \BibitemOpen
  \bibfield  {author} {\bibinfo {author} {\bibfnamefont {L.}~\bibnamefont
  {Santos}}, \bibinfo {author} {\bibfnamefont {T.}~\bibnamefont {Neupert}},
  \bibinfo {author} {\bibfnamefont {C.}~\bibnamefont {Chamon}},\ and\ \bibinfo
  {author} {\bibfnamefont {C.}~\bibnamefont {Mudry}},\ }\bibfield  {title}
  {\bibinfo {title} {Superconductivity on the surface of topological insulators
  and in two-dimensional noncentrosymmetric materials},\ }\href
  {https://doi.org/10.1103/PhysRevB.81.184502} {\bibfield  {journal} {\bibinfo
  {journal} {Phys. Rev. B}\ }\textbf {\bibinfo {volume} {81}},\ \bibinfo
  {pages} {184502} (\bibinfo {year} {2010})}\BibitemShut {NoStop}%
\bibitem [{\citenamefont {Kobayashi}\ and\ \citenamefont
  {Sato}(2015)}]{PhysRevLett.115.187001}%
  \BibitemOpen
  \bibfield  {author} {\bibinfo {author} {\bibfnamefont {S.}~\bibnamefont
  {Kobayashi}}\ and\ \bibinfo {author} {\bibfnamefont {M.}~\bibnamefont
  {Sato}},\ }\bibfield  {title} {\bibinfo {title} {Topological
  superconductivity in dirac semimetals},\ }\href
  {https://doi.org/10.1103/PhysRevLett.115.187001} {\bibfield  {journal}
  {\bibinfo  {journal} {Phys. Rev. Lett.}\ }\textbf {\bibinfo {volume} {115}},\
  \bibinfo {pages} {187001} (\bibinfo {year} {2015})}\BibitemShut {NoStop}%
\bibitem [{\citenamefont {Zhao}\ \emph {et~al.}(2015)\citenamefont {Zhao},
  \citenamefont {Deng}, \citenamefont {Korzhovska}, \citenamefont
  {Begliarbekov}, \citenamefont {Chen}, \citenamefont {Andrade}, \citenamefont
  {Rosenthal}, \citenamefont {Pasupathy}, \citenamefont {Oganesyan},\ and\
  \citenamefont {Krusin-Elbaum}}]{zhao2015emergent}%
  \BibitemOpen
  \bibfield  {author} {\bibinfo {author} {\bibfnamefont {L.}~\bibnamefont
  {Zhao}}, \bibinfo {author} {\bibfnamefont {H.}~\bibnamefont {Deng}}, \bibinfo
  {author} {\bibfnamefont {I.}~\bibnamefont {Korzhovska}}, \bibinfo {author}
  {\bibfnamefont {M.}~\bibnamefont {Begliarbekov}}, \bibinfo {author}
  {\bibfnamefont {Z.}~\bibnamefont {Chen}}, \bibinfo {author} {\bibfnamefont
  {E.}~\bibnamefont {Andrade}}, \bibinfo {author} {\bibfnamefont
  {E.}~\bibnamefont {Rosenthal}}, \bibinfo {author} {\bibfnamefont
  {A.}~\bibnamefont {Pasupathy}}, \bibinfo {author} {\bibfnamefont
  {V.}~\bibnamefont {Oganesyan}},\ and\ \bibinfo {author} {\bibfnamefont
  {L.}~\bibnamefont {Krusin-Elbaum}},\ }\bibfield  {title} {\bibinfo {title}
  {Emergent surface superconductivity in the topological insulator sb2te3},\
  }\href {https://doi.org/10.1038/ncomms9279} {\bibfield  {journal} {\bibinfo
  {journal} {Nature communications}\ }\textbf {\bibinfo {volume} {6}},\
  \bibinfo {pages} {8279} (\bibinfo {year} {2015})}\BibitemShut {NoStop}%
\bibitem [{\citenamefont {Ryu}\ \emph {et~al.}(2009)\citenamefont {Ryu},
  \citenamefont {Mudry}, \citenamefont {Hou},\ and\ \citenamefont
  {Chamon}}]{Ryu_graphene_masses}%
  \BibitemOpen
  \bibfield  {author} {\bibinfo {author} {\bibfnamefont {S.}~\bibnamefont
  {Ryu}}, \bibinfo {author} {\bibfnamefont {C.}~\bibnamefont {Mudry}}, \bibinfo
  {author} {\bibfnamefont {C.-Y.}\ \bibnamefont {Hou}},\ and\ \bibinfo {author}
  {\bibfnamefont {C.}~\bibnamefont {Chamon}},\ }\bibfield  {title} {\bibinfo
  {title} {Masses in graphenelike two-dimensional electronic systems:
  Topological defects in order parameters and their fractional exchange
  statistics},\ }\href {https://doi.org/10.1103/PhysRevB.80.205319} {\bibfield
  {journal} {\bibinfo  {journal} {Phys. Rev. B}\ }\textbf {\bibinfo {volume}
  {80}},\ \bibinfo {pages} {205319} (\bibinfo {year} {2009})}\BibitemShut
  {NoStop}%
\bibitem [{\citenamefont {Yamada}\ \emph {et~al.}(2018)\citenamefont {Yamada},
  \citenamefont {Oshikawa},\ and\ \citenamefont {Jackeli}}]{Masaki_su4}%
  \BibitemOpen
  \bibfield  {author} {\bibinfo {author} {\bibfnamefont {M.~G.}\ \bibnamefont
  {Yamada}}, \bibinfo {author} {\bibfnamefont {M.}~\bibnamefont {Oshikawa}},\
  and\ \bibinfo {author} {\bibfnamefont {G.}~\bibnamefont {Jackeli}},\
  }\bibfield  {title} {\bibinfo {title} {Emergent $\mathrm{SU}(4)$ symmetry in
  $\ensuremath{\alpha}\text{\ensuremath{-}}{\mathrm{zrcl}}_{3}$ and crystalline
  spin-orbital liquids},\ }\href
  {https://doi.org/10.1103/PhysRevLett.121.097201} {\bibfield  {journal}
  {\bibinfo  {journal} {Phys. Rev. Lett.}\ }\textbf {\bibinfo {volume} {121}},\
  \bibinfo {pages} {097201} (\bibinfo {year} {2018})}\BibitemShut {NoStop}%
\bibitem [{\citenamefont {Mondal}\ \emph {et~al.}(2023)\citenamefont {Mondal},
  \citenamefont {Shenoy},\ and\ \citenamefont {Bhattacharjee}}]{basusu8}%
  \BibitemOpen
  \bibfield  {author} {\bibinfo {author} {\bibfnamefont {B.}~\bibnamefont
  {Mondal}}, \bibinfo {author} {\bibfnamefont {V.~B.}\ \bibnamefont {Shenoy}},\
  and\ \bibinfo {author} {\bibfnamefont {S.}~\bibnamefont {Bhattacharjee}},\
  }\bibfield  {title} {\bibinfo {title} {Emergent su(8) dirac semimetal and
  proximate phases of spin-orbit coupled fermions on a honeycomb lattice},\
  }\href {https://doi.org/10.1103/PhysRevB.108.245106} {\bibfield  {journal}
  {\bibinfo  {journal} {Phys. Rev. B}\ }\textbf {\bibinfo {volume} {108}},\
  \bibinfo {pages} {245106} (\bibinfo {year} {2023})}\BibitemShut {NoStop}%
\bibitem [{\citenamefont {Wallace}(1947)}]{PhysRev.71.622}%
  \BibitemOpen
  \bibfield  {author} {\bibinfo {author} {\bibfnamefont {P.~R.}\ \bibnamefont
  {Wallace}},\ }\bibfield  {title} {\bibinfo {title} {The band theory of
  graphite},\ }\href {https://doi.org/10.1103/PhysRev.71.622} {\bibfield
  {journal} {\bibinfo  {journal} {Phys. Rev.}\ }\textbf {\bibinfo {volume}
  {71}},\ \bibinfo {pages} {622} (\bibinfo {year} {1947})}\BibitemShut
  {NoStop}%
\bibitem [{\citenamefont {Kim}\ \emph {et~al.}(2018)\citenamefont {Kim},
  \citenamefont {Wang}, \citenamefont {Nakajima}, \citenamefont {Hu},
  \citenamefont {Ziemak}, \citenamefont {Syers}, \citenamefont {Wang},
  \citenamefont {Hodovanets}, \citenamefont {Denlinger}, \citenamefont {Brydon}
  \emph {et~al.}}]{kim2018beyond}%
  \BibitemOpen
  \bibfield  {author} {\bibinfo {author} {\bibfnamefont {H.}~\bibnamefont
  {Kim}}, \bibinfo {author} {\bibfnamefont {K.}~\bibnamefont {Wang}}, \bibinfo
  {author} {\bibfnamefont {Y.}~\bibnamefont {Nakajima}}, \bibinfo {author}
  {\bibfnamefont {R.}~\bibnamefont {Hu}}, \bibinfo {author} {\bibfnamefont
  {S.}~\bibnamefont {Ziemak}}, \bibinfo {author} {\bibfnamefont
  {P.}~\bibnamefont {Syers}}, \bibinfo {author} {\bibfnamefont
  {L.}~\bibnamefont {Wang}}, \bibinfo {author} {\bibfnamefont {H.}~\bibnamefont
  {Hodovanets}}, \bibinfo {author} {\bibfnamefont {J.~D.}\ \bibnamefont
  {Denlinger}}, \bibinfo {author} {\bibfnamefont {P.~M.}\ \bibnamefont
  {Brydon}}, \emph {et~al.},\ }\bibfield  {title} {\bibinfo {title} {Beyond
  triplet: Unconventional superconductivity in a spin-3/2 topological
  semimetal},\ }\href {https://doi.org/10.1126/sciadv.aao4513} {\bibfield
  {journal} {\bibinfo  {journal} {Science advances}\ }\textbf {\bibinfo
  {volume} {4}},\ \bibinfo {pages} {eaao4513} (\bibinfo {year}
  {2018})}\BibitemShut {NoStop}%
\bibitem [{\citenamefont {Anderson}(1984)}]{anderson1984structure}%
  \BibitemOpen
  \bibfield  {author} {\bibinfo {author} {\bibfnamefont {P.}~\bibnamefont
  {Anderson}},\ }\bibfield  {title} {\bibinfo {title} {Structure of" triplet"
  superconducting energy gaps},\ }\href
  {https://doi.org/10.1103/PhysRevB.30.4000} {\bibfield  {journal} {\bibinfo
  {journal} {Physical Review B}\ }\textbf {\bibinfo {volume} {30}},\ \bibinfo
  {pages} {4000} (\bibinfo {year} {1984})}\BibitemShut {NoStop}%
\bibitem [{\citenamefont {Ramires}(2022)}]{ramires2022nonunitary}%
  \BibitemOpen
  \bibfield  {author} {\bibinfo {author} {\bibfnamefont {A.}~\bibnamefont
  {Ramires}},\ }\bibfield  {title} {\bibinfo {title} {Nonunitary
  superconductivity in complex quantum materials},\ }\href
  {http://dx.doi.org/10.1088/1361-648X/ac6d3a} {\bibfield  {journal} {\bibinfo
  {journal} {Journal of Physics: Condensed Matter}\ }\textbf {\bibinfo {volume}
  {34}},\ \bibinfo {pages} {304001} (\bibinfo {year} {2022})}\BibitemShut
  {NoStop}%
\bibitem [{\citenamefont {Pfleiderer}(2009)}]{RevModPhys.81.1551}%
  \BibitemOpen
  \bibfield  {author} {\bibinfo {author} {\bibfnamefont {C.}~\bibnamefont
  {Pfleiderer}},\ }\bibfield  {title} {\bibinfo {title} {Superconducting phases
  of $f$-electron compounds},\ }\href
  {https://doi.org/10.1103/RevModPhys.81.1551} {\bibfield  {journal} {\bibinfo
  {journal} {Rev. Mod. Phys.}\ }\textbf {\bibinfo {volume} {81}},\ \bibinfo
  {pages} {1551} (\bibinfo {year} {2009})}\BibitemShut {NoStop}%
\bibitem [{\citenamefont {Joynt}\ and\ \citenamefont
  {Taillefer}(2002)}]{RevModPhys.74.235}%
  \BibitemOpen
  \bibfield  {author} {\bibinfo {author} {\bibfnamefont {R.}~\bibnamefont
  {Joynt}}\ and\ \bibinfo {author} {\bibfnamefont {L.}~\bibnamefont
  {Taillefer}},\ }\bibfield  {title} {\bibinfo {title} {The superconducting
  phases of ${\mathrm{upt}}_{3}$},\ }\href
  {https://doi.org/10.1103/RevModPhys.74.235} {\bibfield  {journal} {\bibinfo
  {journal} {Rev. Mod. Phys.}\ }\textbf {\bibinfo {volume} {74}},\ \bibinfo
  {pages} {235} (\bibinfo {year} {2002})}\BibitemShut {NoStop}%
\bibitem [{\citenamefont {Nomoto}\ \emph {et~al.}(2016)\citenamefont {Nomoto},
  \citenamefont {Hattori},\ and\ \citenamefont {Ikeda}}]{PhysRevB.94.174513}%
  \BibitemOpen
  \bibfield  {author} {\bibinfo {author} {\bibfnamefont {T.}~\bibnamefont
  {Nomoto}}, \bibinfo {author} {\bibfnamefont {K.}~\bibnamefont {Hattori}},\
  and\ \bibinfo {author} {\bibfnamefont {H.}~\bibnamefont {Ikeda}},\ }\bibfield
   {title} {\bibinfo {title} {Classification of ``multipole'' superconductivity
  in multiorbital systems and its implications},\ }\href
  {https://doi.org/10.1103/PhysRevB.94.174513} {\bibfield  {journal} {\bibinfo
  {journal} {Phys. Rev. B}\ }\textbf {\bibinfo {volume} {94}},\ \bibinfo
  {pages} {174513} (\bibinfo {year} {2016})}\BibitemShut {NoStop}%
\bibitem [{\citenamefont {Sigrist}\ and\ \citenamefont
  {Ueda}(1991)}]{Ueda_superconductivity}%
  \BibitemOpen
  \bibfield  {author} {\bibinfo {author} {\bibfnamefont {M.}~\bibnamefont
  {Sigrist}}\ and\ \bibinfo {author} {\bibfnamefont {K.}~\bibnamefont {Ueda}},\
  }\bibfield  {title} {\bibinfo {title} {Phenomenological theory of
  unconventional superconductivity},\ }\href
  {https://doi.org/10.1103/RevModPhys.63.239} {\bibfield  {journal} {\bibinfo
  {journal} {Rev. Mod. Phys.}\ }\textbf {\bibinfo {volume} {63}},\ \bibinfo
  {pages} {239} (\bibinfo {year} {1991})}\BibitemShut {NoStop}%
\bibitem [{\citenamefont {Volovik}\ and\ \citenamefont
  {Gor’kov}(1985)}]{volovik1985superconducting}%
  \BibitemOpen
  \bibfield  {author} {\bibinfo {author} {\bibfnamefont {G.}~\bibnamefont
  {Volovik}}\ and\ \bibinfo {author} {\bibfnamefont {L.}~\bibnamefont
  {Gor’kov}},\ }\bibfield  {title} {\bibinfo {title} {Superconducting classes
  in heavy-fermion systems},\ }in\ \href
  {https://doi.org/10.1007/978-94-011-1622-0_14} {\emph {\bibinfo {booktitle}
  {Ten Years of Superconductivity: 1980--1990}}}\ (\bibinfo  {publisher}
  {Springer},\ \bibinfo {year} {1985})\ pp.\ \bibinfo {pages}
  {144--155}\BibitemShut {NoStop}%
\bibitem [{\citenamefont {Yarzhemsky}\ and\ \citenamefont
  {Murav'Ev}(1992)}]{yarzhemsky1992space}%
  \BibitemOpen
  \bibfield  {author} {\bibinfo {author} {\bibfnamefont {V.}~\bibnamefont
  {Yarzhemsky}}\ and\ \bibinfo {author} {\bibfnamefont {E.}~\bibnamefont
  {Murav'Ev}},\ }\bibfield  {title} {\bibinfo {title} {Space group approach to
  the wavefunction of a cooper pair},\ }\href@noop {} {\bibfield  {journal}
  {\bibinfo  {journal} {Journal of Physics: Condensed Matter}\ }\textbf
  {\bibinfo {volume} {4}},\ \bibinfo {pages} {3525} (\bibinfo {year}
  {1992})}\BibitemShut {NoStop}%
\bibitem [{\citenamefont {Blount}(1985)}]{PhysRevB.32.2935}%
  \BibitemOpen
  \bibfield  {author} {\bibinfo {author} {\bibfnamefont {E.~I.}\ \bibnamefont
  {Blount}},\ }\bibfield  {title} {\bibinfo {title} {Symmetry properties of
  triplet superconductors},\ }\href {https://doi.org/10.1103/PhysRevB.32.2935}
  {\bibfield  {journal} {\bibinfo  {journal} {Phys. Rev. B}\ }\textbf {\bibinfo
  {volume} {32}},\ \bibinfo {pages} {2935} (\bibinfo {year}
  {1985})}\BibitemShut {NoStop}%
\bibitem [{\citenamefont {Ozaki}\ and\ \citenamefont
  {Machida}(1989)}]{PhysRevB.39.4145}%
  \BibitemOpen
  \bibfield  {author} {\bibinfo {author} {\bibfnamefont {M.-a.}\ \bibnamefont
  {Ozaki}}\ and\ \bibinfo {author} {\bibfnamefont {K.}~\bibnamefont
  {Machida}},\ }\bibfield  {title} {\bibinfo {title} {Superconducting classes
  of heavy-fermion materials},\ }\href
  {https://doi.org/10.1103/PhysRevB.39.4145} {\bibfield  {journal} {\bibinfo
  {journal} {Phys. Rev. B}\ }\textbf {\bibinfo {volume} {39}},\ \bibinfo
  {pages} {4145} (\bibinfo {year} {1989})}\BibitemShut {NoStop}%
\bibitem [{\citenamefont {Leggett}(1966)}]{leggett1966number}%
  \BibitemOpen
  \bibfield  {author} {\bibinfo {author} {\bibfnamefont {A.}~\bibnamefont
  {Leggett}},\ }\bibfield  {title} {\bibinfo {title} {Number-phase fluctuations
  in two-band superconductors},\ }\href {https://doi.org/10.1143/PTP.36.901}
  {\bibfield  {journal} {\bibinfo  {journal} {Progress of Theoretical Physics}\
  }\textbf {\bibinfo {volume} {36}},\ \bibinfo {pages} {901} (\bibinfo {year}
  {1966})}\BibitemShut {NoStop}%
\bibitem [{\citenamefont {Nanda}\ \emph {et~al.}(2020)\citenamefont {Nanda},
  \citenamefont {Dhochak},\ and\ \citenamefont {Bhattacharjee}}]{Nanda_2020}%
  \BibitemOpen
  \bibfield  {author} {\bibinfo {author} {\bibfnamefont {A.}~\bibnamefont
  {Nanda}}, \bibinfo {author} {\bibfnamefont {K.}~\bibnamefont {Dhochak}},\
  and\ \bibinfo {author} {\bibfnamefont {S.}~\bibnamefont {Bhattacharjee}},\
  }\bibfield  {title} {\bibinfo {title} {Phases and quantum phase transitions
  in an anisotropic ferromagnetic
  kitaev-heisenberg-$\mathrm{\ensuremath{\Gamma}}$ magnet},\ }\bibfield
  {journal} {\bibinfo  {journal} {Physical Review B}\ }\textbf {\bibinfo
  {volume} {102}},\ \href {https://doi.org/10.1103/physrevb.102.235124}
  {10.1103/physrevb.102.235124} (\bibinfo {year} {2020})\BibitemShut {NoStop}%
\bibitem [{\citenamefont {Zhou}(2001)}]{PhysRevLett.87.080401}%
  \BibitemOpen
  \bibfield  {author} {\bibinfo {author} {\bibfnamefont {F.}~\bibnamefont
  {Zhou}},\ }\bibfield  {title} {\bibinfo {title} {Spin correlation and
  discrete symmetry in spinor bose-einstein condensates},\ }\href
  {https://doi.org/10.1103/PhysRevLett.87.080401} {\bibfield  {journal}
  {\bibinfo  {journal} {Phys. Rev. Lett.}\ }\textbf {\bibinfo {volume} {87}},\
  \bibinfo {pages} {080401} (\bibinfo {year} {2001})}\BibitemShut {NoStop}%
\bibitem [{\citenamefont {Ueda}(2014)}]{Ueda_2014}%
  \BibitemOpen
  \bibfield  {author} {\bibinfo {author} {\bibfnamefont {M.}~\bibnamefont
  {Ueda}},\ }\bibfield  {title} {\bibinfo {title} {Topological aspects in
  spinor bose–einstein condensates},\ }\href
  {https://doi.org/10.1088/0034-4885/77/12/122401} {\bibfield  {journal}
  {\bibinfo  {journal} {Reports on Progress in Physics}\ }\textbf {\bibinfo
  {volume} {77}},\ \bibinfo {pages} {122401} (\bibinfo {year}
  {2014})}\BibitemShut {NoStop}%
\bibitem [{\citenamefont {Mukerjee}\ \emph {et~al.}(2006)\citenamefont
  {Mukerjee}, \citenamefont {Xu},\ and\ \citenamefont {Moore}}]{subrotopaper}%
  \BibitemOpen
  \bibfield  {author} {\bibinfo {author} {\bibfnamefont {S.}~\bibnamefont
  {Mukerjee}}, \bibinfo {author} {\bibfnamefont {C.}~\bibnamefont {Xu}},\ and\
  \bibinfo {author} {\bibfnamefont {J.~E.}\ \bibnamefont {Moore}},\ }\bibfield
  {title} {\bibinfo {title} {Topological defects and the superfluid transition
  of the $s=1$ spinor condensate in two dimensions},\ }\href
  {https://doi.org/10.1103/PhysRevLett.97.120406} {\bibfield  {journal}
  {\bibinfo  {journal} {Phys. Rev. Lett.}\ }\textbf {\bibinfo {volume} {97}},\
  \bibinfo {pages} {120406} (\bibinfo {year} {2006})}\BibitemShut {NoStop}%
\bibitem [{\citenamefont {Himeda}\ \emph {et~al.}(2002)\citenamefont {Himeda},
  \citenamefont {Kato},\ and\ \citenamefont {Ogata}}]{PhysRevLett.88.117001}%
  \BibitemOpen
  \bibfield  {author} {\bibinfo {author} {\bibfnamefont {A.}~\bibnamefont
  {Himeda}}, \bibinfo {author} {\bibfnamefont {T.}~\bibnamefont {Kato}},\ and\
  \bibinfo {author} {\bibfnamefont {M.}~\bibnamefont {Ogata}},\ }\bibfield
  {title} {\bibinfo {title} {Stripe states with spatially oscillating
  $\mathit{d}$-wave superconductivity in the two-dimensional
  $\mathit{t}\ensuremath{-}{\mathit{t}}^{\ensuremath{'}}\ensuremath{-}\mathit{J}$
  model},\ }\href {https://doi.org/10.1103/PhysRevLett.88.117001} {\bibfield
  {journal} {\bibinfo  {journal} {Phys. Rev. Lett.}\ }\textbf {\bibinfo
  {volume} {88}},\ \bibinfo {pages} {117001} (\bibinfo {year}
  {2002})}\BibitemShut {NoStop}%
\bibitem [{\citenamefont {Berg}\ \emph {et~al.}(2007)\citenamefont {Berg},
  \citenamefont {Fradkin}, \citenamefont {Kim}, \citenamefont {Kivelson},
  \citenamefont {Oganesyan}, \citenamefont {Tranquada},\ and\ \citenamefont
  {Zhang}}]{PhysRevLett.99.127003}%
  \BibitemOpen
  \bibfield  {author} {\bibinfo {author} {\bibfnamefont {E.}~\bibnamefont
  {Berg}}, \bibinfo {author} {\bibfnamefont {E.}~\bibnamefont {Fradkin}},
  \bibinfo {author} {\bibfnamefont {E.-A.}\ \bibnamefont {Kim}}, \bibinfo
  {author} {\bibfnamefont {S.~A.}\ \bibnamefont {Kivelson}}, \bibinfo {author}
  {\bibfnamefont {V.}~\bibnamefont {Oganesyan}}, \bibinfo {author}
  {\bibfnamefont {J.~M.}\ \bibnamefont {Tranquada}},\ and\ \bibinfo {author}
  {\bibfnamefont {S.~C.}\ \bibnamefont {Zhang}},\ }\bibfield  {title} {\bibinfo
  {title} {Dynamical layer decoupling in a stripe-ordered high-${T}_{c}$
  superconductor},\ }\href {https://doi.org/10.1103/PhysRevLett.99.127003}
  {\bibfield  {journal} {\bibinfo  {journal} {Phys. Rev. Lett.}\ }\textbf
  {\bibinfo {volume} {99}},\ \bibinfo {pages} {127003} (\bibinfo {year}
  {2007})}\BibitemShut {NoStop}%
\bibitem [{\citenamefont {Agterberg}\ \emph {et~al.}(2020)\citenamefont
  {Agterberg}, \citenamefont {Davis}, \citenamefont {Edkins}, \citenamefont
  {Fradkin}, \citenamefont {Van~Harlingen}, \citenamefont {Kivelson},
  \citenamefont {Lee}, \citenamefont {Radzihovsky}, \citenamefont {Tranquada},\
  and\ \citenamefont {Wang}}]{agterberg2020physics}%
  \BibitemOpen
  \bibfield  {author} {\bibinfo {author} {\bibfnamefont {D.~F.}\ \bibnamefont
  {Agterberg}}, \bibinfo {author} {\bibfnamefont {J.~S.}\ \bibnamefont
  {Davis}}, \bibinfo {author} {\bibfnamefont {S.~D.}\ \bibnamefont {Edkins}},
  \bibinfo {author} {\bibfnamefont {E.}~\bibnamefont {Fradkin}}, \bibinfo
  {author} {\bibfnamefont {D.~J.}\ \bibnamefont {Van~Harlingen}}, \bibinfo
  {author} {\bibfnamefont {S.~A.}\ \bibnamefont {Kivelson}}, \bibinfo {author}
  {\bibfnamefont {P.~A.}\ \bibnamefont {Lee}}, \bibinfo {author} {\bibfnamefont
  {L.}~\bibnamefont {Radzihovsky}}, \bibinfo {author} {\bibfnamefont {J.~M.}\
  \bibnamefont {Tranquada}},\ and\ \bibinfo {author} {\bibfnamefont
  {Y.}~\bibnamefont {Wang}},\ }\bibfield  {title} {\bibinfo {title} {The
  physics of pair-density waves: cuprate superconductors and beyond},\ }\href
  {http://dx.doi.org/10.1146/annurev-conmatphys-031119-050711} {\bibfield
  {journal} {\bibinfo  {journal} {Annual Review of Condensed Matter Physics}\
  }\textbf {\bibinfo {volume} {11}},\ \bibinfo {pages} {231} (\bibinfo {year}
  {2020})}\BibitemShut {NoStop}%
\bibitem [{\citenamefont {W.S.Massey}(1973)}]{massey_1973}%
  \BibitemOpen
  \bibfield  {author} {\bibinfo {author} {\bibnamefont {W.S.Massey}},\
  }\bibfield  {title} {\bibinfo {title} {The quotient space of the complex
  projective plane under conjugation is a 4-sphere},\ }\href
  {https://doi.org/10.1007/BF00181480} {\bibfield  {journal} {\bibinfo
  {journal} {Geom Dedicata 2, 371–374}\ } (\bibinfo {year}
  {1973})}\BibitemShut {NoStop}%
\bibitem [{\citenamefont {N.H.Kuiper}(1974)}]{kuiper_1974}%
  \BibitemOpen
  \bibfield  {author} {\bibinfo {author} {\bibnamefont {N.H.Kuiper}},\
  }\bibfield  {title} {\bibinfo {title} {The quotient space of cp(2) by complex
  conjugation is the 4-sphere},\ }\href {https://doi.org/10.1007/BF01432386}
  {\bibfield  {journal} {\bibinfo  {journal} {Math. Ann. 208, 175–177}\ }
  (\bibinfo {year} {1974})}\BibitemShut {NoStop}%
\bibitem [{\citenamefont {Gupta}\ \emph {et~al.}(2023)\citenamefont {Gupta},
  \citenamefont {Mondal}, \citenamefont {Bhattacharjee},\ and\ \citenamefont
  {Saha~Dasgupta}}]{PhysRevResearch.5.043219}%
  \BibitemOpen
  \bibfield  {author} {\bibinfo {author} {\bibfnamefont {M.}~\bibnamefont
  {Gupta}}, \bibinfo {author} {\bibfnamefont {B.}~\bibnamefont {Mondal}},
  \bibinfo {author} {\bibfnamefont {S.}~\bibnamefont {Bhattacharjee}},\ and\
  \bibinfo {author} {\bibfnamefont {T.}~\bibnamefont {Saha~Dasgupta}},\
  }\bibfield  {title} {\bibinfo {title} {Ab initio insights on the fermiology
  of ${d}^{1}$ transition metals on the honeycomb lattice: Hierarchy of hopping
  pathways and spin-orbit coupling},\ }\href
  {https://doi.org/10.1103/PhysRevResearch.5.043219} {\bibfield  {journal}
  {\bibinfo  {journal} {Phys. Rev. Res.}\ }\textbf {\bibinfo {volume} {5}},\
  \bibinfo {pages} {043219} (\bibinfo {year} {2023})}\BibitemShut {NoStop}%
\bibitem [{\citenamefont {Herbut}\ and\ \citenamefont
  {Mandal}(2023)}]{PhysRevB.108.L161108}%
  \BibitemOpen
  \bibfield  {author} {\bibinfo {author} {\bibfnamefont {I.~F.}\ \bibnamefont
  {Herbut}}\ and\ \bibinfo {author} {\bibfnamefont {S.}~\bibnamefont
  {Mandal}},\ }\bibfield  {title} {\bibinfo {title} {$so(8)$ unification and
  the large-$n$ theory of superconductor-insulator transition of
  two-dimensional dirac fermions},\ }\href
  {https://doi.org/10.1103/PhysRevB.108.L161108} {\bibfield  {journal}
  {\bibinfo  {journal} {Phys. Rev. B}\ }\textbf {\bibinfo {volume} {108}},\
  \bibinfo {pages} {L161108} (\bibinfo {year} {2023})}\BibitemShut {NoStop}%
\bibitem [{\citenamefont {Cuozzo}\ \emph {et~al.}(2024)\citenamefont {Cuozzo},
  \citenamefont {Yu}, \citenamefont {Davids}, \citenamefont {Nenoff},
  \citenamefont {Soh}, \citenamefont {Pan},\ and\ \citenamefont
  {Rossi}}]{cuozzo2024leggett}%
  \BibitemOpen
  \bibfield  {author} {\bibinfo {author} {\bibfnamefont {J.~J.}\ \bibnamefont
  {Cuozzo}}, \bibinfo {author} {\bibfnamefont {W.}~\bibnamefont {Yu}}, \bibinfo
  {author} {\bibfnamefont {P.}~\bibnamefont {Davids}}, \bibinfo {author}
  {\bibfnamefont {T.~M.}\ \bibnamefont {Nenoff}}, \bibinfo {author}
  {\bibfnamefont {D.~B.}\ \bibnamefont {Soh}}, \bibinfo {author} {\bibfnamefont
  {W.}~\bibnamefont {Pan}},\ and\ \bibinfo {author} {\bibfnamefont
  {E.}~\bibnamefont {Rossi}},\ }\bibfield  {title} {\bibinfo {title} {Leggett
  modes in a dirac semimetal},\ }\href
  {http://dx.doi.org/10.1038/s41567-024-02412-4} {\bibfield  {journal}
  {\bibinfo  {journal} {Nature Physics}\ }\textbf {\bibinfo {volume} {20}},\
  \bibinfo {pages} {1118} (\bibinfo {year} {2024})}\BibitemShut {NoStop}%
\bibitem [{\citenamefont {Chamon}\ \emph {et~al.}(2012)\citenamefont {Chamon},
  \citenamefont {Hou}, \citenamefont {Mudry}, \citenamefont {Ryu},\ and\
  \citenamefont {Santos}}]{chamon2012masses}%
  \BibitemOpen
  \bibfield  {author} {\bibinfo {author} {\bibfnamefont {C.}~\bibnamefont
  {Chamon}}, \bibinfo {author} {\bibfnamefont {C.-Y.}\ \bibnamefont {Hou}},
  \bibinfo {author} {\bibfnamefont {C.}~\bibnamefont {Mudry}}, \bibinfo
  {author} {\bibfnamefont {S.}~\bibnamefont {Ryu}},\ and\ \bibinfo {author}
  {\bibfnamefont {L.}~\bibnamefont {Santos}},\ }\bibfield  {title} {\bibinfo
  {title} {Masses and majorana fermions in graphene},\ }\href
  {http://dx.doi.org/10.1088/0031-8949/2012/T146/014013} {\bibfield  {journal}
  {\bibinfo  {journal} {Physica Scripta}\ }\textbf {\bibinfo {volume} {2012}},\
  \bibinfo {pages} {014013} (\bibinfo {year} {2012})}\BibitemShut {NoStop}%
\bibitem [{\citenamefont {Scalapino}(2012)}]{scalapino2012common}%
  \BibitemOpen
  \bibfield  {author} {\bibinfo {author} {\bibfnamefont {D.~J.}\ \bibnamefont
  {Scalapino}},\ }\bibfield  {title} {\bibinfo {title} {A common thread: The
  pairing interaction for unconventional superconductors},\ }\href
  {https://link.aps.org/doi/10.1103/RevModPhys.84.1383} {\bibfield  {journal}
  {\bibinfo  {journal} {Reviews of Modern Physics}\ }\textbf {\bibinfo {volume}
  {84}},\ \bibinfo {pages} {1383} (\bibinfo {year} {2012})}\BibitemShut
  {NoStop}%
\bibitem [{\citenamefont {Brydon}\ \emph {et~al.}(2016)\citenamefont {Brydon},
  \citenamefont {Wang}, \citenamefont {Weinert},\ and\ \citenamefont
  {Agterberg}}]{PhysRevLett.116.177001}%
  \BibitemOpen
  \bibfield  {author} {\bibinfo {author} {\bibfnamefont {P.~M.~R.}\
  \bibnamefont {Brydon}}, \bibinfo {author} {\bibfnamefont {L.}~\bibnamefont
  {Wang}}, \bibinfo {author} {\bibfnamefont {M.}~\bibnamefont {Weinert}},\ and\
  \bibinfo {author} {\bibfnamefont {D.~F.}\ \bibnamefont {Agterberg}},\
  }\bibfield  {title} {\bibinfo {title} {Pairing of $j=3/2$ fermions in
  half-heusler superconductors},\ }\href
  {https://doi.org/10.1103/PhysRevLett.116.177001} {\bibfield  {journal}
  {\bibinfo  {journal} {Phys. Rev. Lett.}\ }\textbf {\bibinfo {volume} {116}},\
  \bibinfo {pages} {177001} (\bibinfo {year} {2016})}\BibitemShut {NoStop}%
\bibitem [{\citenamefont {Chiu}\ \emph {et~al.}(2016)\citenamefont {Chiu},
  \citenamefont {Teo}, \citenamefont {Schnyder},\ and\ \citenamefont
  {Ryu}}]{chiu2016classification}%
  \BibitemOpen
  \bibfield  {author} {\bibinfo {author} {\bibfnamefont {C.-K.}\ \bibnamefont
  {Chiu}}, \bibinfo {author} {\bibfnamefont {J.~C.}\ \bibnamefont {Teo}},
  \bibinfo {author} {\bibfnamefont {A.~P.}\ \bibnamefont {Schnyder}},\ and\
  \bibinfo {author} {\bibfnamefont {S.}~\bibnamefont {Ryu}},\ }\bibfield
  {title} {\bibinfo {title} {Classification of topological quantum matter with
  symmetries},\ }\href {http://dx.doi.org/10.1103/RevModPhys.88.035005}
  {\bibfield  {journal} {\bibinfo  {journal} {Reviews of Modern Physics}\
  }\textbf {\bibinfo {volume} {88}},\ \bibinfo {pages} {035005} (\bibinfo
  {year} {2016})}\BibitemShut {NoStop}%
\bibitem [{\citenamefont {Schnyder}\ \emph {et~al.}(2009)\citenamefont
  {Schnyder}, \citenamefont {Ryu}, \citenamefont {Furusaki},\ and\
  \citenamefont {Ludwig}}]{schnyder2009classification}%
  \BibitemOpen
  \bibfield  {author} {\bibinfo {author} {\bibfnamefont {A.~P.}\ \bibnamefont
  {Schnyder}}, \bibinfo {author} {\bibfnamefont {S.}~\bibnamefont {Ryu}},
  \bibinfo {author} {\bibfnamefont {A.}~\bibnamefont {Furusaki}},\ and\
  \bibinfo {author} {\bibfnamefont {A.~W.}\ \bibnamefont {Ludwig}},\ }\bibfield
   {title} {\bibinfo {title} {Classification of topological insulators and
  superconductors},\ }in\ \href {http://dx.doi.org/10.1063/1.3149481} {\emph
  {\bibinfo {booktitle} {AIP conference proceedings}}},\ Vol.\ \bibinfo
  {volume} {1134}\ (\bibinfo {organization} {American Institute of Physics},\
  \bibinfo {year} {2009})\ pp.\ \bibinfo {pages} {10--21}\BibitemShut {NoStop}%
\bibitem [{\citenamefont {Hatsugai}\ and\ \citenamefont
  {Ryu}(2002)}]{PhysRevB.65.212510}%
  \BibitemOpen
  \bibfield  {author} {\bibinfo {author} {\bibfnamefont {Y.}~\bibnamefont
  {Hatsugai}}\ and\ \bibinfo {author} {\bibfnamefont {S.}~\bibnamefont {Ryu}},\
  }\bibfield  {title} {\bibinfo {title} {Topological quantum phase transitions
  in superconductivity on lattices},\ }\href
  {https://doi.org/10.1103/PhysRevB.65.212510} {\bibfield  {journal} {\bibinfo
  {journal} {Phys. Rev. B}\ }\textbf {\bibinfo {volume} {65}},\ \bibinfo
  {pages} {212510} (\bibinfo {year} {2002})}\BibitemShut {NoStop}%
\bibitem [{\citenamefont {Hatsugai}\ \emph {et~al.}(2004)\citenamefont
  {Hatsugai}, \citenamefont {Ryu},\ and\ \citenamefont
  {Kohmoto}}]{PhysRevB.70.054502}%
  \BibitemOpen
  \bibfield  {author} {\bibinfo {author} {\bibfnamefont {Y.}~\bibnamefont
  {Hatsugai}}, \bibinfo {author} {\bibfnamefont {S.}~\bibnamefont {Ryu}},\ and\
  \bibinfo {author} {\bibfnamefont {M.}~\bibnamefont {Kohmoto}},\ }\bibfield
  {title} {\bibinfo {title} {Superconductivity and abelian chiral anomalies},\
  }\href {https://doi.org/10.1103/PhysRevB.70.054502} {\bibfield  {journal}
  {\bibinfo  {journal} {Phys. Rev. B}\ }\textbf {\bibinfo {volume} {70}},\
  \bibinfo {pages} {054502} (\bibinfo {year} {2004})}\BibitemShut {NoStop}%
\bibitem [{\citenamefont {Sato}\ and\ \citenamefont
  {Fujimoto}(2009)}]{PhysRevB.79.094504}%
  \BibitemOpen
  \bibfield  {author} {\bibinfo {author} {\bibfnamefont {M.}~\bibnamefont
  {Sato}}\ and\ \bibinfo {author} {\bibfnamefont {S.}~\bibnamefont
  {Fujimoto}},\ }\bibfield  {title} {\bibinfo {title} {Topological phases of
  noncentrosymmetric superconductors: Edge states, majorana fermions, and
  non-abelian statistics},\ }\href {https://doi.org/10.1103/PhysRevB.79.094504}
  {\bibfield  {journal} {\bibinfo  {journal} {Phys. Rev. B}\ }\textbf {\bibinfo
  {volume} {79}},\ \bibinfo {pages} {094504} (\bibinfo {year}
  {2009})}\BibitemShut {NoStop}%
\bibitem [{\citenamefont {Cornfeld}\ \emph {et~al.}(2021)\citenamefont
  {Cornfeld}, \citenamefont {Rudner},\ and\ \citenamefont
  {Berg}}]{PhysRevResearch.3.013051}%
  \BibitemOpen
  \bibfield  {author} {\bibinfo {author} {\bibfnamefont {E.}~\bibnamefont
  {Cornfeld}}, \bibinfo {author} {\bibfnamefont {M.~S.}\ \bibnamefont
  {Rudner}},\ and\ \bibinfo {author} {\bibfnamefont {E.}~\bibnamefont {Berg}},\
  }\bibfield  {title} {\bibinfo {title} {Spin-polarized superconductivity:
  Order parameter topology, current dissipation, and multiple-period josephson
  effect},\ }\href {https://doi.org/10.1103/PhysRevResearch.3.013051}
  {\bibfield  {journal} {\bibinfo  {journal} {Phys. Rev. Res.}\ }\textbf
  {\bibinfo {volume} {3}},\ \bibinfo {pages} {013051} (\bibinfo {year}
  {2021})}\BibitemShut {NoStop}%
\bibitem [{\citenamefont {Zhang}\ \emph {et~al.}(2024)\citenamefont {Zhang},
  \citenamefont {Hu},\ and\ \citenamefont {Neupert}}]{zhang2024finite}%
  \BibitemOpen
  \bibfield  {author} {\bibinfo {author} {\bibfnamefont {S.-B.}\ \bibnamefont
  {Zhang}}, \bibinfo {author} {\bibfnamefont {L.-H.}\ \bibnamefont {Hu}},\ and\
  \bibinfo {author} {\bibfnamefont {T.}~\bibnamefont {Neupert}},\ }\bibfield
  {title} {\bibinfo {title} {Finite-momentum cooper pairing in proximitized
  altermagnets},\ }\href {http://dx.doi.org/10.1038/s41467-024-45951-3}
  {\bibfield  {journal} {\bibinfo  {journal} {Nature Communications}\ }\textbf
  {\bibinfo {volume} {15}},\ \bibinfo {pages} {1801} (\bibinfo {year}
  {2024})}\BibitemShut {NoStop}%
\bibitem [{\citenamefont {Grover}\ and\ \citenamefont
  {Senthil}(2008)}]{Grover_2008}%
  \BibitemOpen
  \bibfield  {author} {\bibinfo {author} {\bibfnamefont {T.}~\bibnamefont
  {Grover}}\ and\ \bibinfo {author} {\bibfnamefont {T.}~\bibnamefont
  {Senthil}},\ }\bibfield  {title} {\bibinfo {title} {Topological spin hall
  states, charged skyrmions, and superconductivity in two dimensions},\
  }\bibfield  {journal} {\bibinfo  {journal} {Physical Review Letters}\
  }\textbf {\bibinfo {volume} {100}},\ \href
  {https://doi.org/10.1103/physrevlett.100.156804}
  {10.1103/physrevlett.100.156804} (\bibinfo {year} {2008})\BibitemShut
  {NoStop}%
\end{thebibliography}%

%%%%%%%

%%%%%%%%%%%%%%
\end{document}